\documentclass[preprint,amsmath,amssymb,aps,a4]{revtex4-2}
\usepackage{graphicx}
\usepackage{lineno}
\usepackage{bm}
\usepackage{txfonts}
\usepackage{hyperref}
\usepackage{caption}
\usepackage{amsmath}
\usepackage{amssymb}
\usepackage[section]{placeins}
\usepackage{tabularx}
\usepackage{braket}
\usepackage{multirow}
\begin{document}
\title{Photon Leading Twist Transverse Momentum Dependent Parton Distributions}
\author{Satyajit Puhan}
\email{puhansatyajit@gmail.com}
\affiliation{Computational High Energy Physics Lab, Department of Physics, Dr. B R Ambedkar National
	Institute of Technology Jalandhar, Punjab, India, 144008}
\author{Narinder Kumar}
\email{narinderhep@gmail.com}
\affiliation{Computational Theoretical High Energy Physics Lab, Department of Physics, Doaba College, Jalandhar 144004, India}
\affiliation{Computational High Energy Physics Lab, Department of Physics, Dr. B R Ambedkar National
	Institute of Technology Jalandhar, Punjab, India, 144008}
\author{Harleen Dahiya}
\email{dahiyah@nitj.ac.in}
\affiliation{Computational High Energy Physics Lab, Department of Physics, Dr. B.R. Ambedkar National
	Institute of Technology, Jalandhar, 144008, India}

\date{\today}%
\begin{abstract}
In this work, we have calculated the photon leading-twist T-even transverse momentum dependent parton distribution functions (TMDs). For these calculations, we have treated photon as a state of quark anti-quark pair and represent the TMDs in explicit form of helicity amplitudes. We have presented all the T-even TMDs for the photon being massless (real) and massive  (virtual). We have also compared our longitudinally and transversely polarized TMDs results for different photon masses. The collinear parton distribution functions (PDFs) have been predicted for both the real and virtual photon cases. The results for the unpolarized PDF of our calculations are in good agreement with other model predictions. In addition to this, we have also discussed the spin-spin correlations between the quark and photon.

 \vspace{0.1cm}
\end{abstract}
%
\maketitle
%
%
\section{Introduction\label{secintro}}
In the standard model of physics, photon appears as a gauge boson of sub-atomic physics together with $W^{\pm}$ and $Z^0$ bosons. In the theory of strong interaction: quantum chromodynamics (QCD) studies, photon has always been a fascinating object to understand. After a remarkable paper by Witten \cite{witten} on photon PDF, they have been a subject of much interest. Unlike spin-half particles, photon structure can be described by the structure function $F_2^\gamma$ which was initially measured by the PLUTO collaboration  over a wide range of energy \cite{pluto}. Thereafter, it has been experimentally measured by  various collaborations \cite{L3:1998ijt,OPAL:1999rcd,DELPHI:2000uoh,Nisius:1999cv}. The partonic content of photon can be calculated non-perturbatively in  quantum electrodynamics (QED) \cite{Friot:2006mm,ElBeiyad:2008ss} and the point like structure of the photon can be calculated perturbatively \cite{Nisius:1999cv}. However, to understand the three-dimensional structure of the photon one needs to study the structure functions of the photon. While treating photon (spin-$1$ hadron)  in QCD, it behaves as a quark-anti-quark pair in the  Fock-state decomposition \cite{Xiao:2003wf,Pasquini:2024qxn}. Parton distribution functions (PDFs) \cite{pdf,pdf2,pdf3,pdf1,ATLAS:2021qnl} carry only one-dimensional information about the internal structure of the hadrons (for photon in our case) and can be accessed by deep-inelastic scattering (DIS) where  photon is the probe instead of a hadron \cite{dis}. 
PDFs for spin-$0$, spin-$1/2$, and spin-$1$ particles have been explored widely both theoretically and experimentally \cite{Puhan:2023ekt,Meissner:2009ww,Kumano:2021fem,Kumano:2020ijt}, but there are few studies of PDFs for the photon case \cite{Friot:2006mm,Mukherjee:2013yf}. The DIS scattering cross section is used to measure the photon structure function, where photons are probed with high virtuality ($Q^2>>0$) and real photon ($P^2=0$) \cite{Nair:2023lir}. The PDFs do not carry the information about the spatial and transverse structure of photon whereas generalized parton distributions (GPDs) \cite{gpd,gpd2,Belitsky:2005qn,Kaur:2023zhn} provide spatial distribution of partons inside hadron and transverse momentum-dependent parton distributions (TMDs) \cite{tmd,tmd1,tmd2,tmd3} provide distribution of partons in momentum space. 

The GPDs provide a wide range of information about the hadron structure and spin \cite{Belitsky:2005qn}. GPDs being a function of longitudinal momentum fraction ($x$), skewness ($\zeta$) as well as momentum transferred  ($\Delta$) between initial and final state of a hadron, provide the information about charge distribution, form factors, magnetic moment, dipole moment, mechanical properties, charge radii, PDFs etc \cite{Burkert:2023wzr,Kaur:2018ewq}. GPDs of a hadron can be extracted from the deeply virtual Compton scattering (DVCS) \cite{dvcs} and deeply virtual meson production (DVMP) \cite{dvmp1,gpd2} and can be expressed as an off-forward matrix element of light-front bilocal operators. The DVCS ($\gamma^* p \rightarrow \gamma p$) process is used to study the nucleon GPDs with a nucleon target, whereas DVCS ($\gamma^* (Q) \gamma \rightarrow \gamma \gamma$), at high $Q^2$, is used to study the photon GPDs with photon as a target \cite{Friot:2006mm,Mukherjee:2013yf}.  The photon GPDs have been investigated widely in Refs. \cite{Friot:2006mm,ElBeiyad:2008ss,Mukherjee:2013yf,Nair:2023lir}. On the other hand, the generalized distribution amplitude (GDA) is comparatively  less investigated subject \cite{ElBeiyad:2008ss,Kumano:2017lhr,Kawamura:2013wfa}. 

The extended form of collinear PDFs give rise to the three-dimensional hadron TMDs. The spin densities, transverse structure, angular momentum, spin-orbit correlation, PDFs etc. of a hadron can be studied through TMDs \cite{Acharyya:2024enp,Tan:2021osk}. TMDs contain the information in the form of longitudinal momentum fraction ($x$) and transverse momentum ($\textbf{k}^2_{\perp}$) of the constituent quark. The TMDs for hadrons have been studied widely both theoretically and experimentally \cite{tmd,tmd1,tmd2,tmd3, Cerutti:2022lmb}, however,  for photon, it has been investigated in basis light-front quantization (BLFQ) \cite{Nair:2023lir, Nair:2022evk}. Experimentally, TMDs can be extracted from semi-inclusive deep inelastic scattering (SIDIS)\cite{sidis,sidis1,sidis2}, DIS at high energy \cite{dis}, Drell-Yan \cite{drell,drell1,drell2,drell3,drell4}, and $Z^0/W^{\pm}$ production \cite{z1,z2,z3} processes for hadrons. There are total two leading  twist TMDs for spin-$0$ hadrons,  the unpolarized TMD $f_1(x, \textbf{k}^2_{\perp})$ and the Boer-Mulder's TMD $h^\perp_1(x, \textbf{k}^2_{\perp})$ \cite{Meissner:2008ay}. The $f_1(x, \textbf{k}^2_{\perp})$ is the only T-even TMDs for spin-$0$ hadrons. While looking into the spin-$1/2$ nucleon, there are total eight TMDs,  six out of which are T-even TMDs at the leading twist \cite{spt,spt1,spt2}. As photon is a spin-$1$ system, there are total eighteen TMDs at the leading twist. In addition to six T-even TMDs for spin-$1/2$, there are three extra tensor $f_{1LL}(x,{\textbf{k}^2_\perp})$, $f_{1LT}(x, {\textbf{k}^2_\perp})$ and $f_{1TT}(x,{\textbf{k}^2_\perp})$ TMDs that come into the picture  \cite{bse,Kaur:2020emh,Kumano:2021fem,Ninomiya:2017ggn,Puhan:2023hio}. Even in the case of spin-$1$ hadron, there is an extra $f_{1LL}$ PDF as compare to nucleon case. Even though the spin-$1$ TMDs have been explored for the case of $\rho$-meson in  Nambu--Jona-Lasinio (NJL) model \cite{Ninomiya:2017ggn}, light-front quark model (LFQM) \cite{Kaur:2020emh}, light-front holographic model (LFHM) \cite{Kaur:2020emh} and in BSE model \cite{bse}, the case of photon still remains less explored.
 
In this work, we have considered photon as a leading state of quark-anti-quark pair where we have neglected the point-like photon contribution as $|\gamma \rangle = \sum | q \bar{q} \rangle \psi_{q \bar{q}}+....$ \cite{Xiao:2003wf,Pasquini:2024qxn}. The work has been proceeded using the light-front formalism, where the photon wave function is derived using  the light-front expansion of the minimal Fock states \cite{Xiao:2003wf}. The photon wave function is derived following Brodsky {\it et al.} \cite{Brodsky:2000ii} where the wave function was derived for relativistic QED composite systems by giving explicit light-front wave functions for the two-particle Fock states of the electron in QED. These wave functions for photon have been successful in describing the pion-photon and photon-pion transition form factors \cite{Xiao:2003wf}. We have considered both real and virtual photon. For the time-like virtual photon case, the mass square of the photon is taken as positive, while mass square is taken as negative for the space-like virtual photon.

We have calculated all the T-even TMDs for photon by  representing them in the form of light-front helicity amplitude (LFHA), where the explicit form of TMDs have been calculated by introducing different spin polarizations. We have computed the TMDs for both massless (real) photon and massive (virtual) photon cases. In both the cases, the $h^\perp_{1T} (x,\textbf{k}^2_{\perp})$ and tensor $f_{1LL} (x,\textbf{k}^2_{\perp})$ TMDs are coming out to be zero. Due to the factor of photon mass ($M_\gamma$) in the numerator, six out of nine TMDs come out to be zero for real photons. The four PDFs $f_1(x)$, $g_1(x)$, $h_1(x)$ and $f_{1LL}(x)$ have been discussed along with spin densities for real photon case. We have also compared the available model predictions for the case of $f_1(x)$.

This manuscript is arranged as follows. In Section \ref{LFQMe}, we have discussed the spin and momentum space wave functions for photon. We have calculated different photon polarizations. In Section \ref{tmd_defi}, we have defined the leading twist TMDs and the correlation functions for the spin-1 hadrons at lower Fock-state. Further, in Section \ref{tmd_overlap}, the basic formalism of LFWFs has been given. We have represented all possible photon TMDs in the form of light-front amplitude with their explicit overlap forms. In Section \ref{results}, we have given the details of numerical results with the help of three-dimensional and two-dimensional plots for both massless and massive photon. In Section \ref{pdF}, we have discussed the collinear PDFs. We have also compared the PDFs for different photon mass along with other model predictions. Finally,  we have summarized our results in Section \ref{conclude}.

 \section{Methodology}
 \subsection{Light-front formalism for photon}\label{LFQMe}
For the relativistic description of hadrons in terms of quark and gluon degrees of freedom, the light-front formalism provides a convenient framework.
In this work, we have considered photon (in QCD) as a  quark anti-quark pair like any meson \cite{Xiao:2003wf,Pasquini:2024qxn}
\begin{eqnarray}
    |\gamma\rangle = \sum   | q \bar q \rangle \Psi _{q \bar q}+.......
    \label{eigen}
\end{eqnarray}
One can also adopt the photon eigenstate as the excitation of electron positron Fock-state \cite{Nair:2022evk,Nair:2023lir} while treating photon in QED and we have
\begin{eqnarray}
    | \gamma \rangle = | \gamma \rangle \Psi _{\gamma}  + |e^- e^+ \rangle \Psi _{e^- e^+}.
\end{eqnarray}
However, in this work, we have taken Eq. (\ref{eigen}) as the eigenstate for the photon behaving as a spin-$1$ particle for simplification of our calculations. The two-particle Fock state representation for the photon with different isospin ($\Lambda= \pm 1, 0$) with all possibilities of helicities of quark and anti-quark are expressed as \cite{Xiao:2003wf}
\begin{eqnarray}
    |\Psi_{\gamma} (P^+, P_{\perp},h_{q},h_{\bar q})\rangle &=& \int \frac{dx d^2\textbf{k}_{\perp}}{16 \pi^3} [\Psi^{\Lambda} (x,\textbf{k}_{\perp},\uparrow,\downarrow)|(x P^+,\textbf{k}_{\perp},\uparrow,\downarrow)\rangle\nonumber \\
    && + \Psi^{\Lambda} (x,\textbf{k}_{\perp},\uparrow,\uparrow)|(x P^+,\textbf{k}_{\perp},\uparrow,\uparrow)\rangle +\Psi^{\Lambda} (x,\textbf{k}_{\perp},\downarrow,\uparrow)|(x P^+,\textbf{k}_{\perp},\downarrow,\uparrow)\rangle\nonumber \\
    &&+ \Psi^{\Lambda} (x,\textbf{k}_{\perp},\downarrow,\downarrow)|(x P^+,\textbf{k}_{\perp},\downarrow,\downarrow)\rangle].
    \label{photon}
\end {eqnarray}
Here, $P=(P^+, P^-,{\bf P_{\perp}})$ and $k=(k^+, k^-,k_{\perp})$ are the four vector momentum of the photon and active quark respectively. $x={k^+}/{P^+}$ is the longitudinal momentum fraction carried by the quark, $h_{q(\bar q)}$ are the different helicity possibilities of quark (anti-quark) respectively. In the above equation, $\Psi^{\Lambda} (x,\textbf{k}_{\perp},h_{q},h_{\bar q})$ is the dynamic spin wave functions and can be expressed as \cite{Xiao:2003wf}
\begin{eqnarray}
    \Psi^{\Lambda} (x,\textbf{k}_{\perp},h_{q},h_{\bar q}) &=& \chi^{\Lambda} (x,\textbf{k}_{\perp},h_{q},h_{\bar q}) \phi_{\gamma}(x, \textbf{k}_{\perp}),
    \label{wave}
\end{eqnarray}
where $\phi_{\gamma}(x, \textbf{k}_{\perp})$ in Eq. (\ref{wave}) is the momentum space wave function for photon can be expressed as \cite{Xiao:2003wf,Brodsky:2000ii}
\begin{eqnarray}
	\phi_{\gamma}(x, \textbf{k}_{\perp})=\frac{e_{q}}{M_{\gamma}-(\textbf{k}^2_{\perp}+m^2)/x -(\textbf{k}^2_{\perp}+m^2)/(1-x)}.
	\label{mome}
\end{eqnarray}
Here $e^2_{q}= 4 \pi \alpha $, with $\alpha$  being the strong coupling constant having value $1/137$. The first term on the RHS of Eq. (\ref{wave}) $\chi^{\Lambda} (x,\textbf{k}_{\perp},h_{q},h_{\bar q})$ is the Lorentz invariant spin structure of the photon expressed  by accounting the photon-quark-anti-quark vertex as \cite{Kaur:2020emh,Xiao:2003wf}
\begin{eqnarray}
\chi^{\Lambda}(x,{\bf k}_\perp,h_q, h_{\bar{q}})= \frac{\bar{u}_{h_q}(k^+,{\bf k}_\perp)}{\sqrt{x}} \, \epsilon_\Lambda \cdot \gamma \, \frac{v_{h_{\bar{q}}}(k^{\prime +},{\bf k}^{\prime}_\perp)}{\sqrt{1-x}}\,.
\label{spin-structure}
\end{eqnarray} 
Here, $k$ and $k^{\prime}$ denote the four momenta of quark and anti-quark respectively, $(1-x)$ is the momentum fraction carried by the anti-quark from photon and $\epsilon_\Lambda$ is the polarization vector of the photon. 

For longitudinally polarized photon $\Lambda=L$ and we have
\begin{eqnarray}
    \epsilon_L=\left(\frac{P^+}{M_{\gamma}},-\frac{M_{\gamma}}{P^+},0,0\right)\,,
    \label{long}
\end{eqnarray}
where $M_{\gamma}$ is the mass of the photon. For real photon, $M^2_{\gamma}=0$, whereas $M^2_{\gamma}>0$ and $M^2_{\gamma}<0$ for the case of time-like and space like virtual photon respectively \cite{Nair:2023lir}.
Similarly, for the transversely polarized photon $\Lambda=T(\pm)$ and we have
\begin{eqnarray}
    \epsilon^{\pm}_\perp=\mp\frac{1}{\sqrt{2}}\left(0,0,1,\pm i\right).
    \label{trans}
\end{eqnarray}

Substituting Eq. (\ref{long}) in Eq. (\ref{spin-structure}), we have obtained the spin wave function for the longitudinally polarized photon $\Lambda=L$ as \cite{Xiao:2003wf}

\begin{eqnarray}
    \chi^{L}(x,{\bf k}_\perp,\uparrow, \uparrow) &=& -\frac{\sqrt{2}m}{x(1-x)},\ \ \  \ \ \ [L_{z}=-1] \\
    \chi^{L}(x,{\bf k}_\perp,\uparrow, \downarrow) &=& -\frac{\sqrt{2} k_{R}}{1-x}, \ \ \ \ \ \ \ [L_{z}=0] \\
    \chi^{L}(x,{\bf k}_\perp,\downarrow, \uparrow) &=& \frac{\sqrt{2} k_{R}}{x}, \ \ \ \ \ \ \ [L_{z}=0] \\
    \chi^{L}(x,{\bf k}_\perp,\downarrow, \downarrow) &=& 0. \ \ \ \ \ \ \ [L_{z}=1]
    \label{zero}
\end{eqnarray}
Here $k_{R(L)}= k^1\pm i k^2$, $L_{z}$ is the possible orbital angular momentum (OAM) of the photon and $m$ is the quark (anti-quark) mass.

For the transverse polarization of photon $\Lambda=T(+)$, the spin wave function can be expressed by substituting Eq. (\ref{trans}) in Eq. (\ref{spin-structure}) as
\begin{eqnarray}
    \chi^{T(+)}(x,{\bf k}_\perp,\uparrow, \uparrow) &=& -\frac{\sqrt{2}m}{x(1-x)},\ \ \  \ \ \ [L_{z}=0] \\
    \chi^{T(+)}(x,{\bf k}_\perp,\uparrow, \downarrow) &=& -\frac{\sqrt{2} k_{R}}{1-x}, \ \ \ \ \ \ \ [L_{z}=1] \\
    \chi^{T(+)}(x,{\bf k}_\perp,\downarrow, \uparrow) &=& \frac{\sqrt{2} k_{R}}{x}, \ \ \ \ \ \ \ [L_{z}=1] \\
    \chi^{T(+)}(x,{\bf k}_\perp,\downarrow, \downarrow) &=& 0. \ \ \ \ \ \ \ [L_{z}=2]
    \label{plusone}
\end{eqnarray}
Similarly, for $\Lambda=T(-)$, the spin wave function can be expressed as 
\begin{eqnarray}
    \chi^{T(-)}(x,{\bf k}_\perp,\uparrow, \uparrow) &=& 0,\ \ \  \ \ \ [L_{z}=-2] \\
    \chi^{T(-)}(x,{\bf k}_\perp,\uparrow, \downarrow) &=& -\frac{\sqrt{2} k_{L}}{x}, \ \ \ \ \ \ \ [L_{z}=-1] \\
    \chi^{T(-)}(x,{\bf k}_\perp,\downarrow, \uparrow) &=& \frac{\sqrt{2} k_{L}}{1-x}, \ \ \ \ \ \ \ [L_{z}=-1] \\
    \chi^{T(-)}(x,{\bf k}_\perp,\downarrow, \downarrow) &=& -\frac{\sqrt{2}m}{x(1-x)}. \ \ \ \ \ \ \ [L_{z}=2]
    \label{minusone}
\end{eqnarray}
It is important to mention here that the above spin wave functions for the polarized photon satisfy the spin sum rule  $\Lambda=h_{q}+h_{\bar q}+L_{z}$. The different possibilities of OAM for different $\Lambda$ and $h_{q (\bar q)}$ are presented in Table \ref{table1}. For the case of photon, different configurations of
the light-front wave function with the OAM  $L_z = 0$, $\pm1$ and $\pm2$ correspond to the S, P and
D wave components respectively.
\begin{table}
\centering
\begin{tabular}{|c|c|c|c|}
\hline
 $\Lambda$ & $h_{ q}$ & $h_{\bar q}$ & $L_{z}$ \\
(L, T($\pm$)) & (quark helicity) & (anti-quark helicity) & (OAM)\\
\hline
 & 1/2 & 1/2 &  0 \\
\cline{2-4}
 & 1/2 & $-$1/2 &  +1 \\
 \cline{2-4}
T=+1 & $-$1/2 & +1/2 &  +1 \\
 \cline{2-4}
 & $-$1/2 & $-$1/2 &  +2 \\
 \hline
  & 1/2 & 1/2 &  $-$2 \\
\cline{2-4}
 & 1/2 & $-$1/2 &  $-$1 \\
 \cline{2-4}
T=$-$1 & $-$1/2 & +1/2 &  $-$1 \\
 \cline{2-4}
 & $-$1/2 & $-$1/2 &  0 \\
 \hline
  & 1/2 & 1/2 &  $-$1 \\
\cline{2-4}
 & 1/2 & $-$1/2 &  0 \\
 \cline{2-4}
L & $-$1/2 & +1/2 &  0 \\
 \cline{2-4}
 & $-$1/2 & $-$1/2 &  +1 \\
 \hline
\end{tabular}
\caption{All possible orbital angular momentum $L_{z}$ values with different quark (anti-quark) spin projections $h_{q (\bar q)}$ of photon polarizations $\Lambda$.}
\label{table1}
\end{table}

\subsection{Transverse momentum-dependent parton distributions for spin-1 particles}\label{tmd_defi}
 The transverse momentum-dependent quark  correlation function for spin-$1$ particle TMDs can be expressed as \cite{drell2,bse,tmd11,tmd12,tmd13,tmd14,tmd15}
\begin{eqnarray}
\Theta_{i j}^{(\Lambda)_{\bf \mathcal{S}}}(x, {\bf k}^2_\perp)&\equiv& \epsilon^*_{\Lambda (\mu)}(P) \ \Theta_{ij}^{\mu\nu}(x,{\bf k}_\perp) \ \epsilon_{\Lambda (\nu)}(P)\nonumber\\
&=&
\int \frac{{\rm d}z^- \, {\rm d}^2 {\bf z}_\perp}{(2\pi)^3} \, 
e^{\iota (k \cdot z)} \ {}_{\bf {\mathcal{S}}}\langle \Psi_{\gamma} (P^+, {\bf P}_{\perp}, \Lambda) | \bar \psi_{j}(0) \ \mathcal{W}(0,z) \
\psi_{i}(z) | \Psi_{\gamma} (P^+, {\bf P}_{\perp}, \Lambda) \rangle_{\bf {\mathcal{S}}, z^+=0}, \nonumber\\&&
\label{phi1}
\end{eqnarray}
where ${\bf k_{\perp}}$ and $k^+$ are the transverse and longitudinal momentum carried by the active quark, $z$ is the position  four vector  and is expressed as, $z=(z^+,z^-,{\bf z_{\perp}})$ in light-front framework. Here, $\psi_{i(j)}$ is the flavor SU(2) quark field operator with \textit{i} and \textit{j} being the the Dirac indices. $\mathcal{W}(0,z)$ is the gauge link \cite{tmd14,tmd15}. For simplicity, we have taken the lowest order gauge link as unity to study the T-even TMDs. $P$ is the four-vector momentum of photon and is expressed as
\begin{eqnarray*}
  P=\left(P^+,P^-,{\bf P}_\perp \right)=\left(P^+, \frac{M_{\gamma}}{P^+},{\bf 0}_\perp\right). 
\end{eqnarray*}
 In Eq. (\ref{phi1}), $\Theta_{ij}^{\mu\nu}$ is the polarization-independent Lorentz tensor matrix, $\epsilon_{\mu(\nu)}$ is the polarization four vector and the $| \Psi_{\gamma} (P^+, P_{\perp}, \Lambda) \rangle_{\bf {\mathcal{S}}}$ state indicates that the spin projection of the target photon in  in $S=(S_{L},S_{T})$ direction with helicities $\Lambda=\pm 1,0$. There are total nine T-even TMDs at the leading twist which are expressed with unpolarized ($U$), transversely polarized ($T$) and longitudinally polarized ($L$) target. We have
\begin{eqnarray}
\epsilon^{*}_{\Lambda(\mu)} (P)~\langle \gamma^+ \rangle^{\mu \nu}_{\mathcal{S}} (x,{\bf k^2_\perp})~\epsilon_{\Lambda(\nu)}(P) &=&
f_1(x,{\bf k^2_{\perp}}) + S_{LL}f_{1LL}(x,\bf{k}^2_{\perp})\nonumber\\
&&+\frac{{\bf \mathcal{S}}_{LT}\cdot{\bf k_\perp}}{M_\gamma}\, f_{1LT}(x, {\bf k^2_{\perp}})
+ \frac{ \bf{k}_{\perp} \cdot {\bf \mathcal{S}}_{TT}\cdot{\bf k_\perp} }{M_\gamma^2}f_{1TT}(x, {\bf k^2_{\perp}}),\nonumber\\
&& \label{f1} \\
\epsilon^{*}_{\Lambda(\mu)} (P)~\langle \gamma^+ \gamma_5 \rangle^{\mu \nu}_{\mathcal{S}} (x,{\bf k}^2_\perp)~\epsilon_{\Lambda(\nu)}(P)&=&
\mathcal{S}_L\,g_{1L}(x,{\bf k}^2_\perp)+ \frac{{\bf k}_\perp \cdot {\mathcal{S}}_\perp}{M_\gamma} g_{1T}(x,{\bf k}^2_\perp),\\
\epsilon^{*}_{\Lambda(\mu)} (P)~\langle \gamma^+ \gamma^i \gamma_5 \rangle^{\mu \nu}_{\mathcal{S}} (x,{\bf k}^2_\perp)~\epsilon_{\Lambda(\nu)}(P)
&=& \mathcal{S}_\perp^i h_1(x,{\bf k}^2_\perp)+\mathcal{S}_L\frac{k_\perp^i}{M_\gamma} h_{1L}^\perp (x,{\bf k}^2_\perp) \nonumber\\
&+& \frac{1}{2 M_\gamma^2}\bigg(2\,k_\perp^i {\bf k}_\perp \cdot {\mathcal{S}}_\perp - \mathcal{S}_\perp^i~{\bf k}_\perp^2\bigg) \ h_{1T}^{\perp}(x,{\bf k}_\perp^2),
\label{f3}
\end{eqnarray}
with 
\begin{eqnarray*}
\mathcal{S}_{LL}&=&\left(3 \Lambda^2-2 \right)\left(\frac{1}{6}-\frac{1}{2} \mathcal{S}_L^2\right),\\
\mathcal{S} ^i_{LT}&=&\left(3 \Lambda^2-2 \right)\mathcal{S}_L \mathcal{S}^i_\perp,\\
\mathcal{S}_{TT}^{ij}&=&\left(3 \Lambda^2-2 \right)(\mathcal{S}_\perp^i \mathcal{S}_\perp^j-\frac{1}{2}\mathcal{S}_\perp^2~ \delta^{ij})\,.
\end{eqnarray*}
The Lorentz tensor is defined as
\begin{eqnarray*}
\epsilon^{*}_{\Lambda(\mu)} (P)~\langle \Gamma\rangle^{\mu \nu}(x,{\bf k}^2_\perp)~\epsilon_{\Lambda(\nu)}(P)=\frac{1}{2} {\rm Tr}_{D}\left(\Gamma \Theta^{(\Lambda)_{\bf \mathcal{S}}}(x, {\bf k}^2_\perp)\right).
\label{ap}
\end{eqnarray*}
In the above equations, $\mathcal{S}^{i(j)}_\perp$
symbolizes the transverse polarization of the target meson in the direction $i(j)$ or $x(y)$. The functions $\textit{f}$, $\textit{g}$ and $\textit{h}$ denote the unpolarized, longitudinally polarized and transversely polarized quark within the photon.
The Dirac matrix $\Gamma$ for leading twist are $\gamma^+$, $\gamma^+\gamma_5$ and $ \gamma^+\gamma^i\gamma_5$ with $\textit{i}=(1,2)$. The subscript $1$ in the functions $\textit{f}$, $\textit{g}$, $\textit{h}$ (Eqs. (\ref{f1})-(\ref{f3})) denotes the leading  twist TMDs. The longitudinal and transverse photon polarizations have been represented as $L$ and $T$ respectively. When compared with the spin-$\frac{1}{2}$ case of nucleons, the spin-1 photon has three extra  tensor polarized TMD functions $f_{1LT}$, ${S}_{LT}$ and $f_{1TT}$. The $f_{1} (x,\textbf{k}^2_{\perp})$, $g_{1L} (x,\textbf{k}^2_{\perp})$ and $h_{1} (x,\textbf{k}^2_{\perp})$ functions are unpolarized, longitudinally polarized and transversely polarized photon TMDs.

\subsection{Overlap form of the LFWFs} \label{tmd_overlap}
All these T-even TMDs can be expressed in the form of LF helicity amplitudes. By considering different quark (anti-quark) spin and photon polarization, the overlap form of the LFWFs in terms of the LF helicity amplitudes can be expressed as \cite{bse,Kaur:2020emh,Puhan:2023hio}
\begin{eqnarray}
\mathcal{A}_{h^\prime_q \Lambda^\prime, h_q \Lambda}(x,{\bf k}^2_\perp)&=&\frac{1}{(2 \pi)^3} \sum_{ h_{\bar{q}}} \Psi^{\Lambda^\prime *}_{h^\prime_q, h_{\bar{q}}}(x,{\bf k}^2_\perp)\,\Psi^{\Lambda}_{h_q, h_{\bar{q}}}(x,{\bf k}^2_\perp)\, .
\label{O1}
\end{eqnarray}
The explicit overlap form for all the T-even TMDs in terms of helicity amplitudes are given as \cite{bse,Puhan:2023hio}
\begin{eqnarray}
f_{1}(x,\mathbf{k}_{\perp}^2) &=& \frac{1}{6}(\mathcal{A}_{\uparrow L,\uparrow L}+\mathcal{A}_{\downarrow L,\downarrow L}+\mathcal{A}_{\uparrow T(+),\uparrow T(+)}+\mathcal{A}_{\downarrow T(+),\downarrow T(+)}+\mathcal{A}_{\uparrow T(-),\uparrow T(-)}+ \nonumber\\
&& \mathcal{A}_{\downarrow T(-),\downarrow T(-)}), \\
g_{1L}(x,\mathbf{k}_{\perp}^2) &=& \frac{1}{4}(\mathcal{A}_{\uparrow T(+),\uparrow T(+)}-\mathcal{A}_{\downarrow T(+),\downarrow T(+)}-\mathcal{A}_{\uparrow T(-),\uparrow T(-)}+\mathcal{A}_{\downarrow T(-),\downarrow T(-)}), \\
g_{1T}(x,\mathbf{k}_{\perp}^2) &=& \frac{M_{\gamma}}{4\sqrt{2}\mathbf{k}_{\perp}^2}\left(k_R(\mathcal{A}_{\uparrow T(+),\uparrow L}-\mathcal{A}_{\downarrow T(+),\downarrow L}+\mathcal{A}_{\uparrow L,\uparrow T(-)}-\mathcal{A}_{\downarrow L,-\downarrow T(-)}) \right. \nonumber \\
&&\left. + k_L(\mathcal{A}_{\uparrow L,\uparrow T (+)}-\mathcal{A}_{\downarrow L,\downarrow T(+)}+\mathcal{A}_{\uparrow T(-),\uparrow L}-\mathcal{A}_{\downarrow T(-),\downarrow L})\right), \\
h_{1}(x,\mathbf{k}_{\perp}^2) &=& \frac{1}{4\sqrt{2}}(\mathcal{A}_{\uparrow T(+),\downarrow L}+\mathcal{A}_{\downarrow L,\uparrow T(+)}+\mathcal{A}_{\uparrow L,\downarrow T(-)}+\mathcal{A}_{\downarrow T(-),\uparrow L}), \\
h_{1L}^{\perp}(x,\mathbf{k}_{\perp}^2) &=& \frac{M_{\gamma}}{4\mathbf{k}_{\perp}^2}\left(k_R(\mathcal{A}_{\downarrow T(+),\uparrow T(+)}-\mathcal{A}_{\downarrow T(-),\uparrow T(-)}) + k_L(\mathcal{A}_{\uparrow T(+),\downarrow T(+)}-\mathcal{A}_{\uparrow T(-),\downarrow T(-)})\right), \\
h_{1T}^{\perp}(x,\mathbf{k}_{\perp}^2) &=& \frac{M_{\gamma}^2}{2\sqrt{2}\mathbf{k}_{\perp}^4}\left(k_R^2(\mathcal{A}_{\downarrow T(+),\uparrow L}+A_{\downarrow L,\uparrow T(-)}) + k_L^2(\mathcal{A}_{\uparrow L,\downarrow T(+)}+\mathcal{A}_{\uparrow T(-),\downarrow L})\right), \\
f_{1LL}(x,\mathbf{k}_{\perp}^2) &=& \frac{1}{2}\mathcal{A}_{\uparrow L,\uparrow L}+\mathcal{A}_{\downarrow L,\downarrow L}-\frac{1}{4}(\mathcal{A}_{\uparrow T(+),\uparrow T(+)}+\mathcal{A}_{\downarrow T(+),\downarrow T(+)}+\mathcal{A}_{\uparrow T(-),\uparrow T(-)}+\mathcal{A}_{\downarrow T(-),\downarrow T(-)}), \\
f_{1LT}(x,\mathbf{k}_{\perp}^2) &=& \frac{M_{\gamma}}{4\sqrt{2}\mathbf{k}_{\perp}^2}\left(k_R(\mathcal{A}_{\uparrow T(+),\uparrow L}+\mathcal{A}_{\downarrow T(+),\downarrow L}-\mathcal{A}_{\uparrow L,\uparrow T(-)}-\mathcal{A}_{\downarrow L,\downarrow T(-)}) \right. \nonumber \\
&&\left. + k_L(\mathcal{A}_{\uparrow L,\uparrow T(+)}+\mathcal{A}_{\downarrow L,\downarrow T(-)}-\mathcal{A}_{\uparrow T(-),\uparrow L}-\mathcal{A}_{\downarrow T(-),\downarrow L})\right), \\
f_{1TT}(x,\mathbf{k}_{\perp}^2) &=& \frac{M_{\gamma}^2}{4\sqrt{2}\mathbf{k}_{\perp}^4}\left(k_R^2(\mathcal{A}_{\uparrow T(+),\uparrow T(-)}+\mathcal{A}_{\downarrow T(+),\downarrow T(-)}) + k_L^2(\mathcal{A}_{\uparrow T(-),\uparrow T(+)}+\mathcal{A}_{\downarrow T(-),\downarrow T(+)})\right).
\label{a12}
\end{eqnarray}
Introducing spin and momentum wave functions corresponding to each overlap form of T-even TMDs helicity amplitudes, we get the expressions as
\begin{eqnarray}
    f_{1}(x,\mathbf{k}_{\perp}^2) &=& \frac{1}{(2\pi)^3} \bigg(\frac{m^2+ (x^2 +(1-x)^2) \mathbf{k}_{\perp}^2}{x^2 (1-x)^2}\bigg)\phi^2_{\gamma}(x,\mathbf{k}_{\perp}),
    \\
    g_{1L}(x,\mathbf{k}_{\perp}^2) &=& \frac{1}{(2\pi)^3} \bigg(\frac{m^2+ (x^2 -(1-x)^2) \mathbf{k}_{\perp}^2}{x^2 (1-x)^2}\bigg)\phi^2_{\gamma}(x,\mathbf{k}_{\perp}),
     \label{f2}
    \\
    \end{eqnarray}
    \begin{eqnarray}
    g_{1T}(x,\mathbf{k}_{\perp}^2) &=& \frac{M_\gamma}{\sqrt{2}(2\pi)^3} \bigg(\frac{2x-1}{x^2 (1-x)^2}\bigg) \phi^2_{\gamma}(x,\mathbf{k}_{\perp}),\\
     h_{1}(x,\mathbf{k}_{\perp}^2) &=& -\frac{m}{\sqrt{2
     }(2\pi)^3} \bigg(\frac{1}{x^2 (1-x)}\bigg)\phi^2_{\gamma}(x,\mathbf{k}_{\perp}),
      \label{f4}
      \\
      h^{\perp}_{1L}(x,\mathbf{k}_{\perp}^2) &=& -\frac{2}{(2\pi)^3} \bigg(\frac{m M_\gamma}{x^2 (1-x)}\bigg) \phi^2_{\gamma}(x,\mathbf{k}_{\perp}),
       \label{f5}\\
      h^{\perp}_{1T}(x,\mathbf{k}_{\perp}^2) &=& 0,
       \label{f6}
     \\
      f_{1LL}(x,\mathbf{k}_{\perp}^2) &=& 0,
        \label{f7}\\
        f_{1LT}(x,\mathbf{k}_{\perp}^2) &=& \frac{M_\gamma}{\sqrt{2}(2 \pi)^3} \frac{(2x-1)^2}{x^2 (1-x)^2}\phi^2_{\gamma}(x,\mathbf{k}_{\perp}),
         \label{f8}\\
     f_{1TT}(x,\mathbf{k}_{\perp}^2) &=& \frac{M^2_\gamma}{(2 \pi)^3} \frac{(2x^2-2x+1)}{x^2 (1-x)^2}\phi^2_{\gamma}(x,\mathbf{k}_{\perp}).
      \label{f9}
\end{eqnarray}
The OAM along z-axis is conserved for all the TMDs. The functions $f_1$, $g_{1L}$, $h^{\perp}_{1L}$ and $f_{1LL}$ have zero OAM transfer from initial state to final state photon. There is one unit OAM transfer in the case of $h^\perp_{1L}$, $g_{1T}$ and $f_{1LT}$ TMDs, However, in the case of $f_{1TT}$ TMDs, two units of OAM transferred. The $h_{1T}$ and $f_{1LL}$ TMDs vanish in our case. 
\subsection{Numerical results}\label{results}
\begin{figure}[ht]
	\centering
	\begin{minipage}[c]{1\textwidth}\begin{center}
	(a)\includegraphics[width=.45\textwidth]{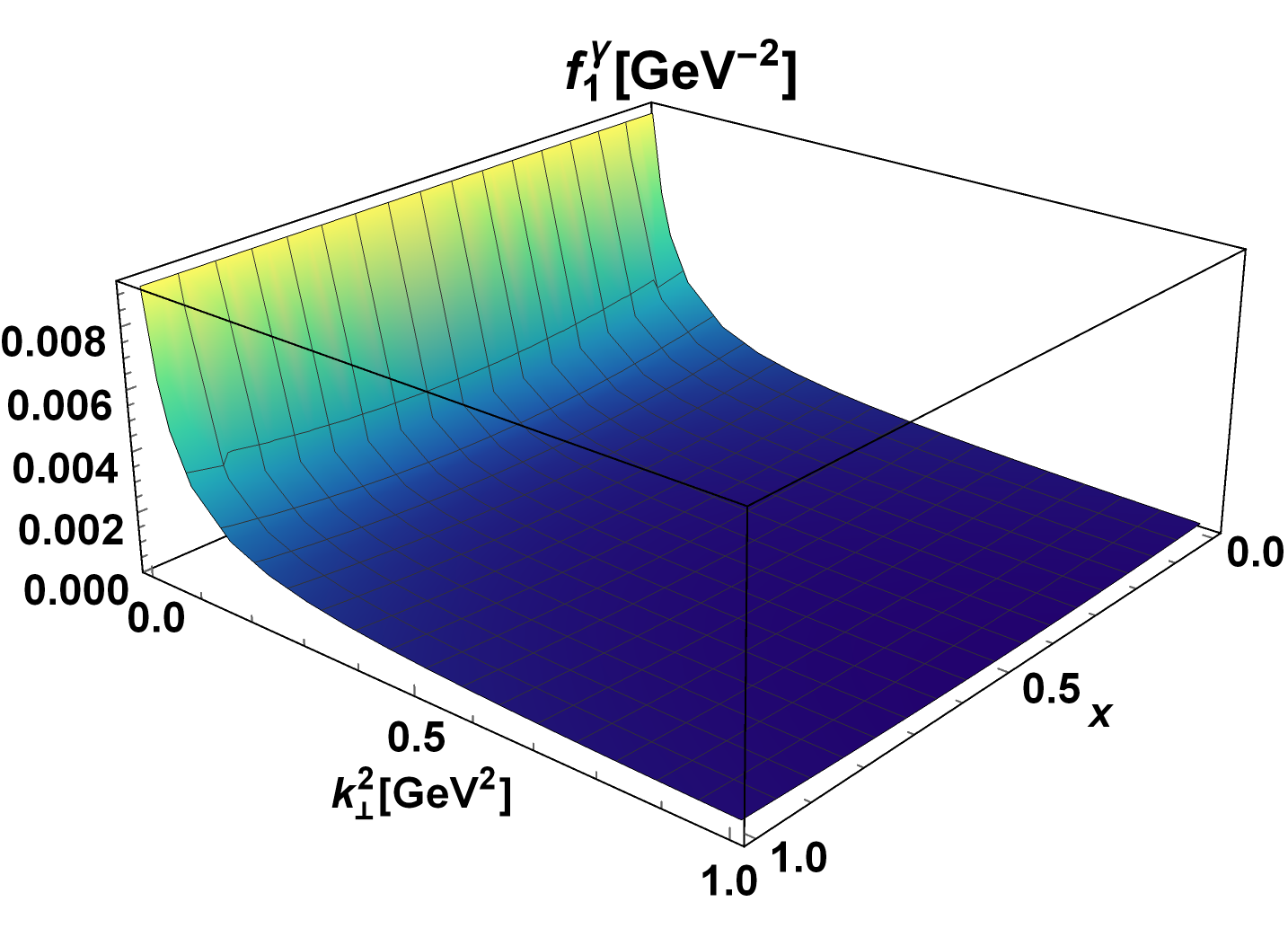}
		(b)\includegraphics[width=.45\textwidth]{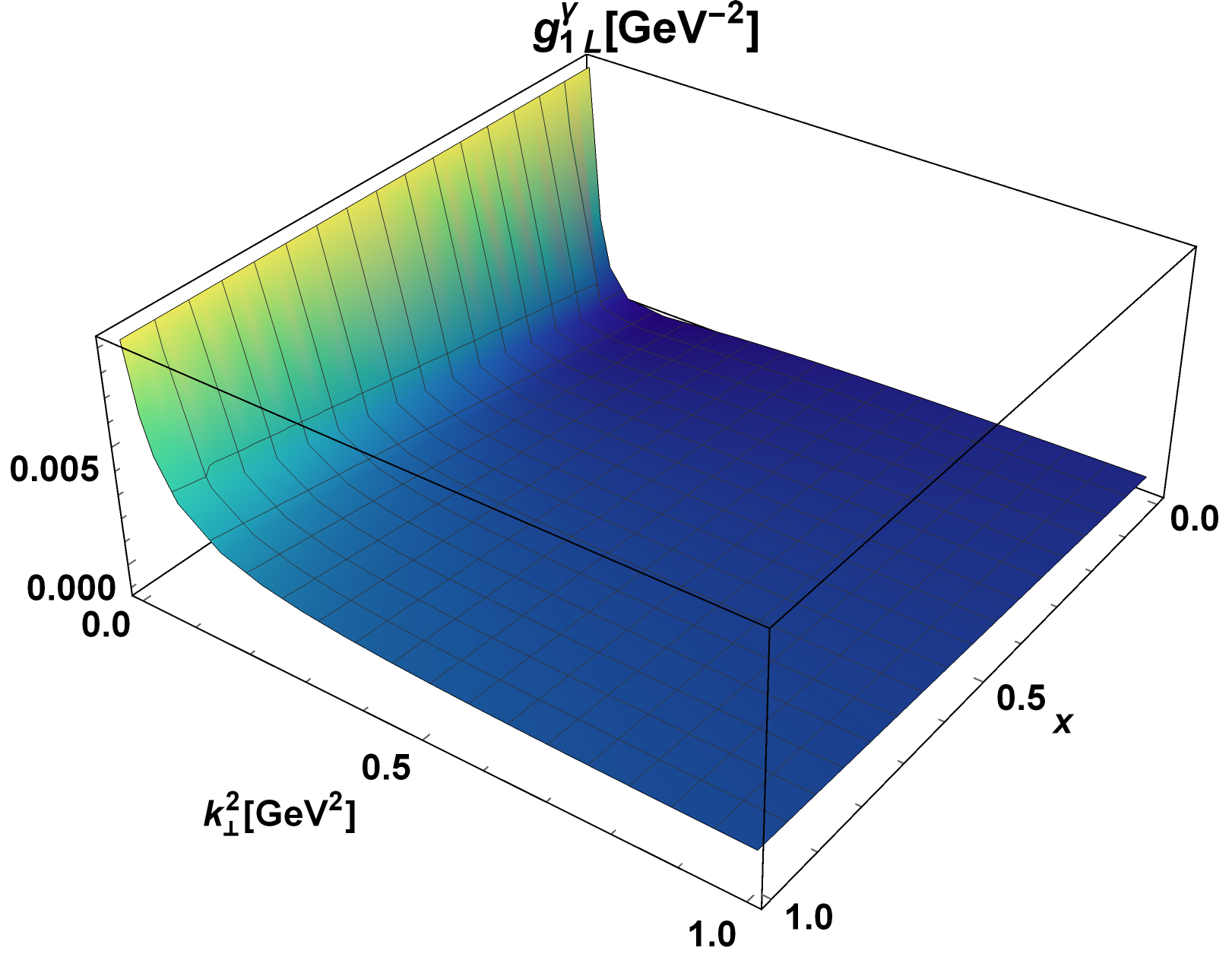}
   \end{center}
	\end{minipage}
	\begin{minipage}[c]{1\textwidth}\begin{center}
			(c)\includegraphics[width=.45\textwidth]{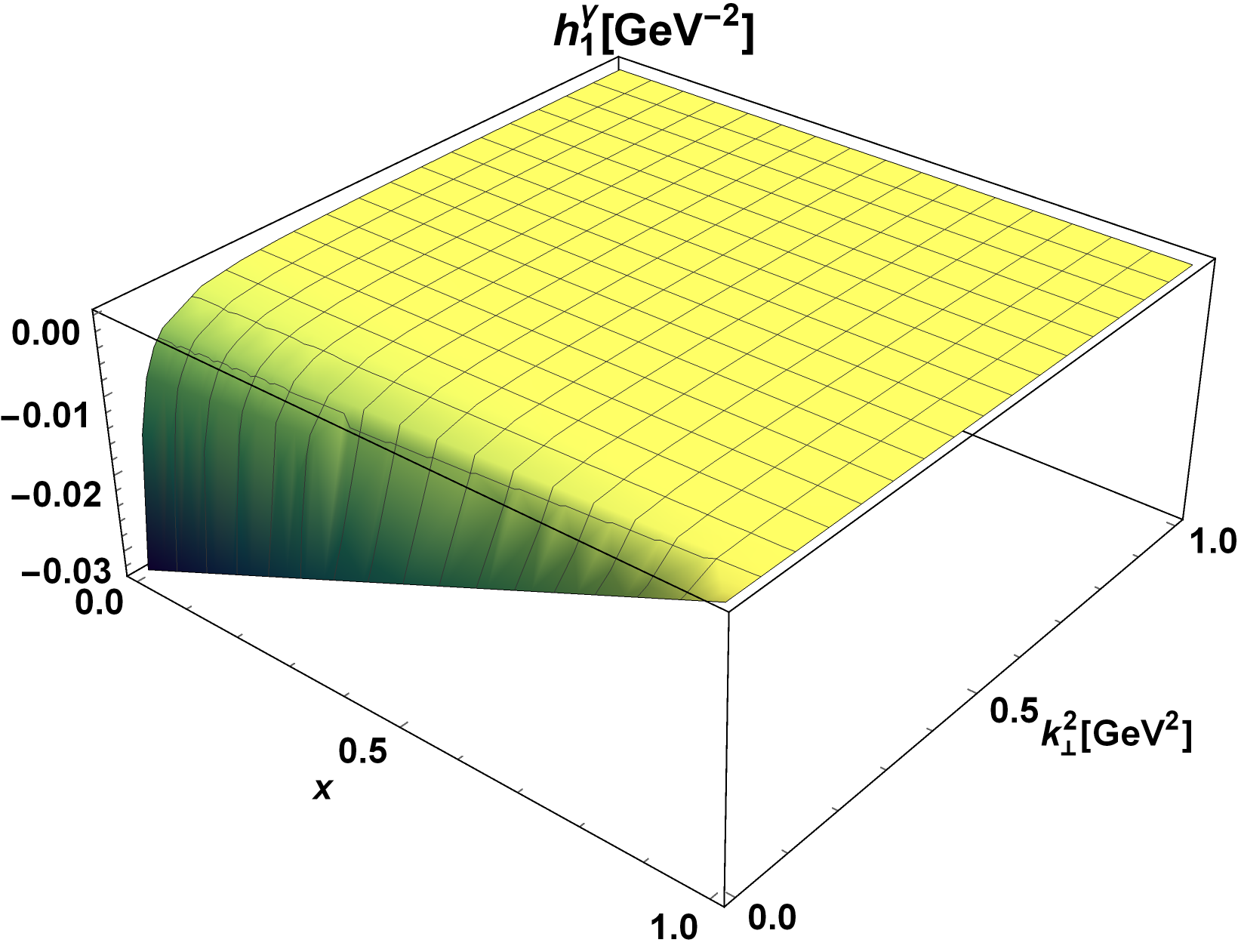}
			\end{center}
	\end{minipage}
	\caption{Plots of (a) $f_1(x,{\bf k}^2_\perp)$, (b) $g_{1L}(x,{\bf k}^2_\perp)$ and (c) $h_{1}(x,{\bf k}^2_\perp)$ T-even TMDs with respect to $x$ and ${\bf k}^2_\perp$ for the case of real photon ($M^2_\gamma=0$ GeV$^2$). }
	\label{realtmds}
\end{figure}

For numerical predictions of photon leading twist TMDs, we have taken the quark mass ($m$) and real photon mass ($M^2_\gamma$) equal to $0.2$ GeV$^2$ and $0$ GeV$^2$ respectively \cite{Xiao:2003wf}. For time-like (space-like) virtual photon, we have taken  $M^2_{\gamma}=0.1(-0.1)$ GeV$^2$. In Fig. \ref{realtmds}, we have plotted the three-dimensional unpolarized $f_{1}^\gamma(x,\textbf{k}^2_{\perp})$,  longitudinally polarized $g_{1}^\gamma(x,\textbf{k}^2_{\perp})$ and transversity $h_{1}^\gamma(x,\textbf{k}^2_{\perp})$ TMDs with respect to longitudinal momentum fraction $x$ and transverse momenta $\textbf{k}^2_{\perp}$ for real photon ($M^2_{\gamma}=0$). There are only three non-vanishing TMDs for the case of real photon whereas $h^{\perp}_{1T}(x,\textbf{k}^2_{\perp})$ and $f_{1LL}(x,\textbf{k}^2_{\perp})$ are already zero for photon TMDs. In addition to this, due to the presence of factor $M_{\gamma}$ in the numerator, other TMDs vanish. The $f_{1}^\gamma(x,\textbf{k}^2_{\perp})$ and $g_{1L}(x,\textbf{k}^2_{\perp})$ TMDs shows a positive distribution and have maximum distribution around lower $\textbf{k}^2_{\perp}$ values. They are also symmetric under $x \leftrightarrow (1-x)$ while in case of $h_{1}^\gamma(x,\textbf{k}^2_{\perp})$ TMD, providing correlation  between transverse spin of the quark and transverse momentum of the quark inside the photon, shows negative distribution along $x$. 
All these three TMDs shows maxima around $\textbf{k}^2_{\perp} \sim 0$ GeV$^2$ beyond which the magnitude decreases with increasing value of $\textbf{k}^2_{\perp}$. All these TMDs vanish for $\textbf{k}^2_{\perp}\ge 1$ GeV$^2$.
\begin{figure}[ht]
\centering
\begin{minipage}[c]{1\textwidth}
\begin{center}
(a)\includegraphics[width=.45\textwidth]{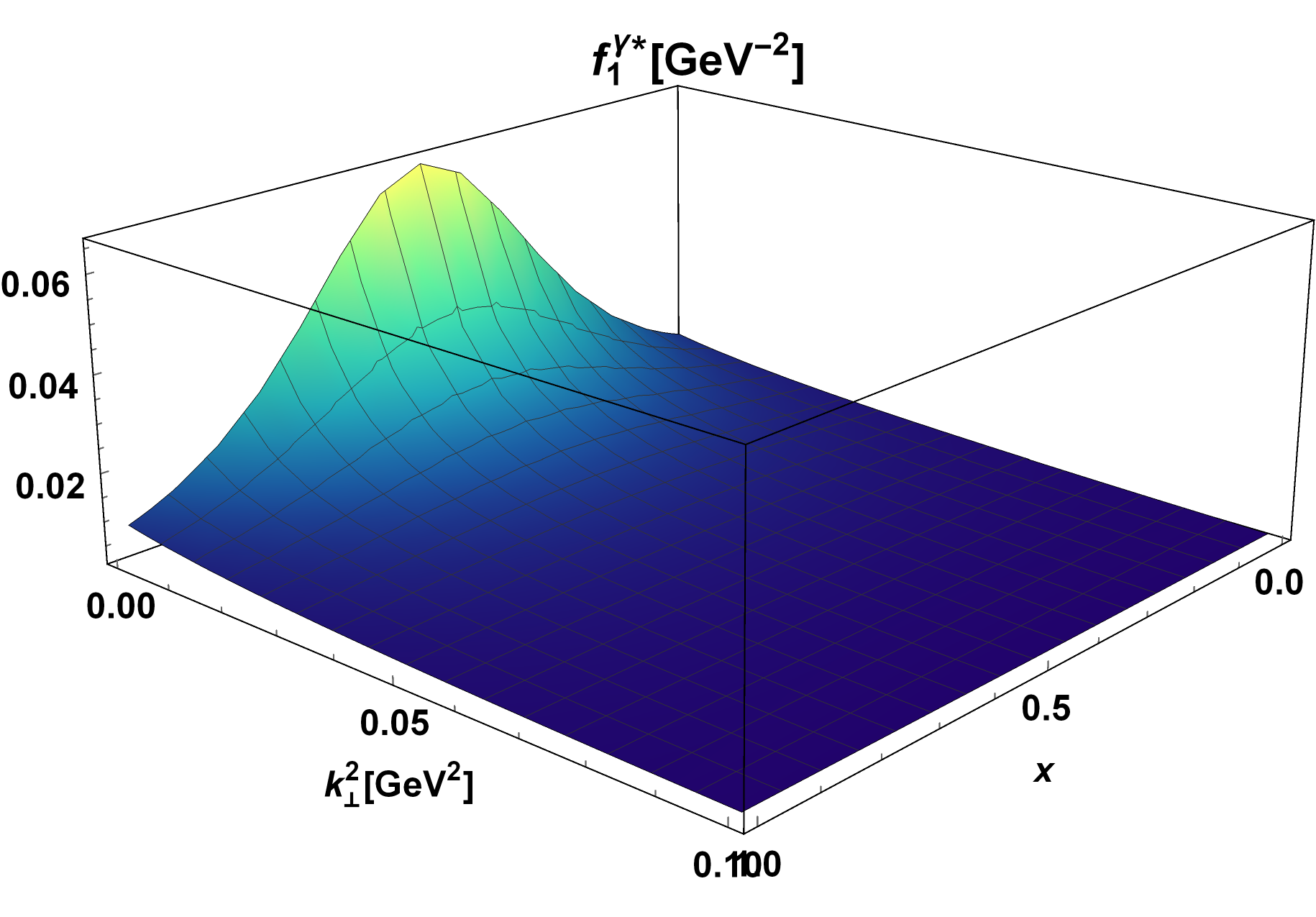}\hspace{0.35cm}
(b)\includegraphics[width=.45\textwidth]{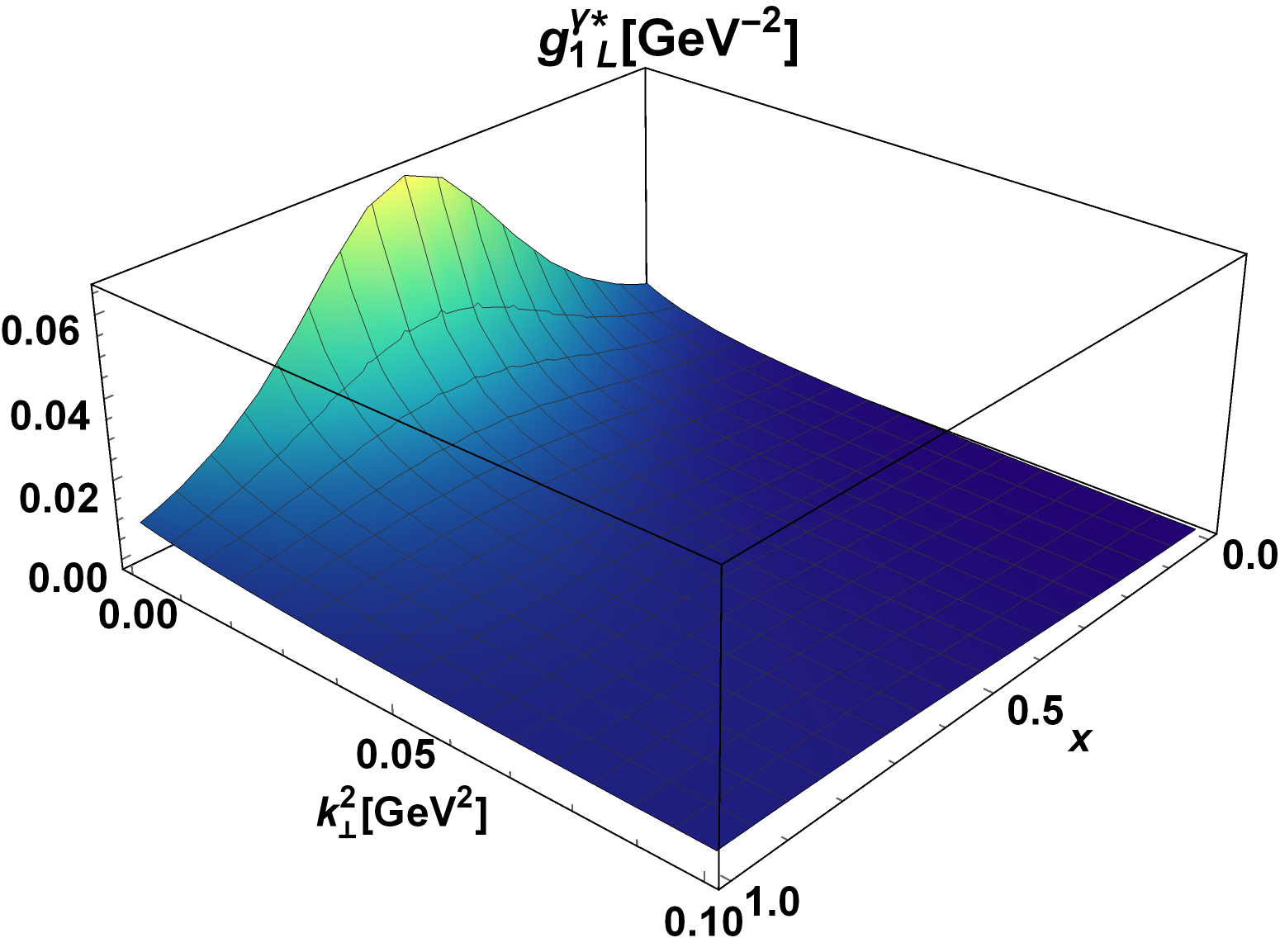}
\end{center}
\end{minipage}
\begin{minipage}[c]{1\textwidth}
\begin{center}
(c)\includegraphics[width=.45\textwidth]{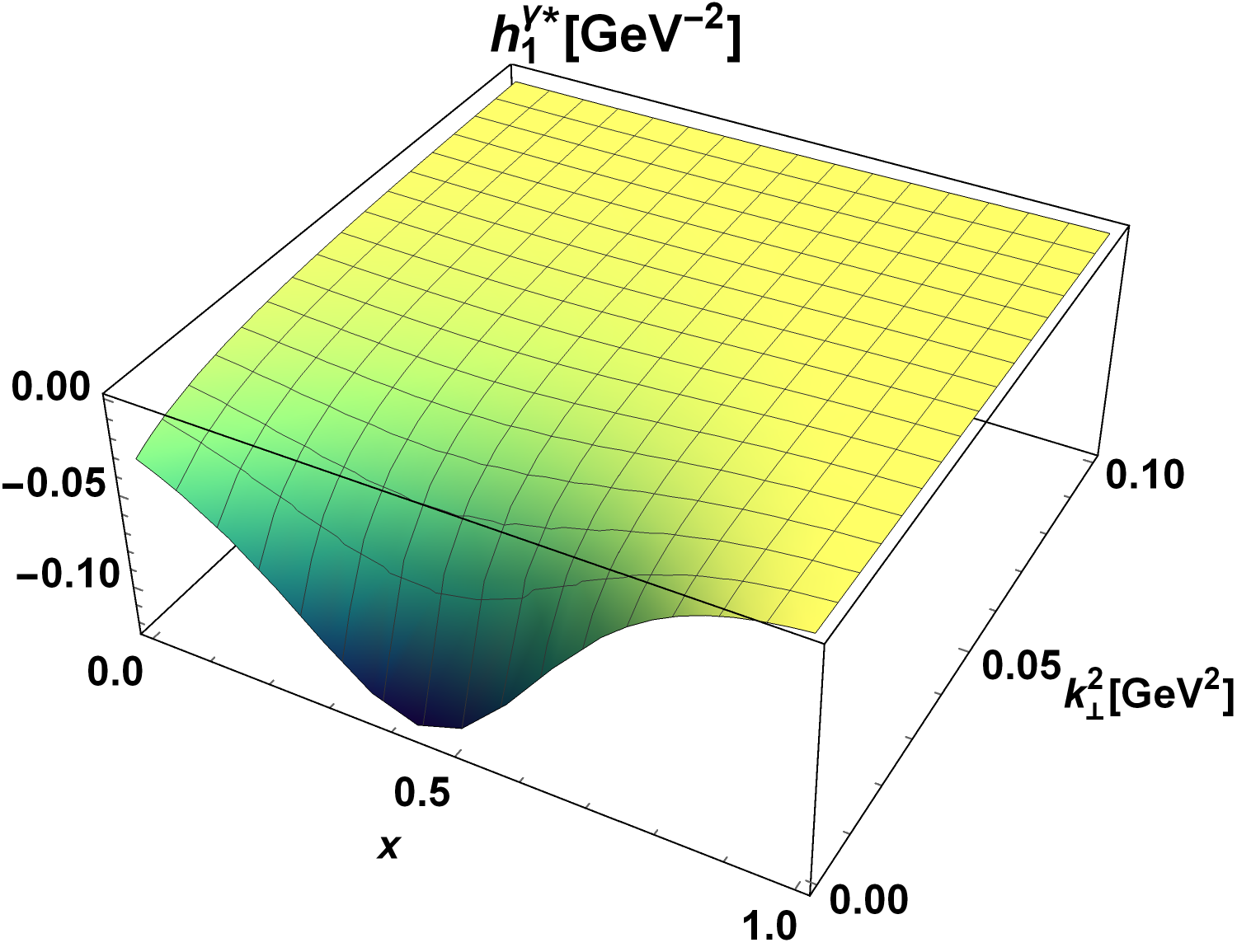}\hspace{0.35cm}
(d)\includegraphics[width=.45\textwidth]{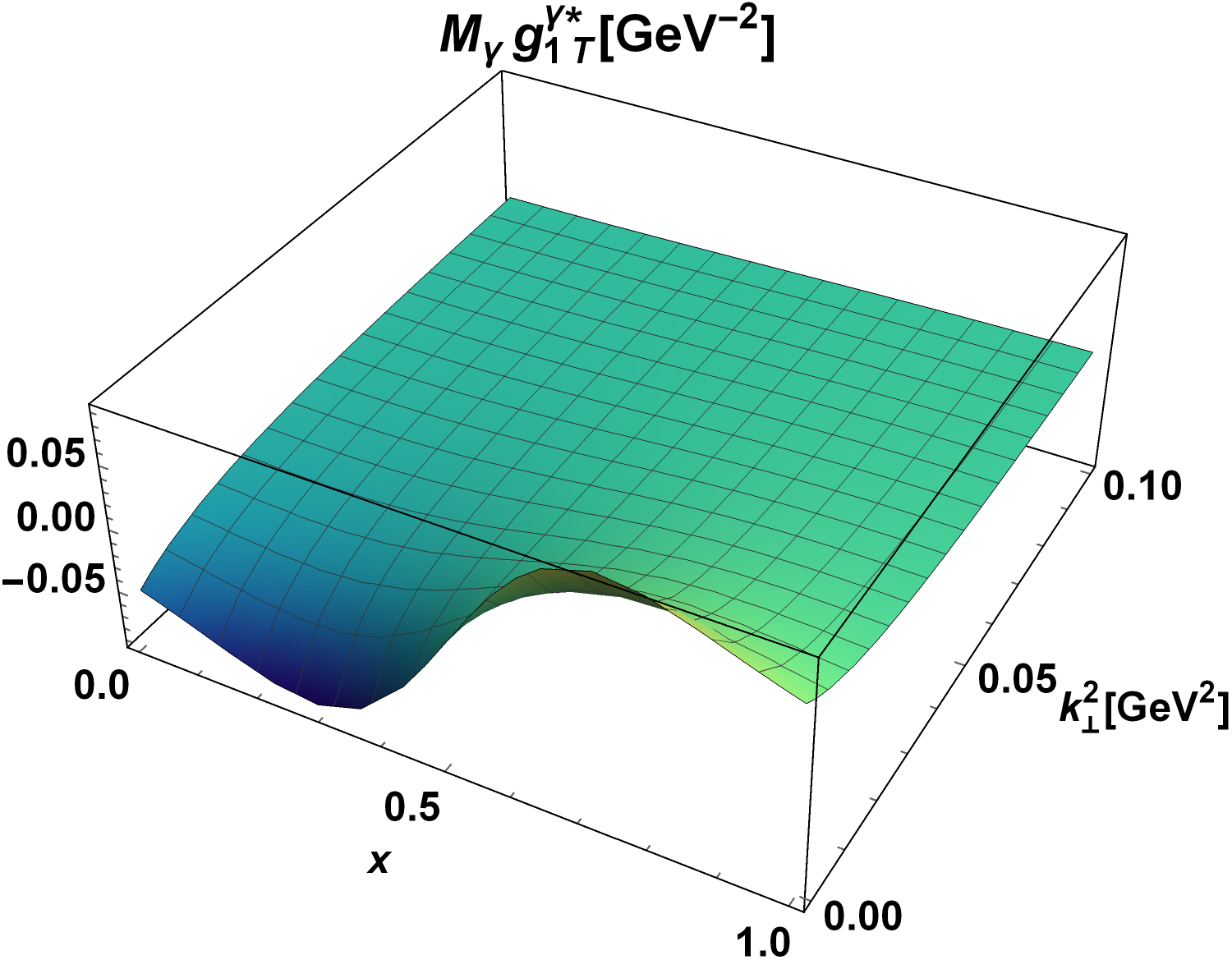}
\end{center}
\end{minipage}
\caption{Plots of T-even TMDs (a) $f_1^{\gamma^*}(x,{\bf k}^2_\perp)$, (b)$g_{1L}^{\gamma^*}(x,{\bf k}^2_\perp)$, (c) $h_1^{\gamma^*}(x,{\bf k}^2_\perp)$, and (d) $g_{1T}^{\gamma^*}(x,{\bf k}^2_\perp)$ for time-like virtual photon as a function of $x$ and ${\bf k}^2_\perp$ at $M^2_{\gamma}=0.1$ GeV$^2$.}
\label{Fig_2}
\end{figure}
\begin{figure}[ht]
\centering
\begin{minipage}[c]{1\textwidth}
\begin{center}
(a)\includegraphics[width=.45\textwidth]{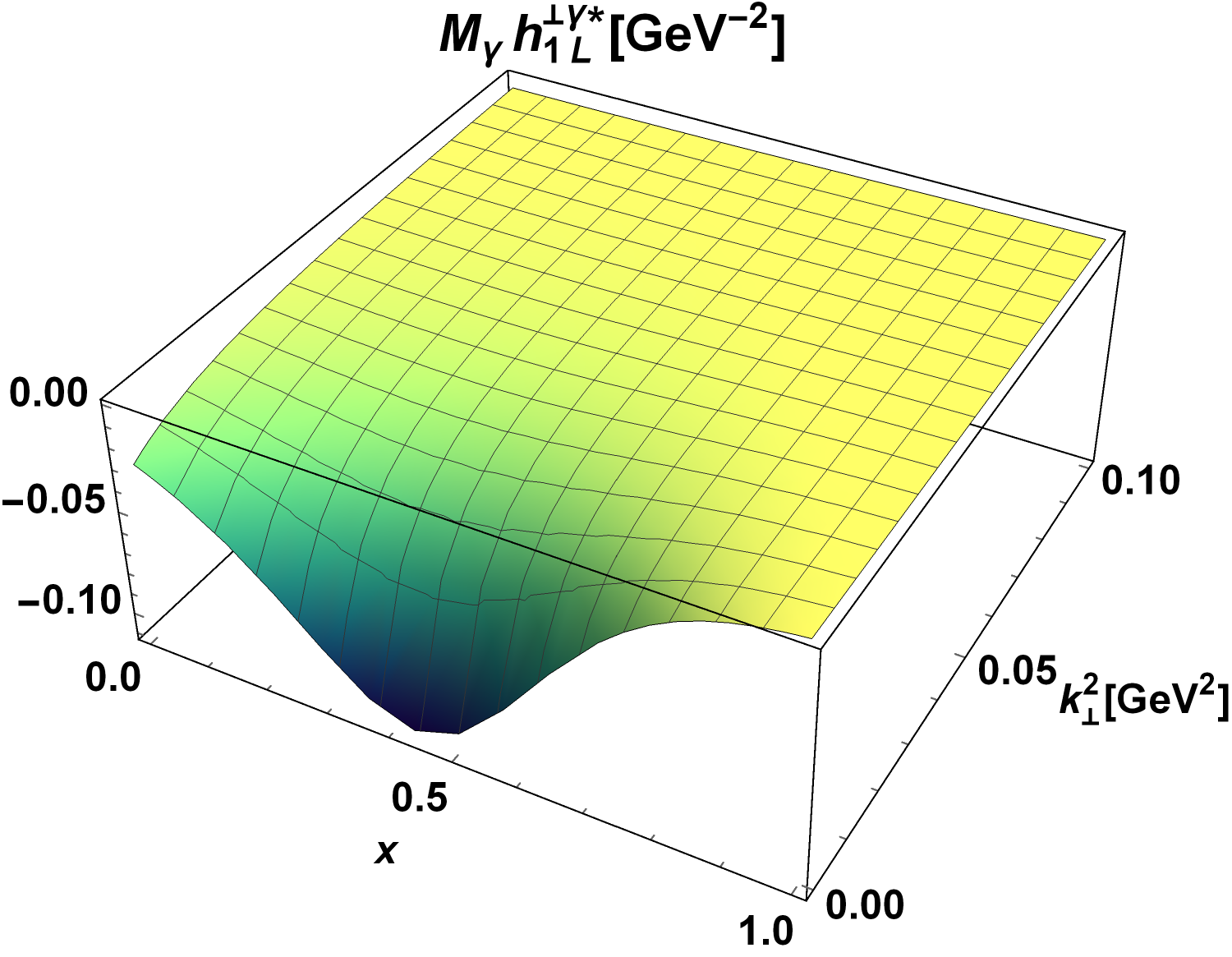}\hspace{0.35cm}
(b)\includegraphics[width=.45\textwidth]{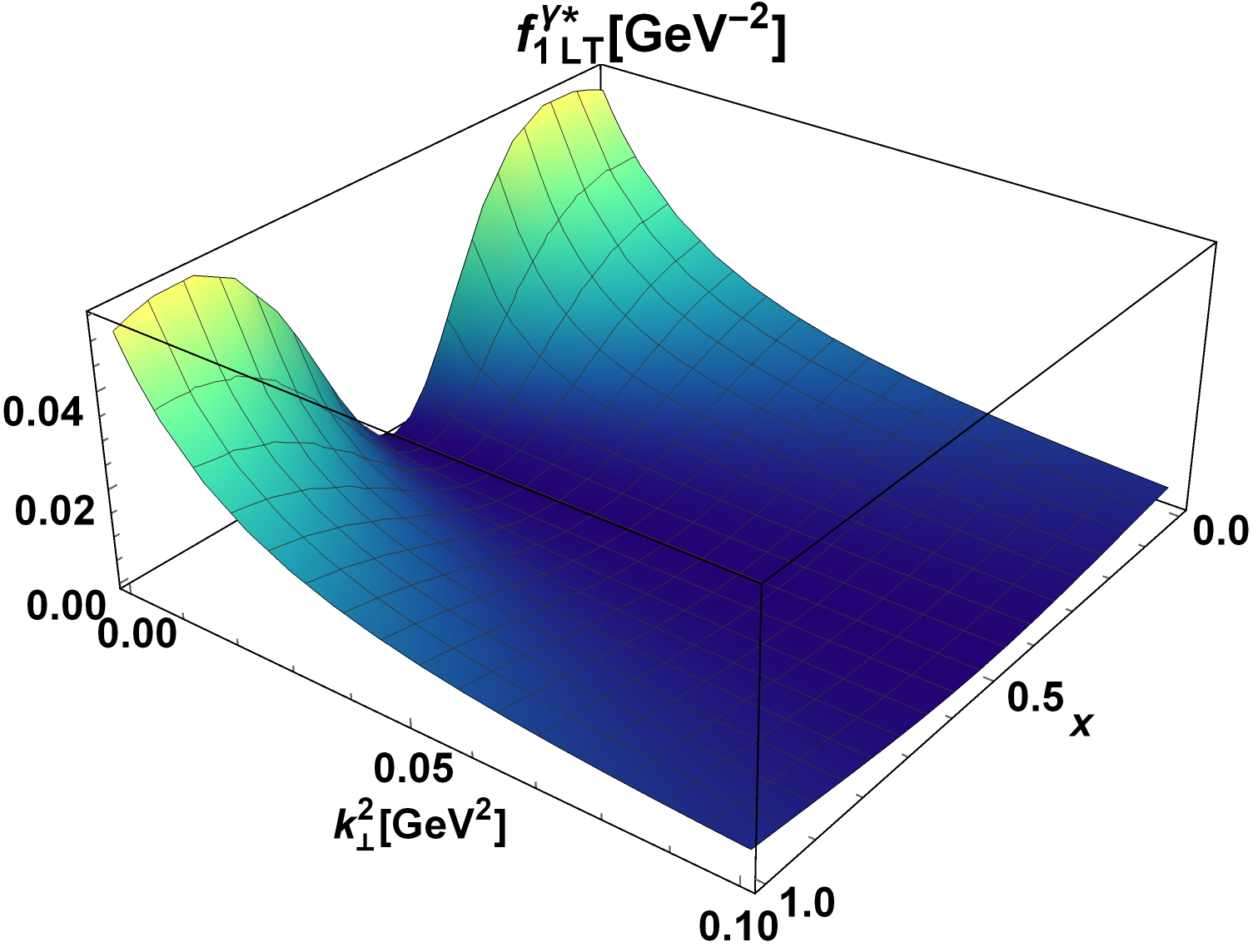}
\end{center}
\end{minipage}
\begin{minipage}[c]{1\textwidth}
\begin{center}
(c)\includegraphics[width=.45\textwidth]{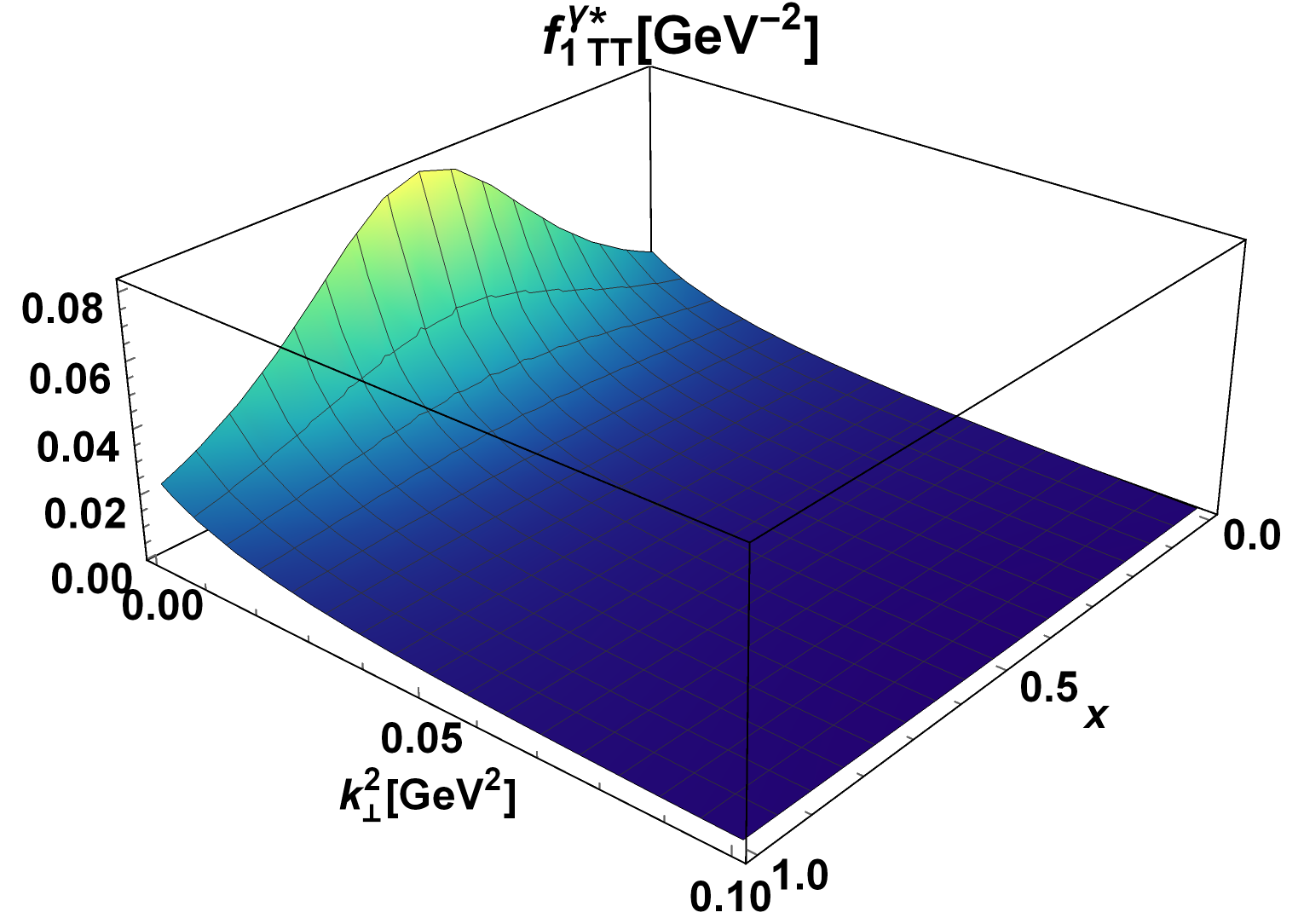}\hspace{0.45cm}
\end{center}
\end{minipage}
\caption{ Plots of T-even TMDs (a) $h_{1L}^{\perp \gamma^*}(x,{\bf k}^2_\perp)$, (b)$f_{1LT}^{\gamma^*}(x,{\bf k}^2_\perp)$, and (c) $f_{1TT}^{\gamma^*}(x,{\bf k}^2_\perp)$ for time-like virtual photon as a function of $x$ and ${\bf k}^2_\perp$ at $M^2_{\gamma}=0.1$ GeV$^2$.}
\label{Fig_22}
\end{figure}

In Figs. \ref{Fig_2} and \ref{Fig_22}, we have plotted all the T-even TMDs with respect to $x$ and $\textbf{k}^2_{\perp}$ for time-like virtual photon at $M^2_{\gamma}=0.1$ GeV$^2$. The unpolarized $f_{1}^{\gamma*}(x,\textbf{k}^2_{\perp})$ TMD for both space-like and time-like virtual photon are symmetric around $x=0.5$ and have lower (higher) peak than the real photon for space-like (time-like) virtual photon respectively. Further, $f_{1}^{\gamma*}(x,\textbf{k}^2_{\perp})$, $g_{1L}^{\gamma*}(x,\textbf{k}^2_{\perp})$, $f_{1LT}^{\gamma*}(x,\textbf{k}^2_{\perp})$ and $f_{1TT}^{\gamma*}(x,\textbf{k}^2_{\perp})$ show positive distributions, whereas $h_{1}^{\gamma*}(x,\textbf{k}^2_{\perp})$ and $h^{\perp \gamma*}_{1L}(x,\textbf{k}^2_{\perp})$ show negative distributions. All the TMDs in  Figs. \ref{Fig_2} and \ref{Fig_22} shows maximum distributions below $\textbf{k}^2_{\perp} \leq 0.1$ GeV$^2$. The worm gear $2$ distribution, $g_{1T}^{\gamma*}(x,\textbf{k}^2_{\perp})$ shows both negative and positive distributions with anti-symmetric behavior under $x \longleftrightarrow (1-x)$ which is clear from Fig. \ref{Fig_2} (d). This TMD vanishes at around $x=0.5$ and exhibits positive and negative distributions for $x > 0.5$ and $x < 0.5$ respectively. The $g_{1T}^{\gamma*}(x,\textbf{k}^2_{\perp})$ describes the momentum distribution of longitudinally polarized quark in the transversely
polarized photon. The positive and negative distribution is due to the mixing result of S-wave and P-wave LFWFs. Except for $g_{1T}^{\gamma*}(x,\textbf{k}^2_{\perp})$ TMD, other TMDs show symmetry under $x \leftrightarrow (1-x)$. The tensor $f_{1LT}^{ \gamma*} (x,\textbf{k}^2_{\perp})$ shows symmetry  under $x \leftrightarrow (1-x)$ with double peak structure at $x=0$ and $1$. The distributions of photon TMDs are more centralized around $x=0.5$ than the real photon TMDs except for the cases of $g_{1T}^{\gamma*}(x,\textbf{k}^2_{\perp})$ and $f_{1LT}^{\gamma*}(x,\textbf{k}^2_{\perp})$ TMDs with addition of photon mass. 

\begin{figure}[ht]
\centering
\begin{minipage}[c]{1\textwidth}
\begin{center}
(a)\includegraphics[width=.43\textwidth]{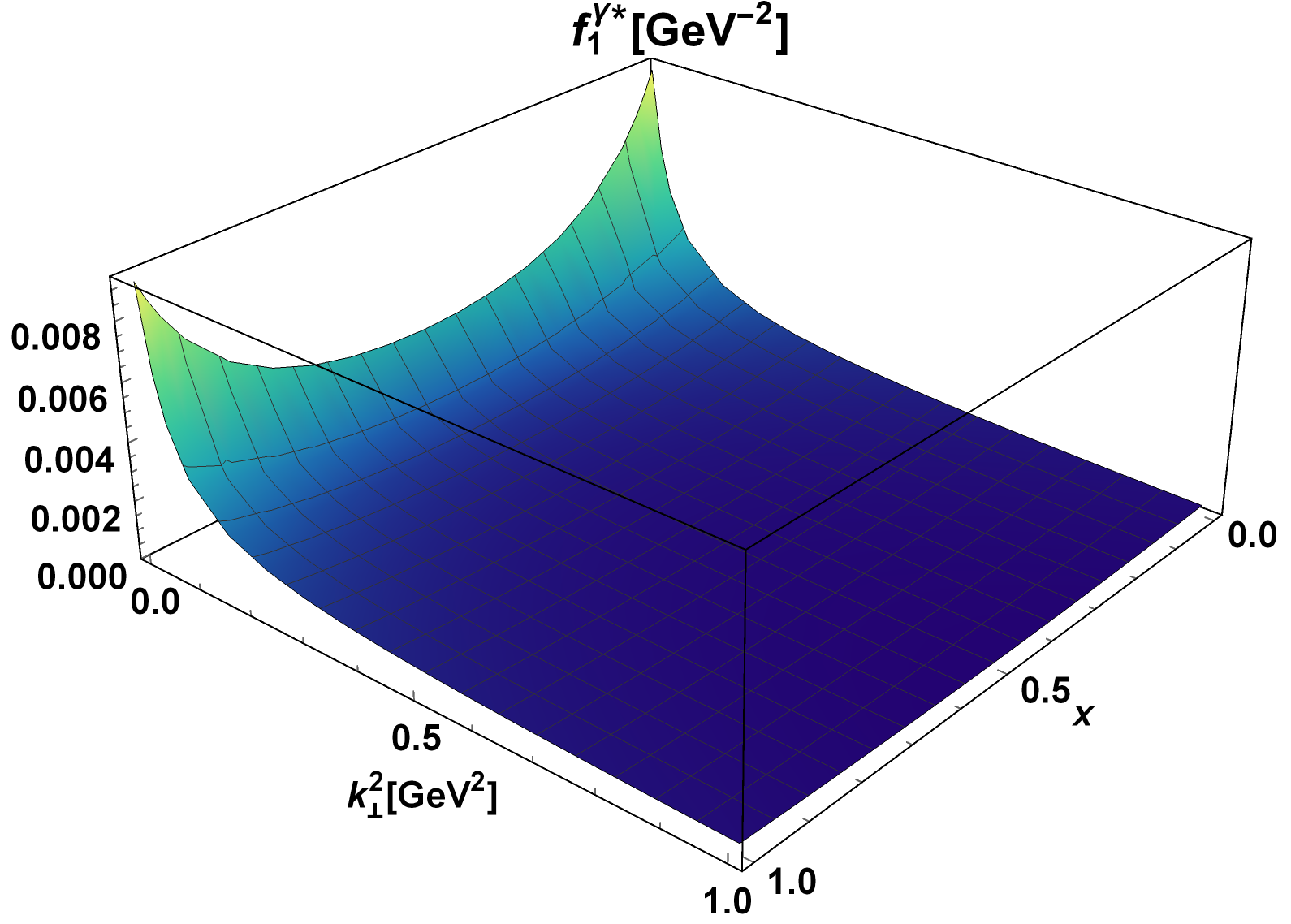}\hspace{0.5cm}
(b)\includegraphics[width=.44\textwidth]{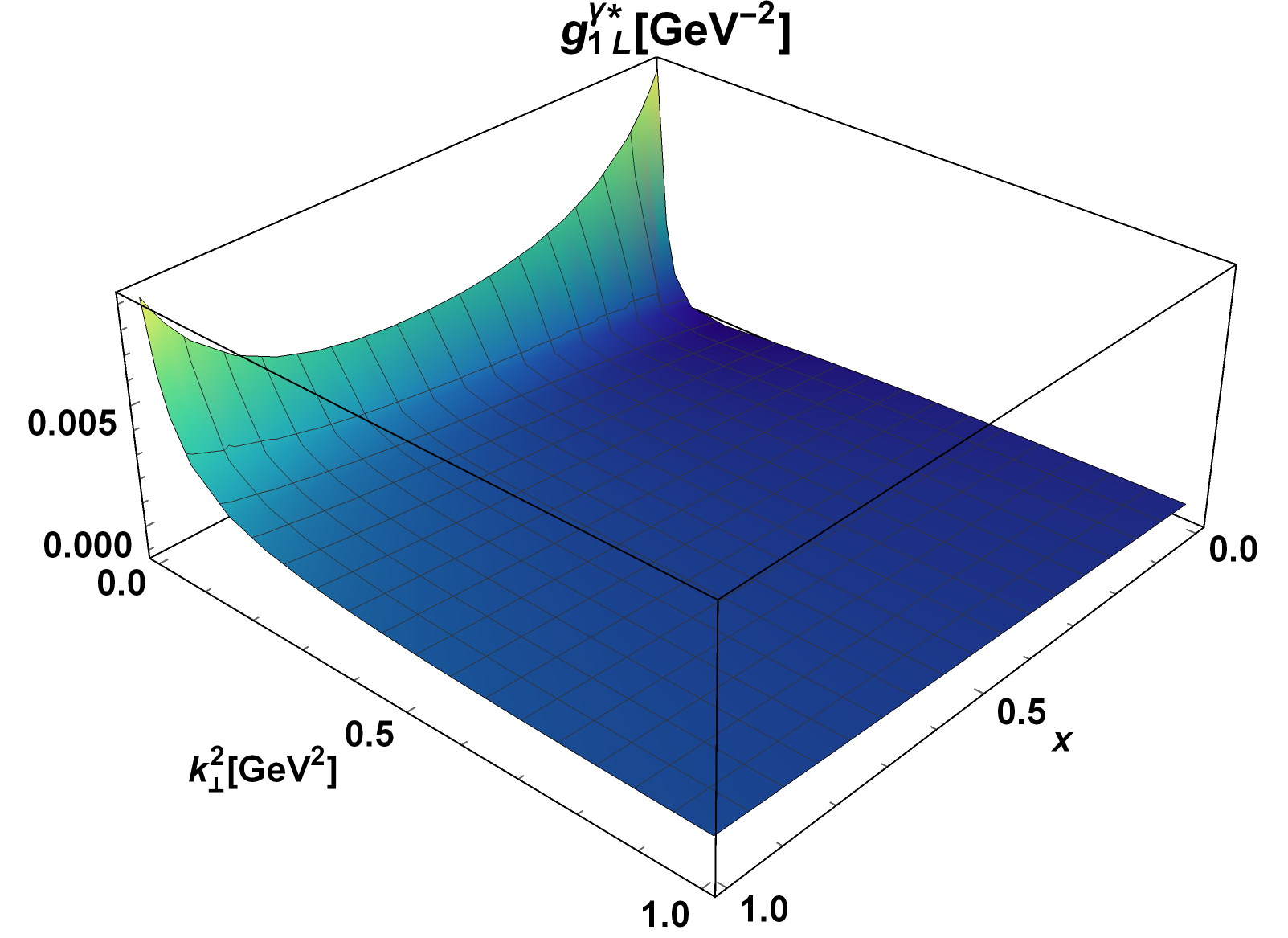}
\end{center}
\end{minipage}
\begin{minipage}[c]{1\textwidth}
\begin{center}
(c)\includegraphics[width=.44\textwidth]{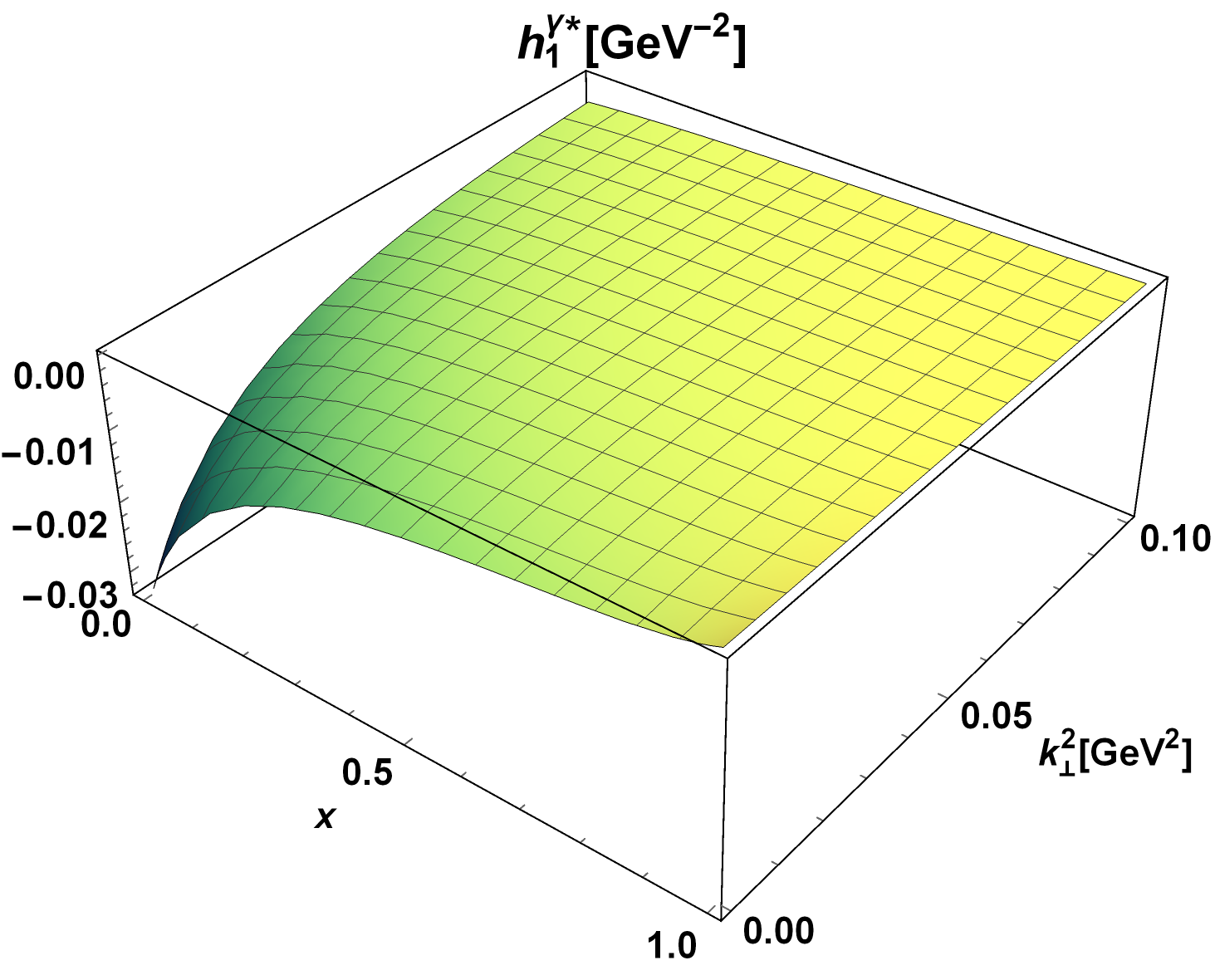}\hspace{0.5cm}
(d)\includegraphics[width=.44\textwidth]{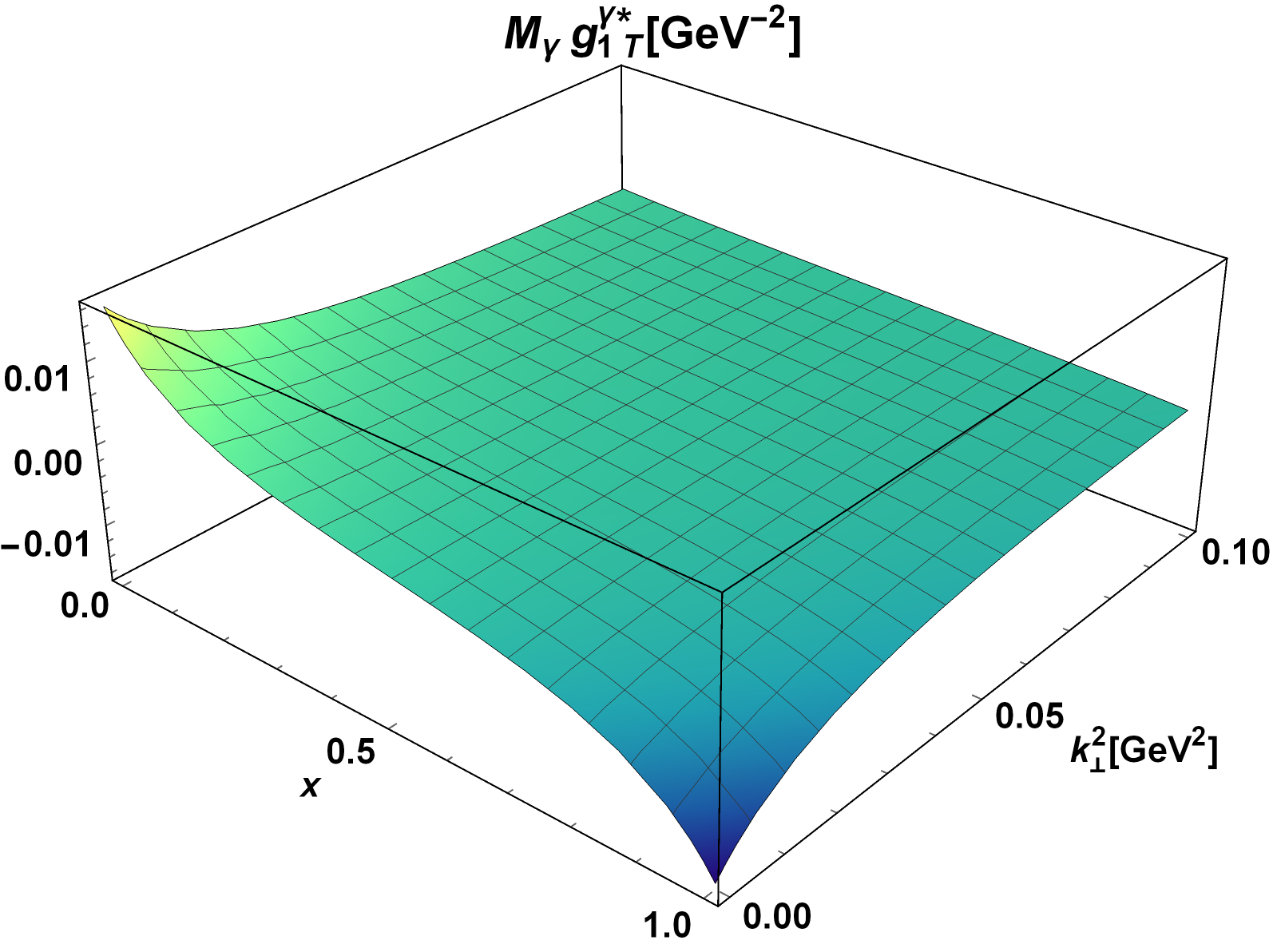}
\end{center}
\end{minipage}
\caption{Plots of T-even TMDs (a) $f_1^{\gamma^*}(x,{\bf k}^2_\perp)$, (b) $g_{1L}^{\gamma^*}(x,{\bf k}^2_\perp)$, (c) $h_1^{\gamma^*}(x,{\bf k}^2_\perp)$, and (d) $g_{1T}^{\gamma^*}(x,{\bf k}^2_\perp)$ for space-like virtual photon as a function of $x$ and ${\bf k}^2_\perp$ at $M^2_{\gamma}=-0.1$ GeV$^2$.}
\label{space_like}
\end{figure}

\begin{figure}
\begin{minipage}[c]{1\textwidth}
\begin{center}
(a)\includegraphics[width=.45\textwidth]{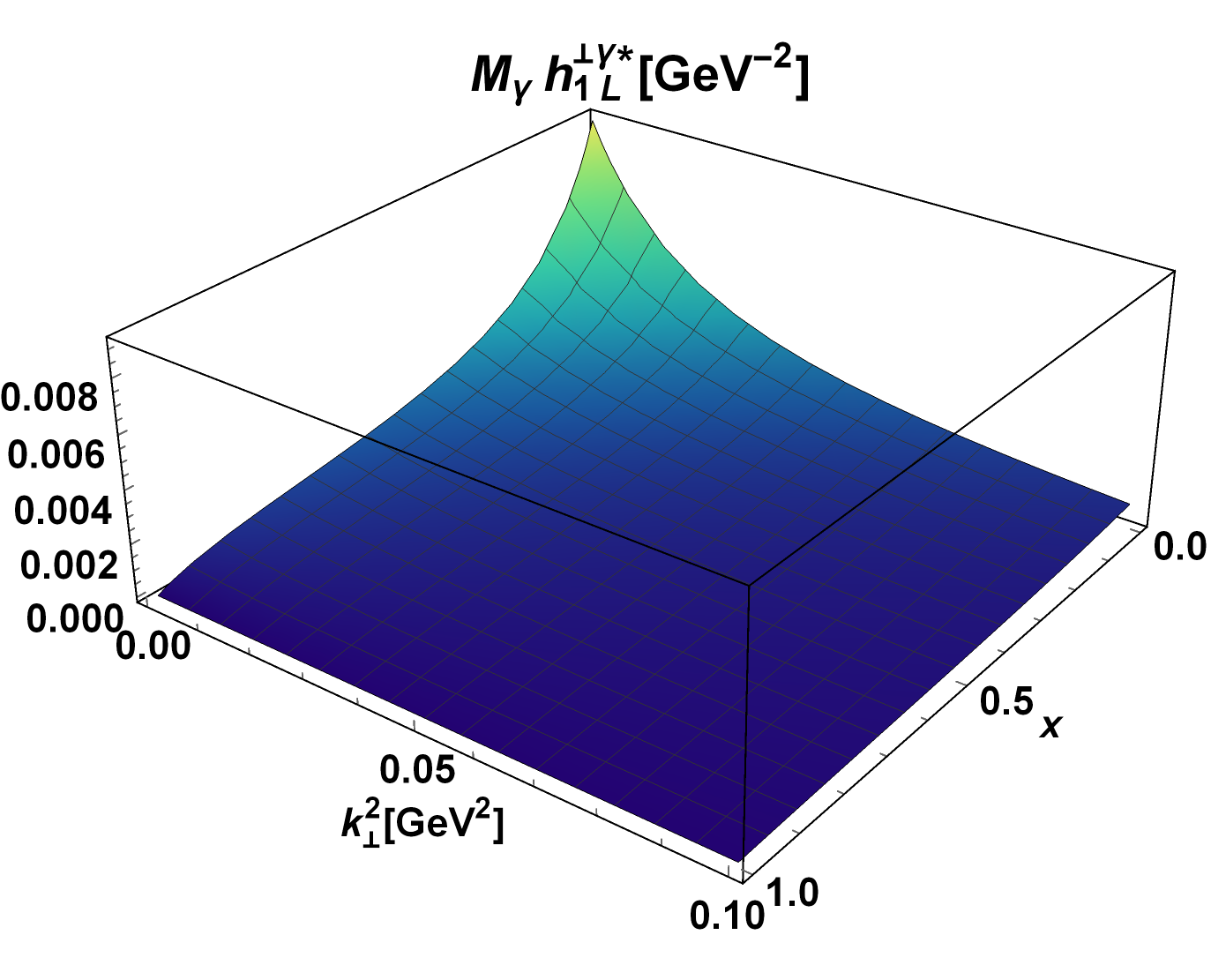}\hspace{0.5cm}
(b)\includegraphics[width=.45\textwidth]{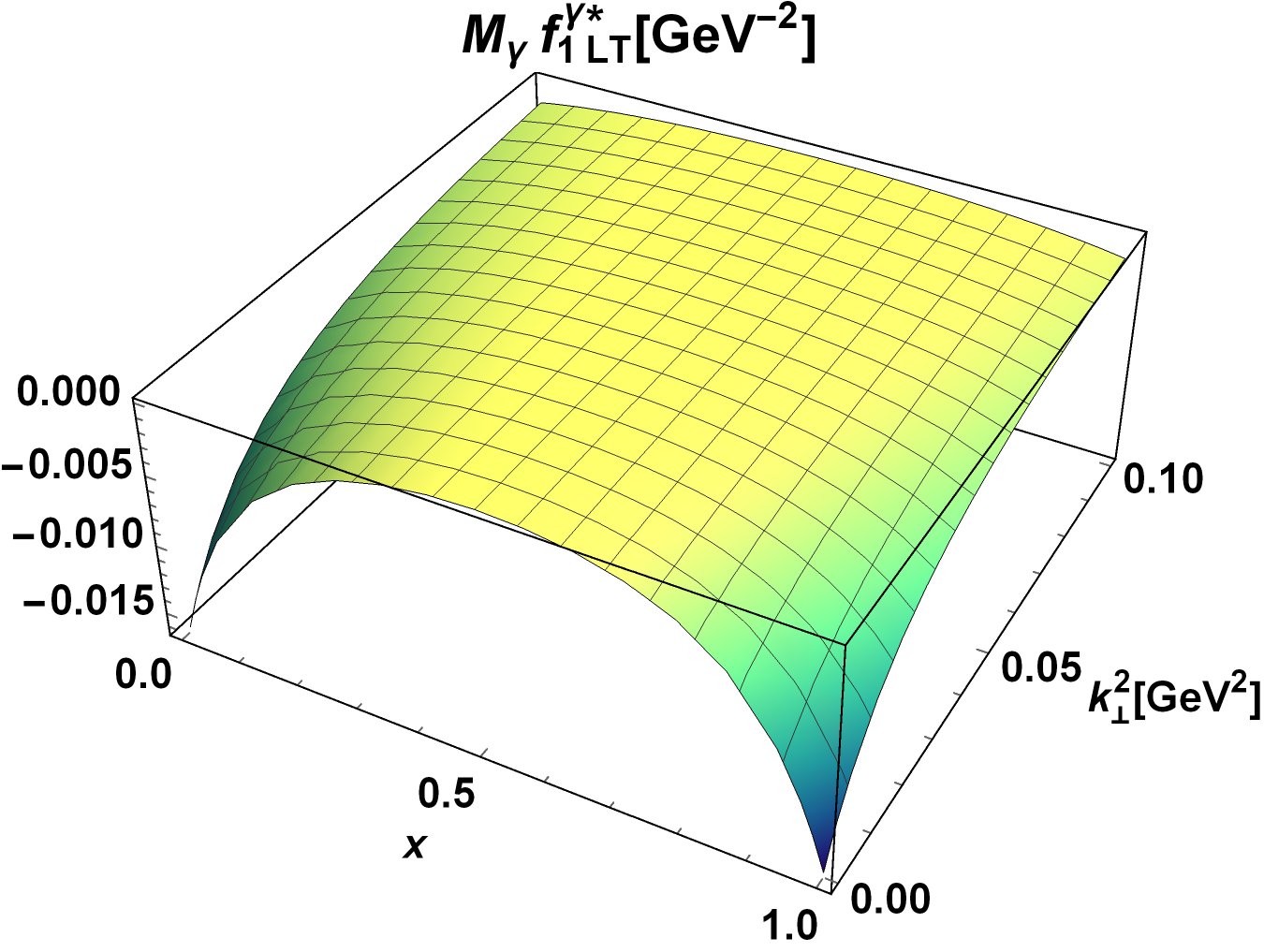}
\end{center}
\end{minipage}
\begin{minipage}[c]{1\textwidth}
\begin{center}
(c)\includegraphics[width=.45\textwidth]{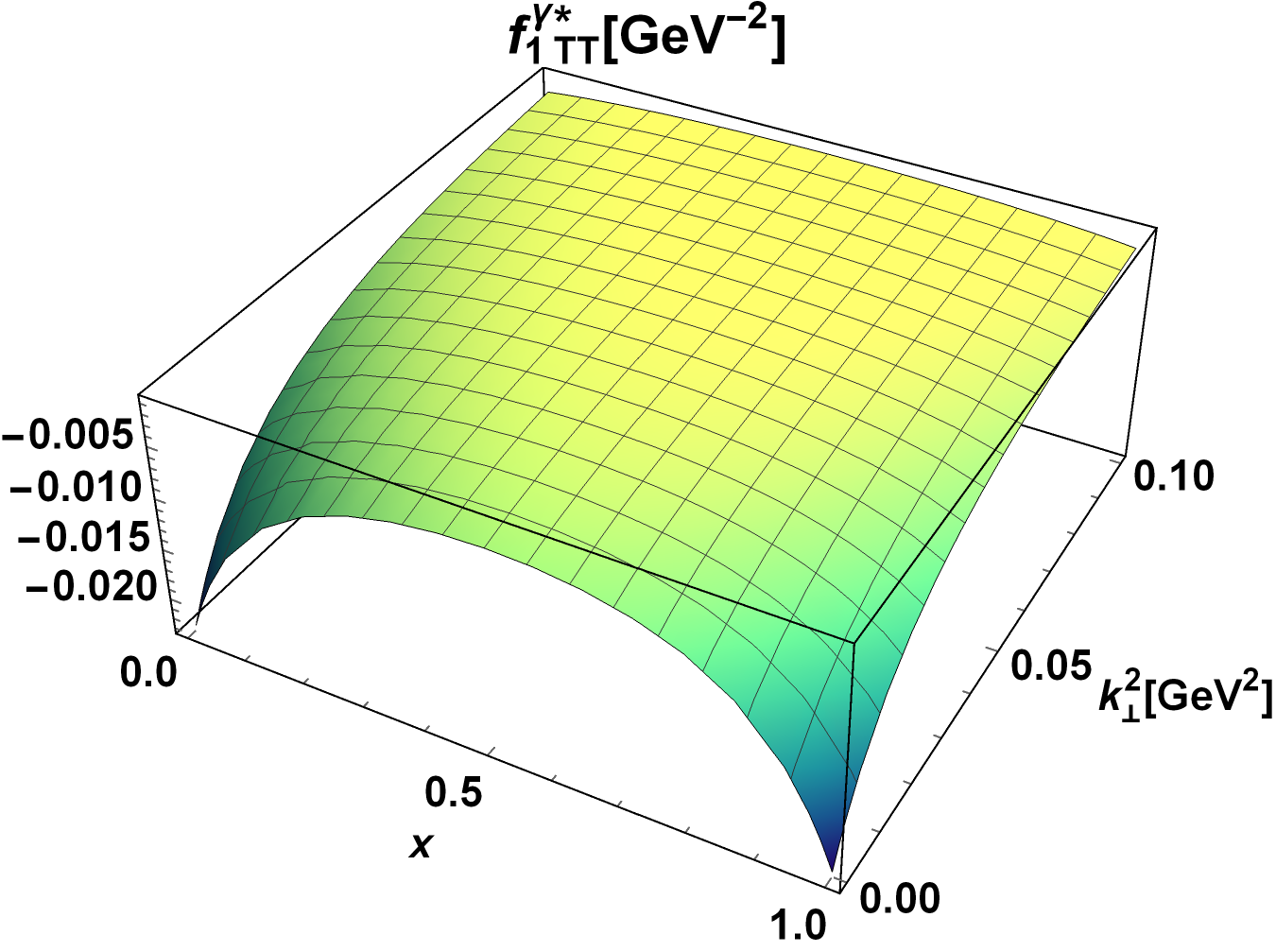}\hspace{0.5cm}
\end{center}
\end{minipage}
\caption{ Plots of T-even TMDs (a) $M_\gamma h_{1L}^{\perp \gamma^*}(x,{\bf k}^2_\perp)$, (b) $M_\gamma f_{1LT}^{\gamma^*}(x,{\bf k}^2_\perp)$, and (c) $f_{1TT}^{\gamma^*}(x,{\bf k}^2_\perp)$ for space-like virtual photon as a function of $x$ and ${\bf k}^2_\perp$ at $M^2_{\gamma}=-0.1$ GeV$^2$.}
\label{space_like_1}
\end{figure}

In Fig. \ref{space_like}, we have presented the results for the space-like virtual TMDs of the photon.  In Fig. \ref{space_like}(a) and (b) we have presented the  results for $f_1^{\gamma*}(x,\textbf{k}^2_{\perp})$ and $g_{1L}^{\gamma*}(x,\textbf{k}^2_{\perp})$ respectively. It is clear from the plots that both TMDs show a double peak structure. In Fig. \ref{space_like}(c) and (d), we have presented the results for $h_{1}^{\gamma*}(x,\textbf{k}^2_{\perp})$ and $M_\gamma g_{1T}^{\gamma*}(x,\textbf{k}^2_{\perp})$ respectively where the former gives a negative distribution whereas the latter depicts a dipole structure. In Fig. \ref{space_like_1}, we have presented the results for $M_\gamma h_{1L}^{\perp \gamma^*}(x,{\bf k}^2_\perp)$, $M_\gamma f_{1LT}^{\gamma^*}(x,{\bf k}^2_\perp)$, and $f_{1TT}^{\gamma^*}(x,{\bf k}^2_\perp)$ space-like virtual TMDs. In Fig. \ref{space_like_1}(a), the distribution is positive and peaks at $x={\bf k_\perp^2}=0$, whereas $M_\gamma f_{1LT}^{\gamma^*}(x,{\bf k}^2_\perp)$ and $f_{1TT}^{\gamma^*}(x,{\bf k}^2_\perp)$ show negative distributions  depicting a double peak structure.
 \begin{figure}[ht]
	\centering
	\begin{minipage}[c]{1\textwidth}\begin{center}
			(a)\includegraphics[width=.43\textwidth]{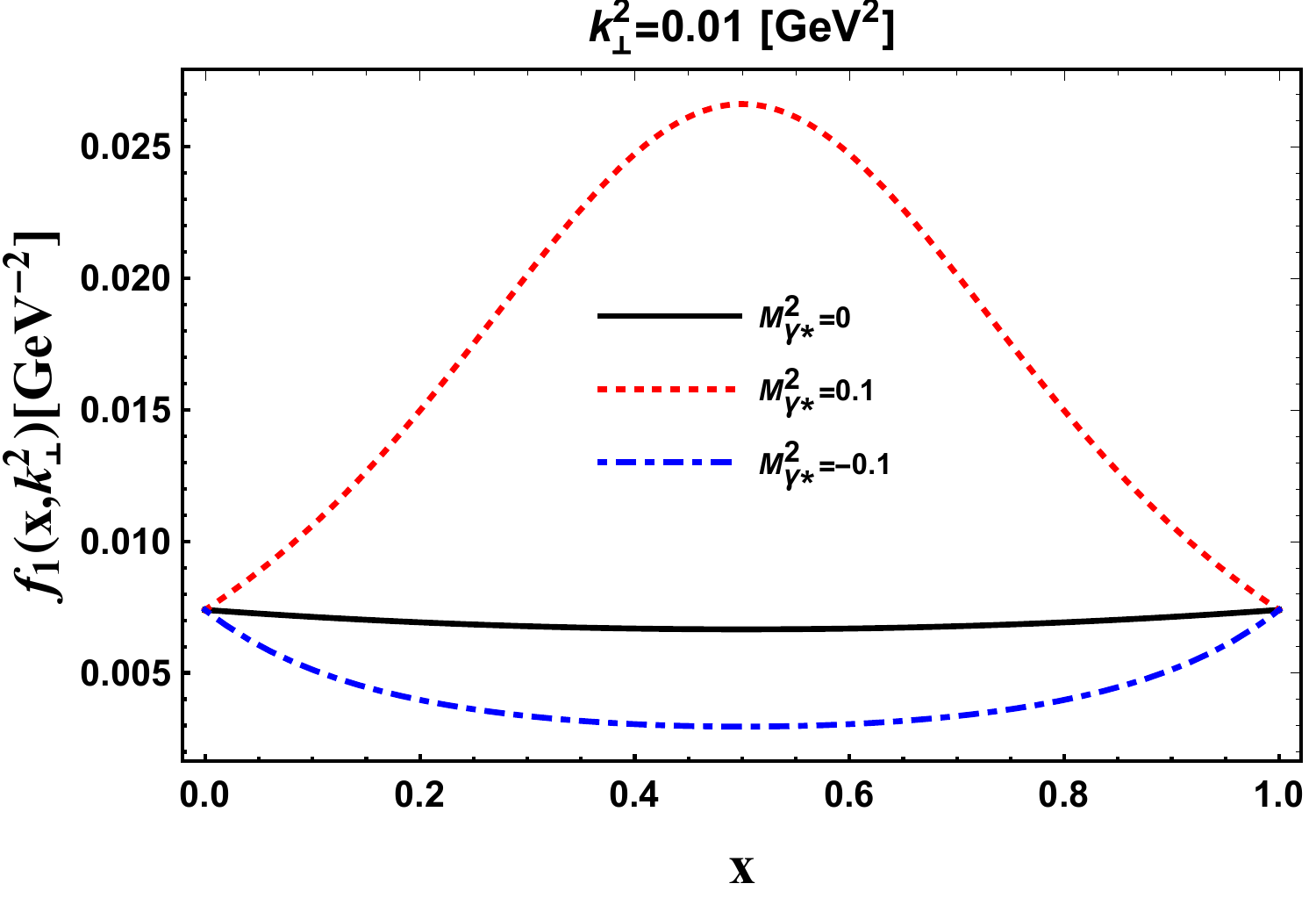}
			(b)\includegraphics[width=.45\textwidth]{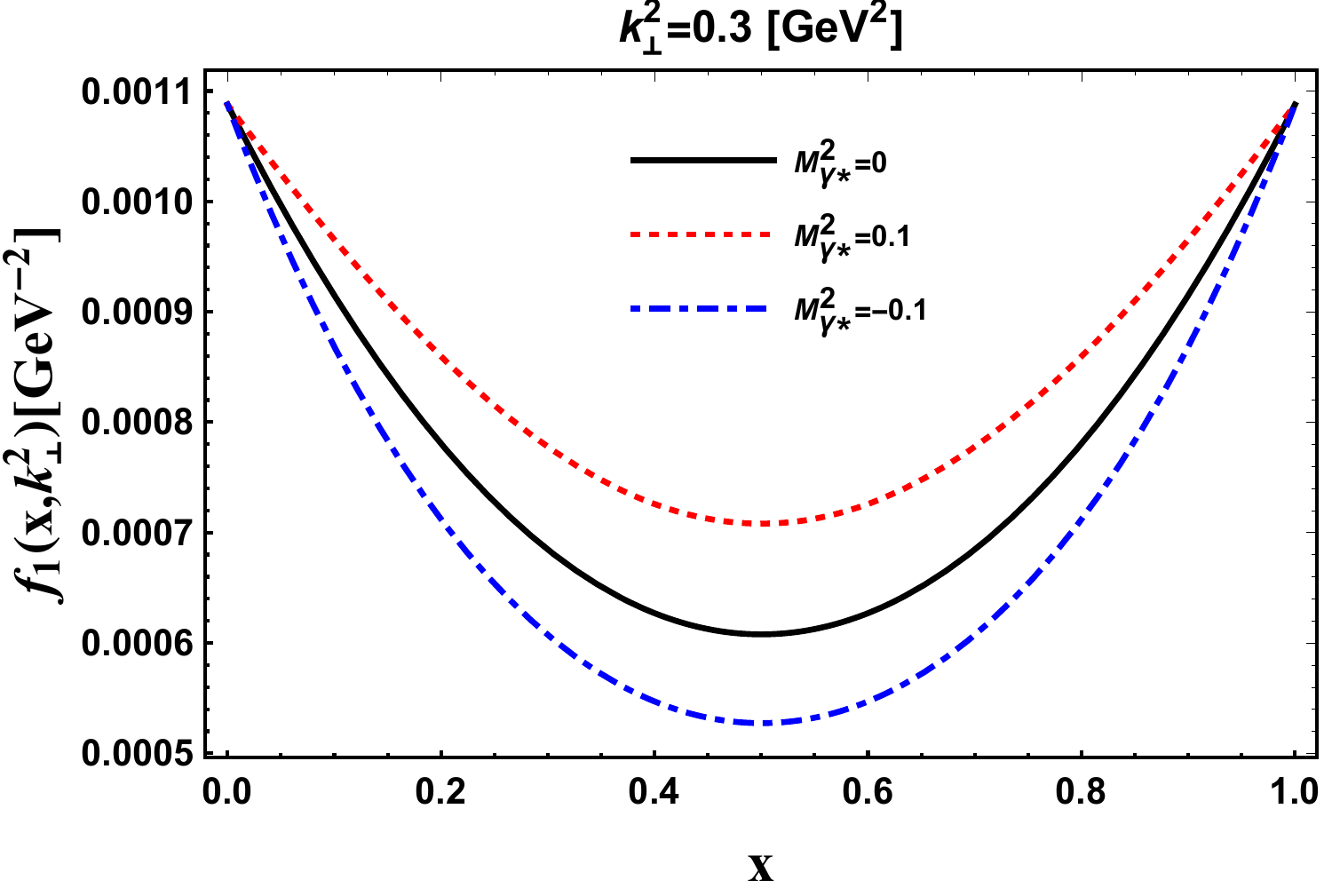}\end{center}
	\end{minipage}
	\begin{minipage}[c]{1\textwidth}\begin{center}
			(c)\includegraphics[width=.45\textwidth]{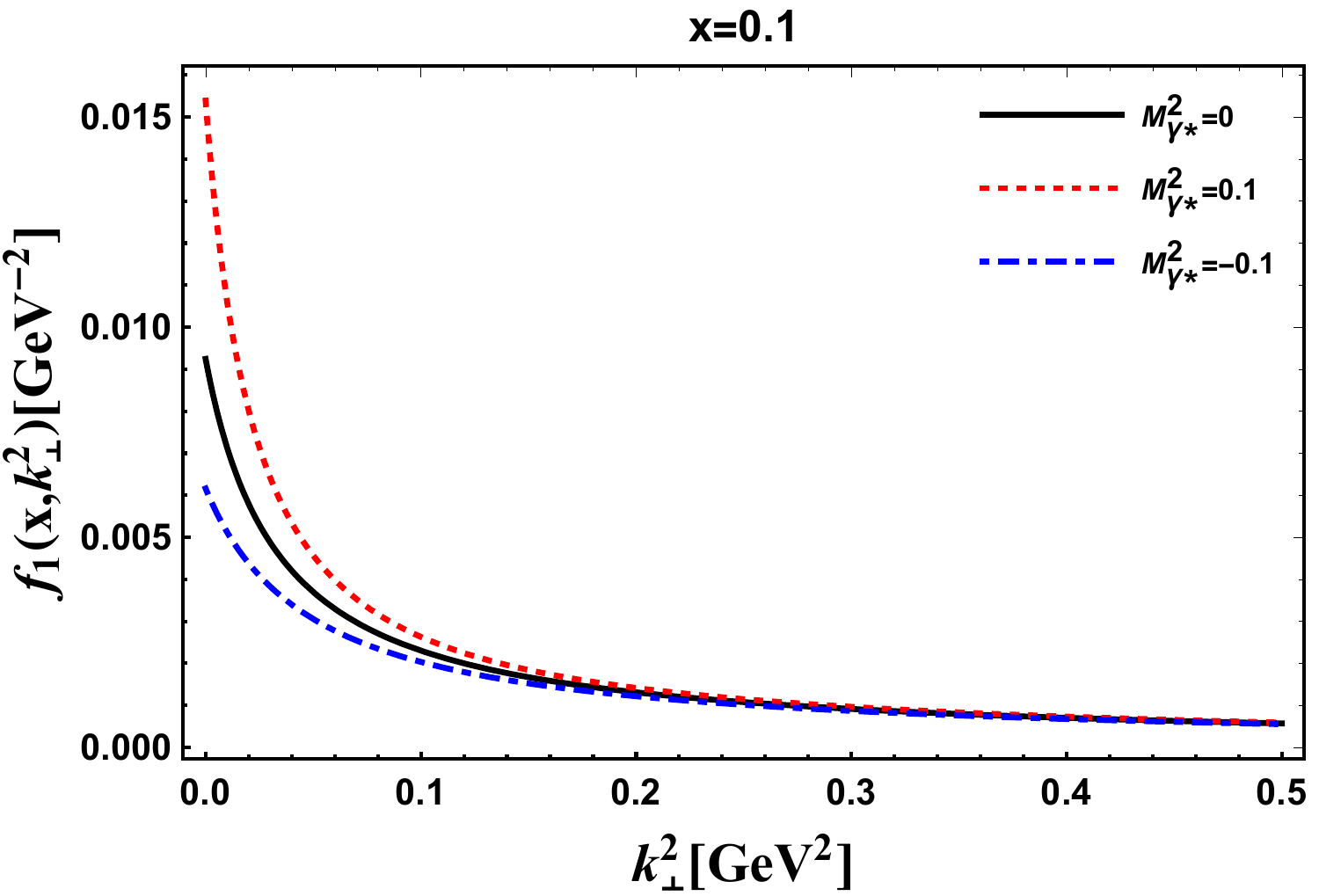}
			(d)\includegraphics[width=.445\textwidth]{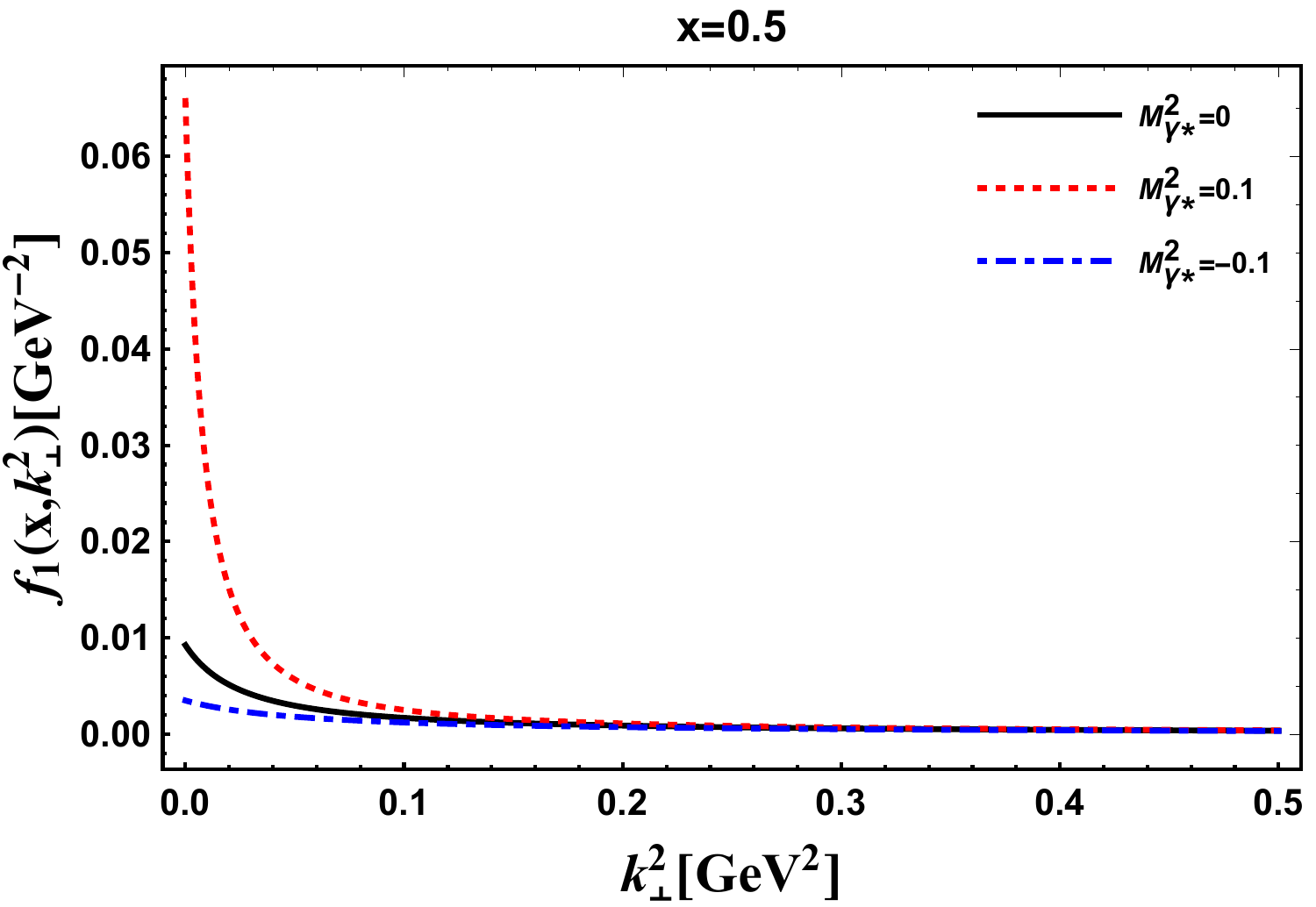}
   \end{center}
	\end{minipage}
	\caption{(Color online) $f_1(x,{\bf k}^2_\perp)$ has been plotted with fixed $x$ and $\textbf{k}_{\perp}$ for different photon masses $M^2_\gamma=0$, $0.1$ and $-0.1$ GeV$^2$. In the upper panel $f_1(x,{\bf k}^2_\perp)$ has been plotted w.r.t $x$ for a fixed values of $\textbf{k}_{\perp}^2$ as $\textbf{k}_{\perp}^2=0.01$ GeV$^2$ and $\textbf{k}_{\perp}^2=0.3$ GeV$^2$. In lower panel, $f_1(x,{\bf k}^2_\perp)$ has been plotted w.r.t $\textbf{k}_{\perp}^2$ for fixed values of $x$ at $x=0.1$ and $x=0.5$. The blue, red and black color curves are for real photon, time-like and space-like virtual photon respectively.}
	\label{f12d}
\end{figure}
 \begin{figure}[ht]
	\centering
	\begin{minipage}[c]{1\textwidth}\begin{center}
			(a)\includegraphics[width=.43\textwidth]{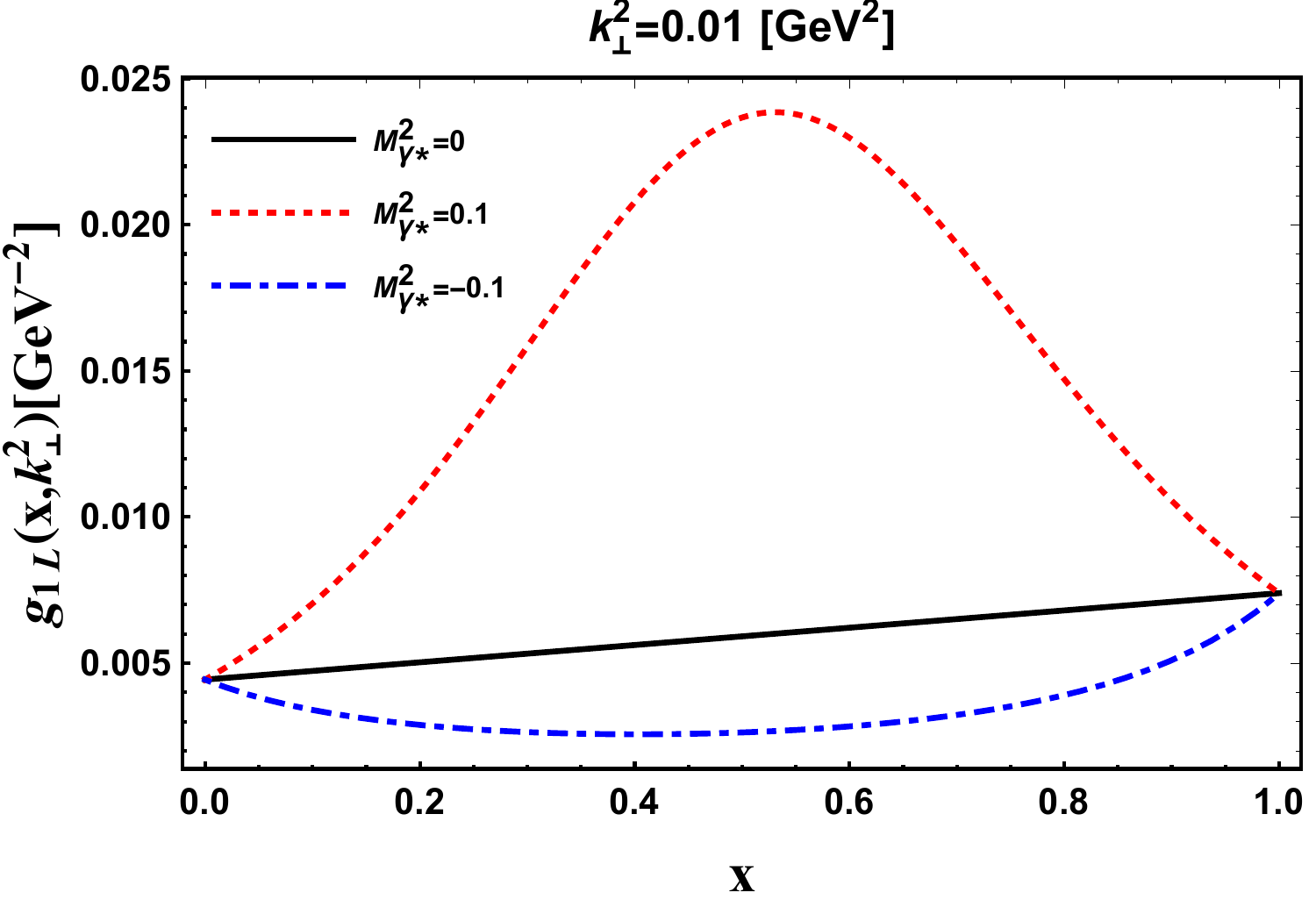}
			(b)\includegraphics[width=.445\textwidth]{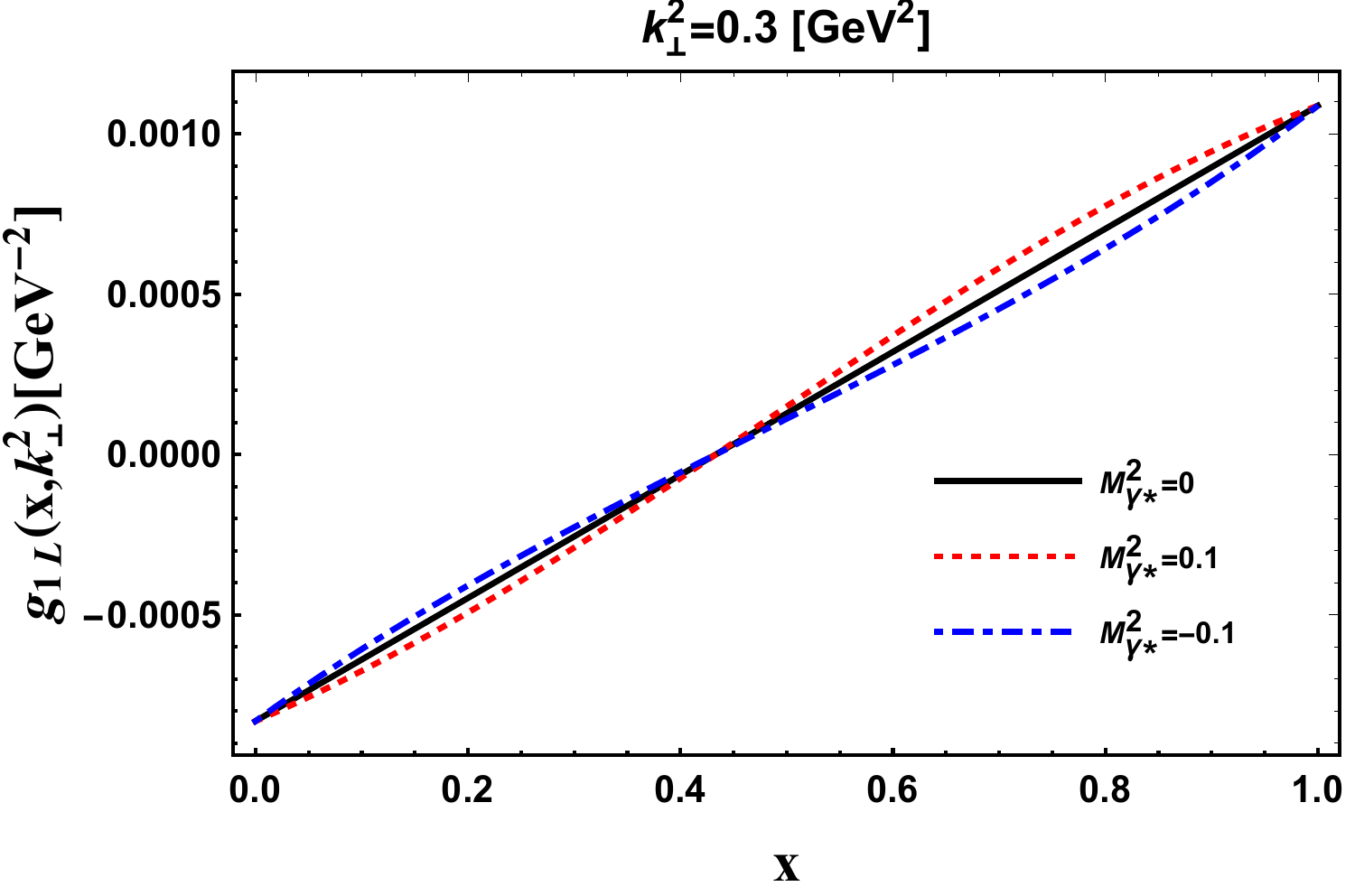}\end{center}
	\end{minipage}
	\begin{minipage}[c]{1\textwidth}\begin{center}
			(c)\includegraphics[width=.43\textwidth]{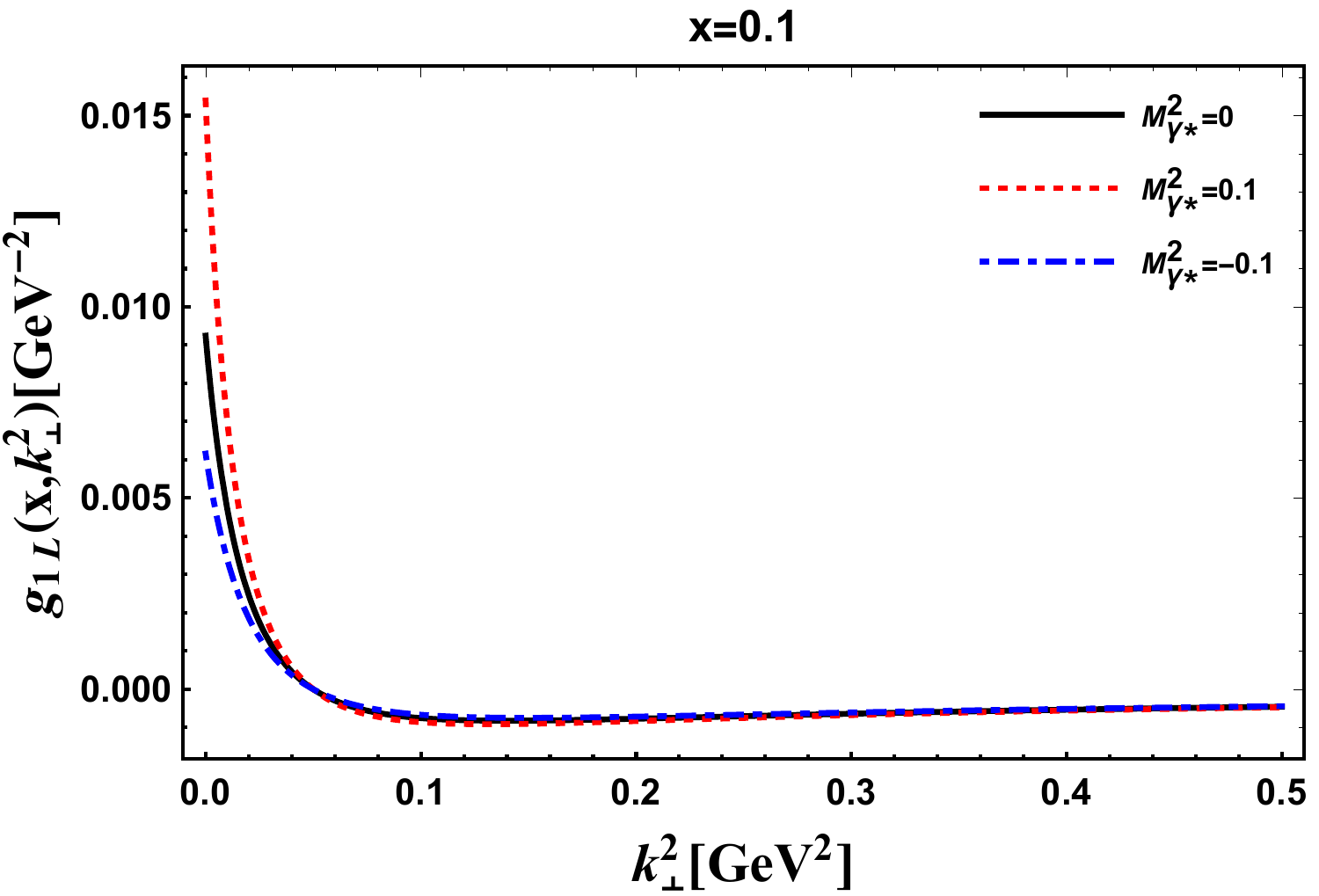}
			(d)\includegraphics[width=.445\textwidth]{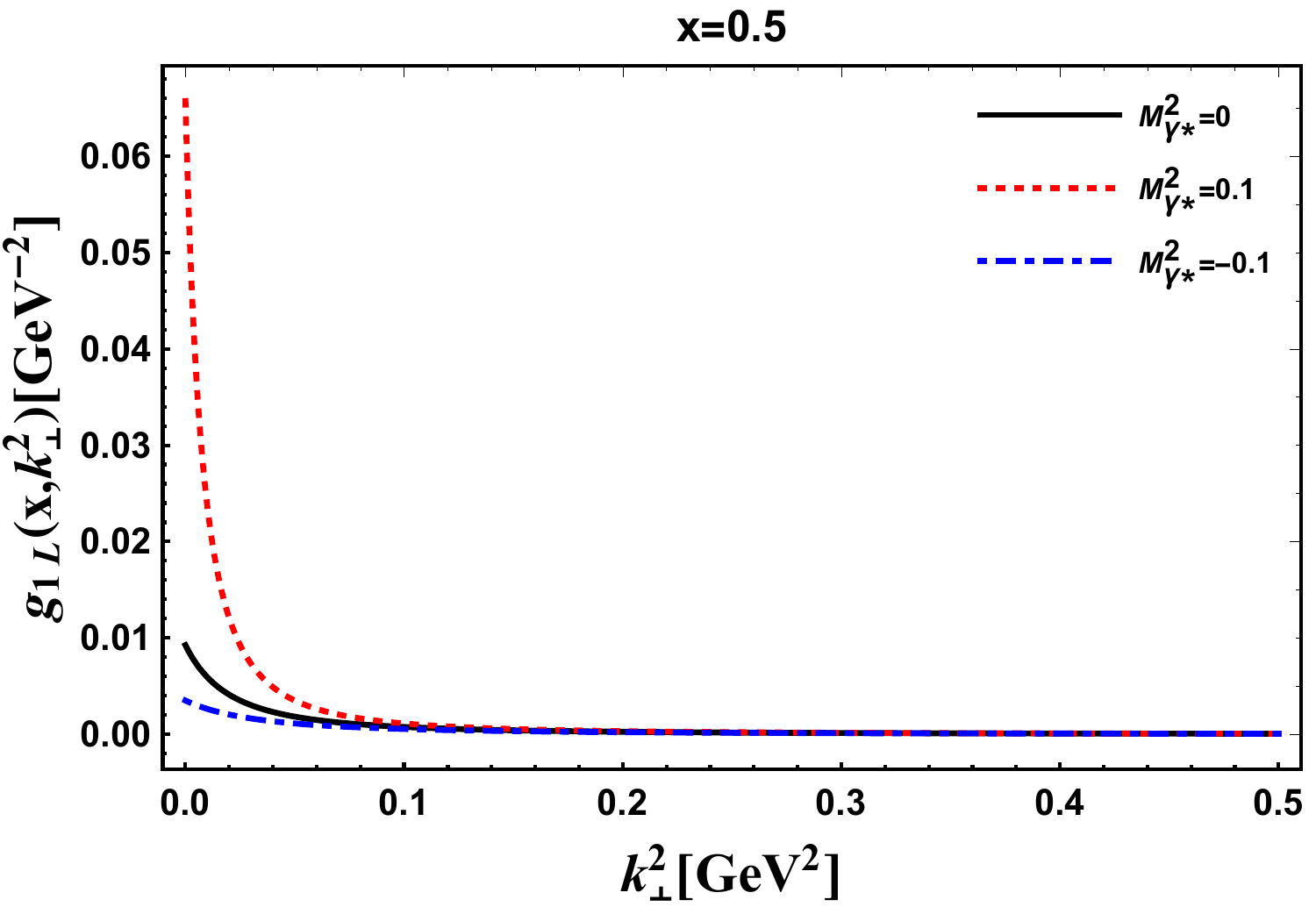}\end{center}
	\end{minipage}
	
	\caption{(Color online) $g_{1L}(x,{\bf k}^2_\perp)$ has been plotted with different $x$ and $\textbf{k}_{\perp}$ for different photon masses $M^2_\gamma=0$, $0.1$ and $-0.1$ GeV$^2$. In the upper panel $g_{1L}(x,{\bf k}^2_\perp)$ has been plotted w.r.t $x$  for a fixed values of $\textbf{k}_{\perp}^2$ as $\textbf{k}_{\perp}^2=0.01$ GeV$^2$ and $\textbf{k}_{\perp}^2=0.3$ GeV$^2$. In lower panel, $g_{1L}(x,{\bf k}^2_\perp)$ has been plotted w.r.t $\textbf{k}_{\perp}^2$ for fixed values of $x$ at $x=0.1$ and $x=0.5$.
		The solid, dashed, and dot-dashed lines represent the real photon, time-like virtual photon, and space-like virtual photon respectively.}
	\label{g1l2d}
\end{figure}

 \begin{figure}[ht]
	\centering
	\begin{minipage}[c]{1\textwidth}\begin{center}
			(a)\includegraphics[width=.43\textwidth]{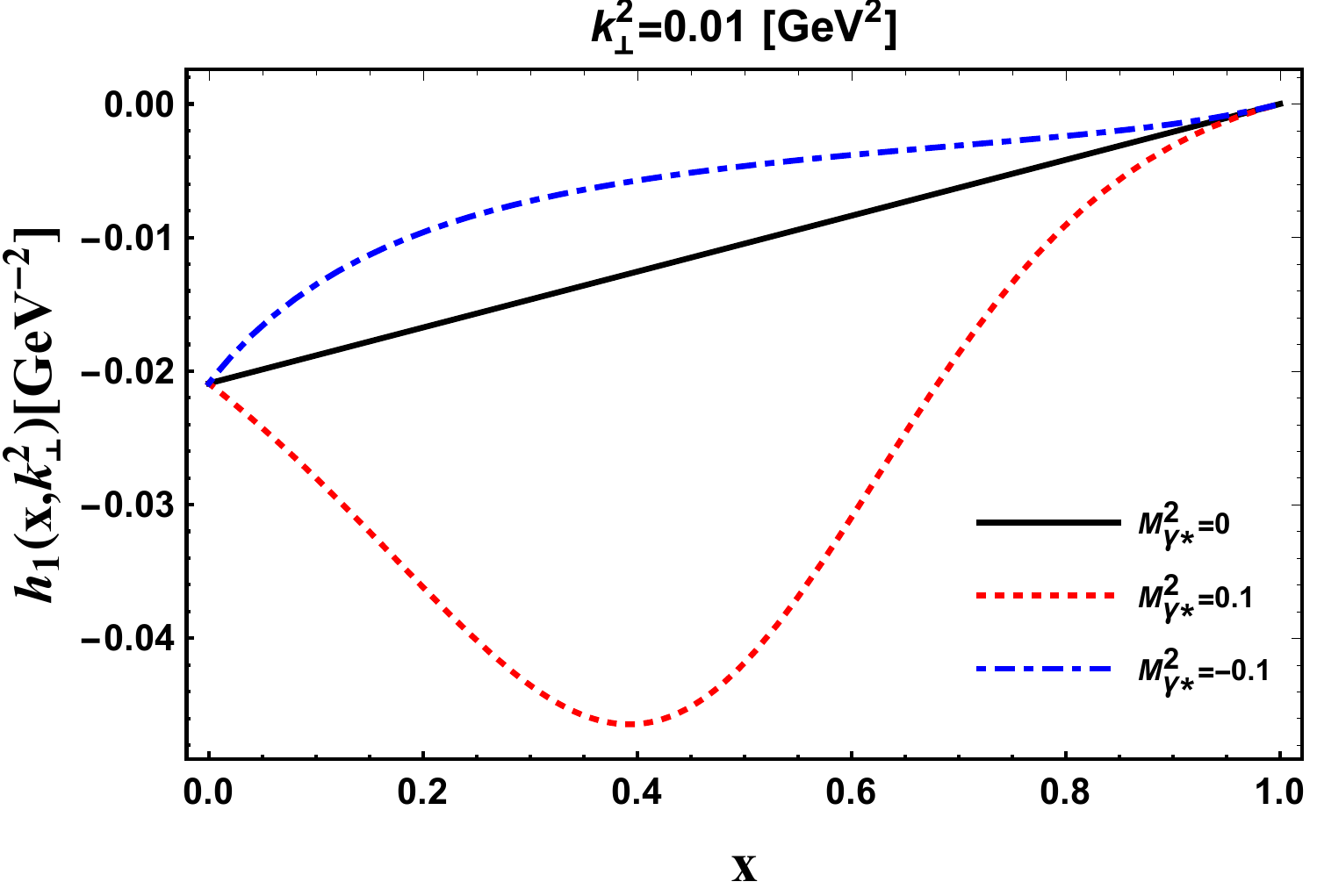}
			(b)\includegraphics[width=.445\textwidth]{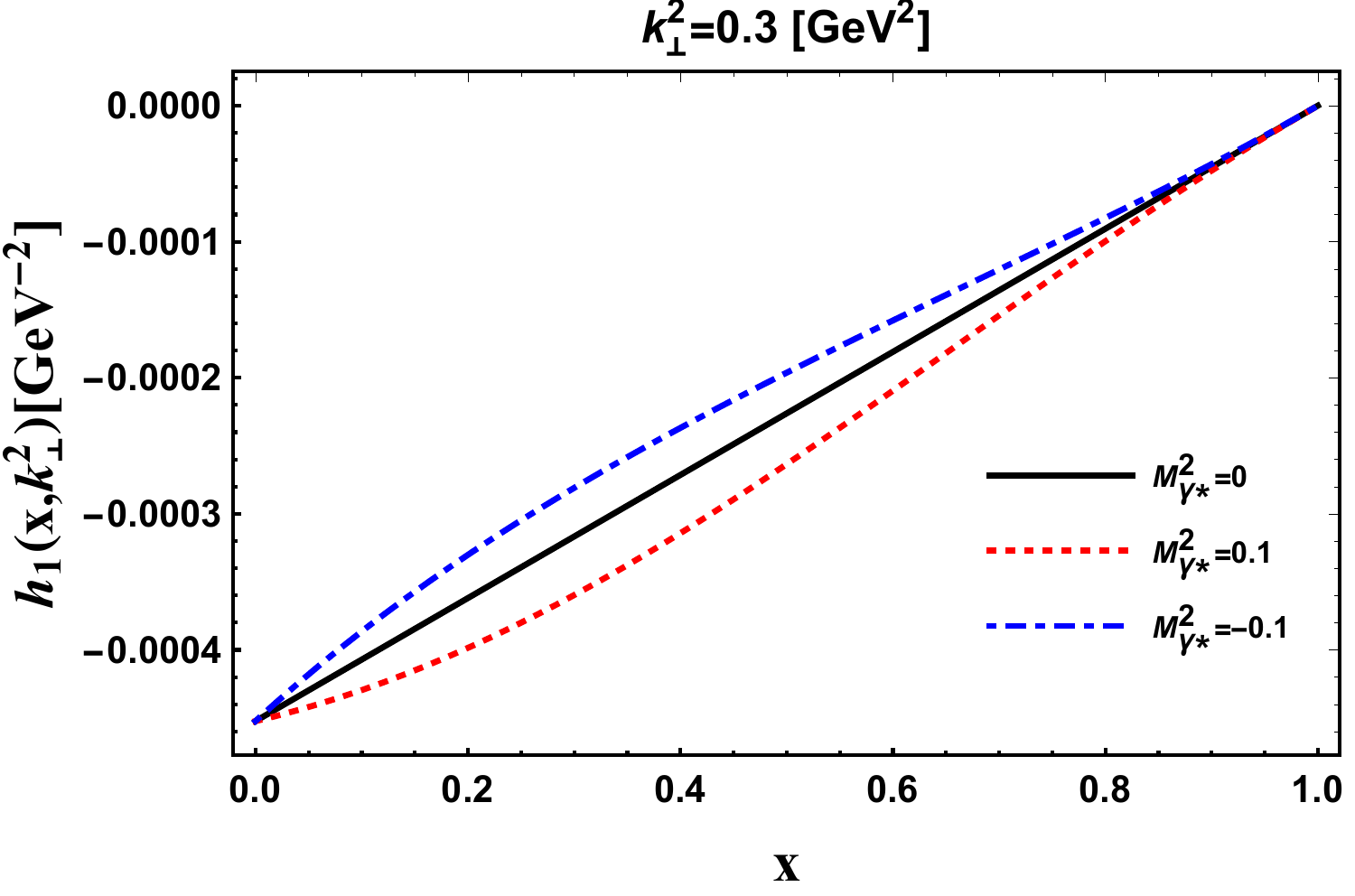}\end{center}
	\end{minipage}
	\begin{minipage}[c]{1\textwidth}\begin{center}
			(c)\includegraphics[width=.43\textwidth]{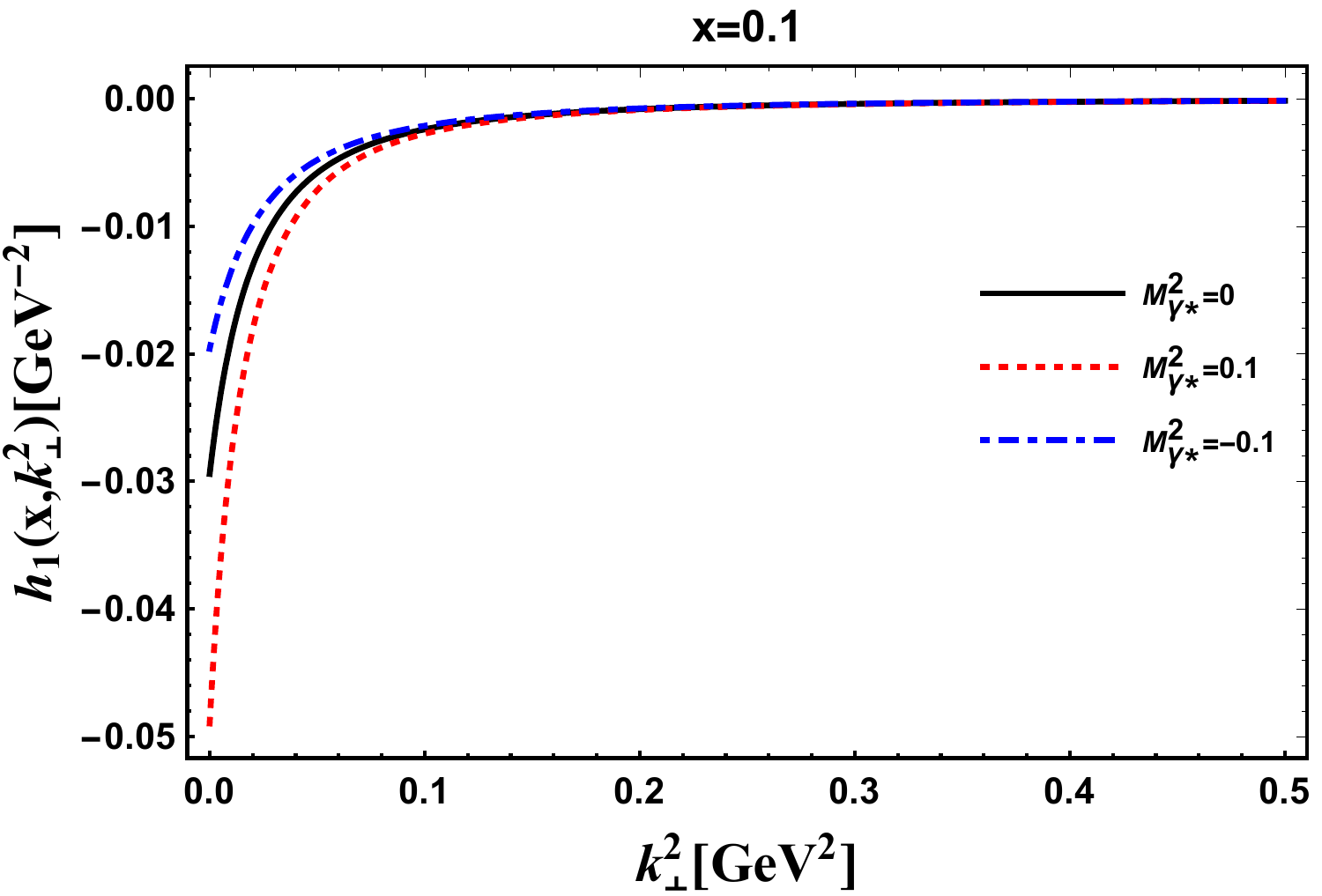}
			(d)\includegraphics[width=.445\textwidth]{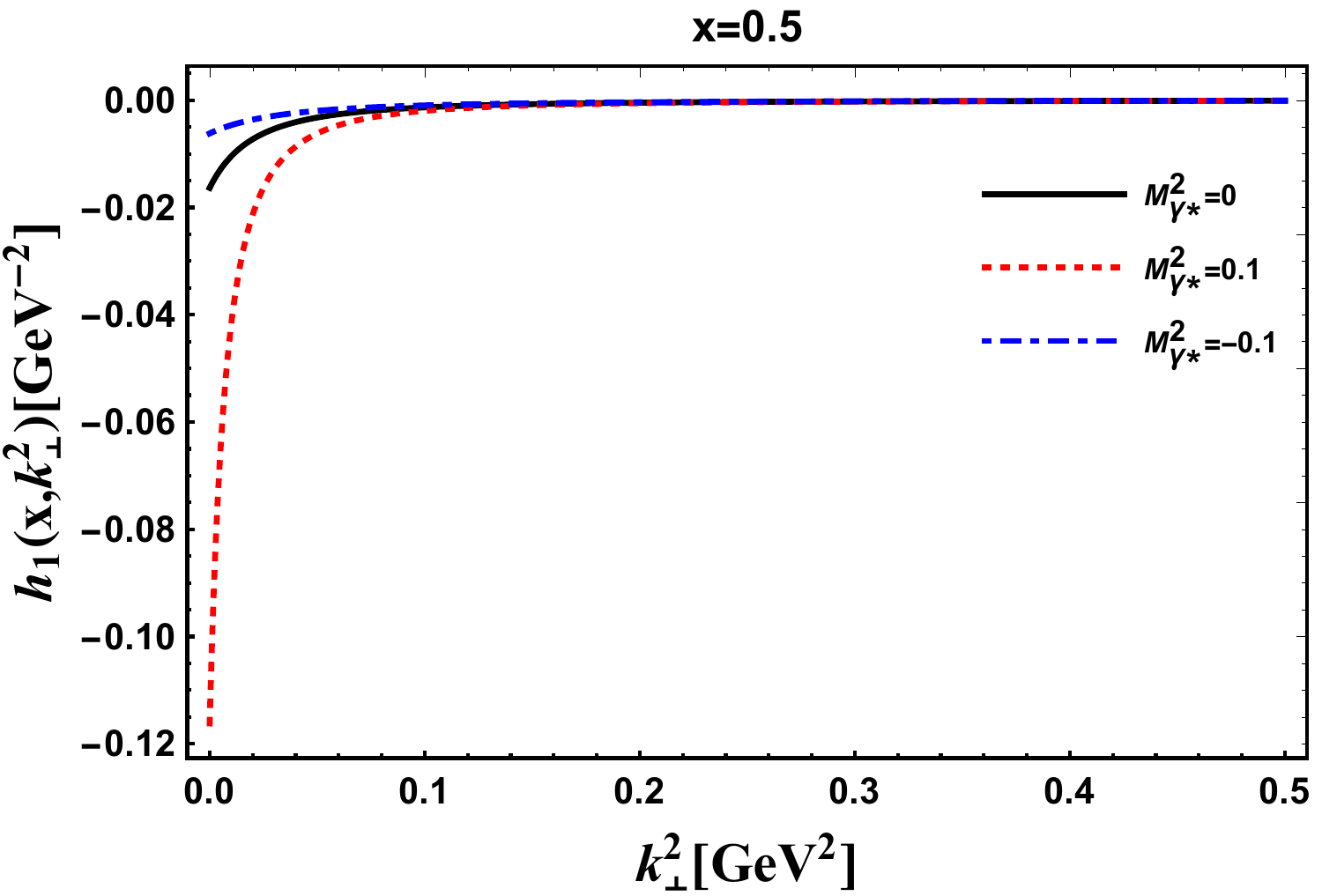}\end{center}
	\end{minipage}
	\caption{(Color online) $h_1(x,{\bf k}^2_\perp)$ has been plotted with different $x$ and $\textbf{k}_{\perp}$ for different photon masses $M^2_\gamma=0$, $0.1$ and $-0.1$ GeV$^2$. In the upper panel $h_1(x,{\bf k}^2_\perp)$ has been plotted w.r.t $x$ for a fixed values of $\textbf{k}_{\perp}^2$ as $\textbf{k}_{\perp}^2=0.01$ GeV$^2$ and $\textbf{k}_{\perp}^2=0.3$ GeV$^2$. In lower panel,
	 $h_1(x,{\bf k}^2_\perp)$	 has been plotted w.r.t $\textbf{k}_{\perp}^2$ for fixed values of $x$ at $x=0.1$ and $x=0.5$. The blue, red and black color curves are for real photon, time-like and space-like virtual photon respectively.}
	\label{h12d}
\end{figure}

For more understanding between real and virtual photon, we have presented $f_{1}(x,\textbf{k}^2_{\perp})$, $g_{1L}(x,\textbf{k}^2_{\perp})$ and $h_{1}(x,\textbf{k}^2_{\perp})$ TMDs with respect to $x$ at fixed values of $\textbf{k}_\perp^2$ = 0.01 and  0.3 GeV and also with respect to $\textbf{k}^2_\perp$ at fixed values of $x$ = 0.1 and 0.5 in Figs. (\ref{f12d}), (\ref{g1l2d}) and (\ref{h12d}) respectively. At $M^2_{\gamma}= 0.1$ GeV$^2$, for fixed $\textbf{k}^2_{\perp}$, the unpolarized $f_{1}(x,\textbf{k}^2_{\perp})$ TMD shows a peak around $x=0.5$ as compared to other mass values. However, as we increase the value of $\textbf{k}_\perp^2=0.3$, the unpolarized $f_{1}(x,\textbf{k}^2_{\perp})$ TMD approaches its maximum value when either the quark or the anti-quark carries most of the longitudinal momentum fraction. This occurs at the end points of $x$. Meanwhile, $f_{1}(x,\textbf{k}^2_{\perp})$ approaches its minimum value at $x = 0.5$, i.e., when the quark anti-quark pair share exactly equal momenta. From Fig. \ref{g1l2d}(b), it can be seen that the $g_{1L}(x,\textbf{k}^2_{\perp})$ TMD vanishes at around $x=0.45$ for all the photon masses and breaks the symmetry. The TMD $g_{1L}(x,\textbf{k}^2_{\perp})$ becomes negative for certain range of $x$ and $\textbf{k}^2_{\perp}$.
 \begin{figure}[ht]
	\centering
	\begin{minipage}[c]{1\textwidth}
		\begin{center}
			(a)\includegraphics[width=.40\textwidth]{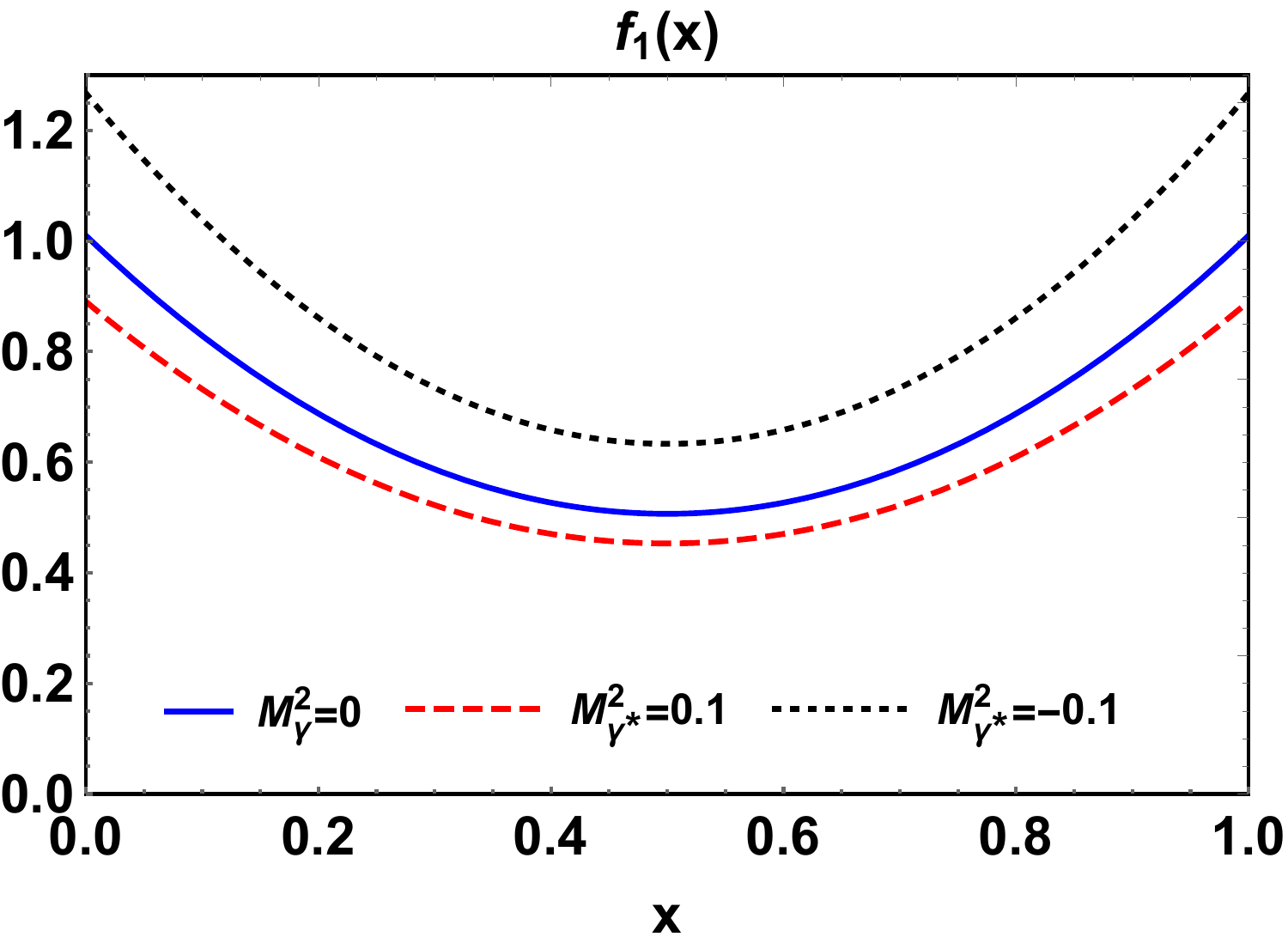}\hspace{1cm}
			(b)\includegraphics[width=.40\textwidth]{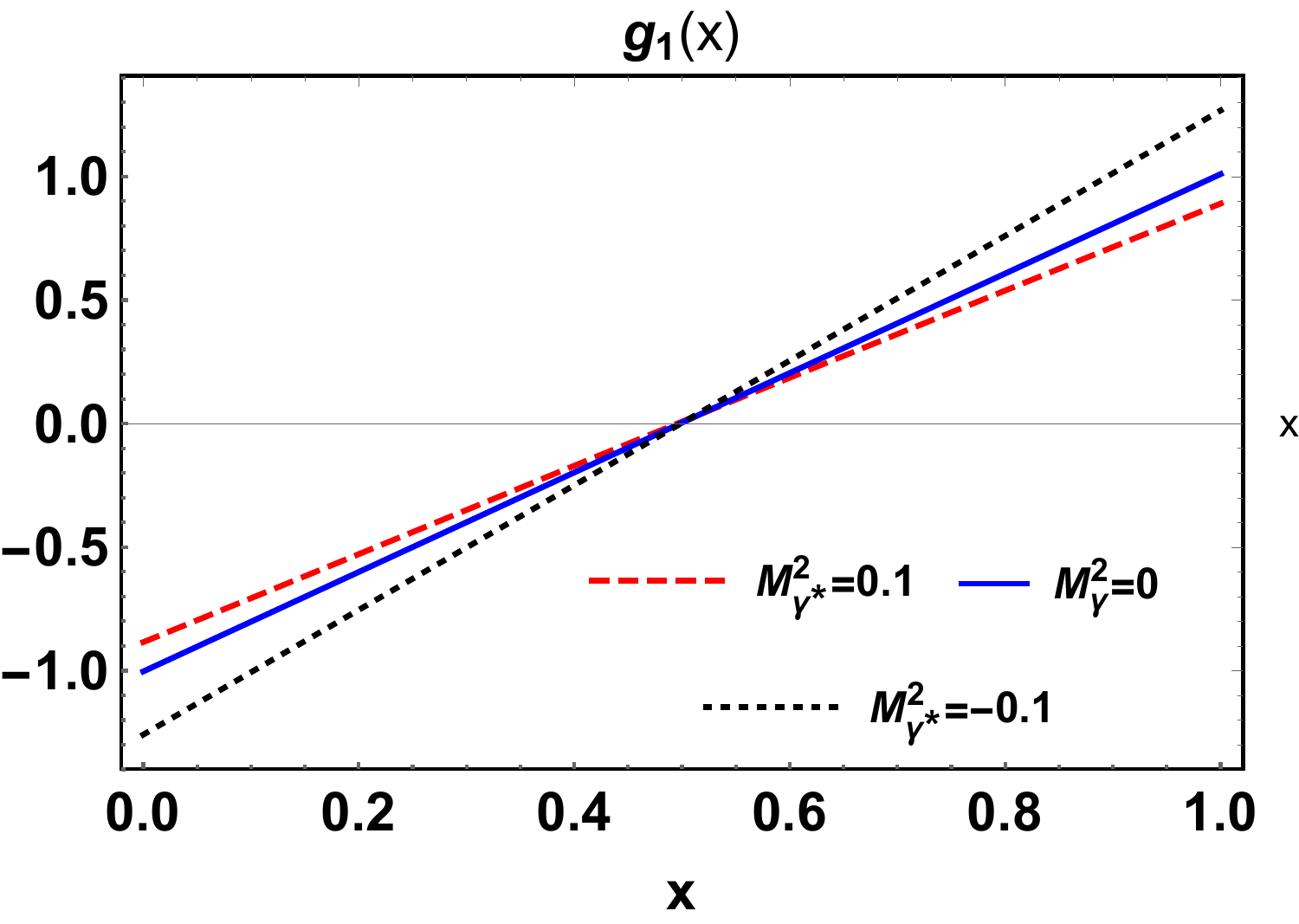}
		\end{center}
	\end{minipage}
	\begin{minipage}[c]{1\textwidth}
		\begin{center}
			(c)\includegraphics[width=.41\textwidth]{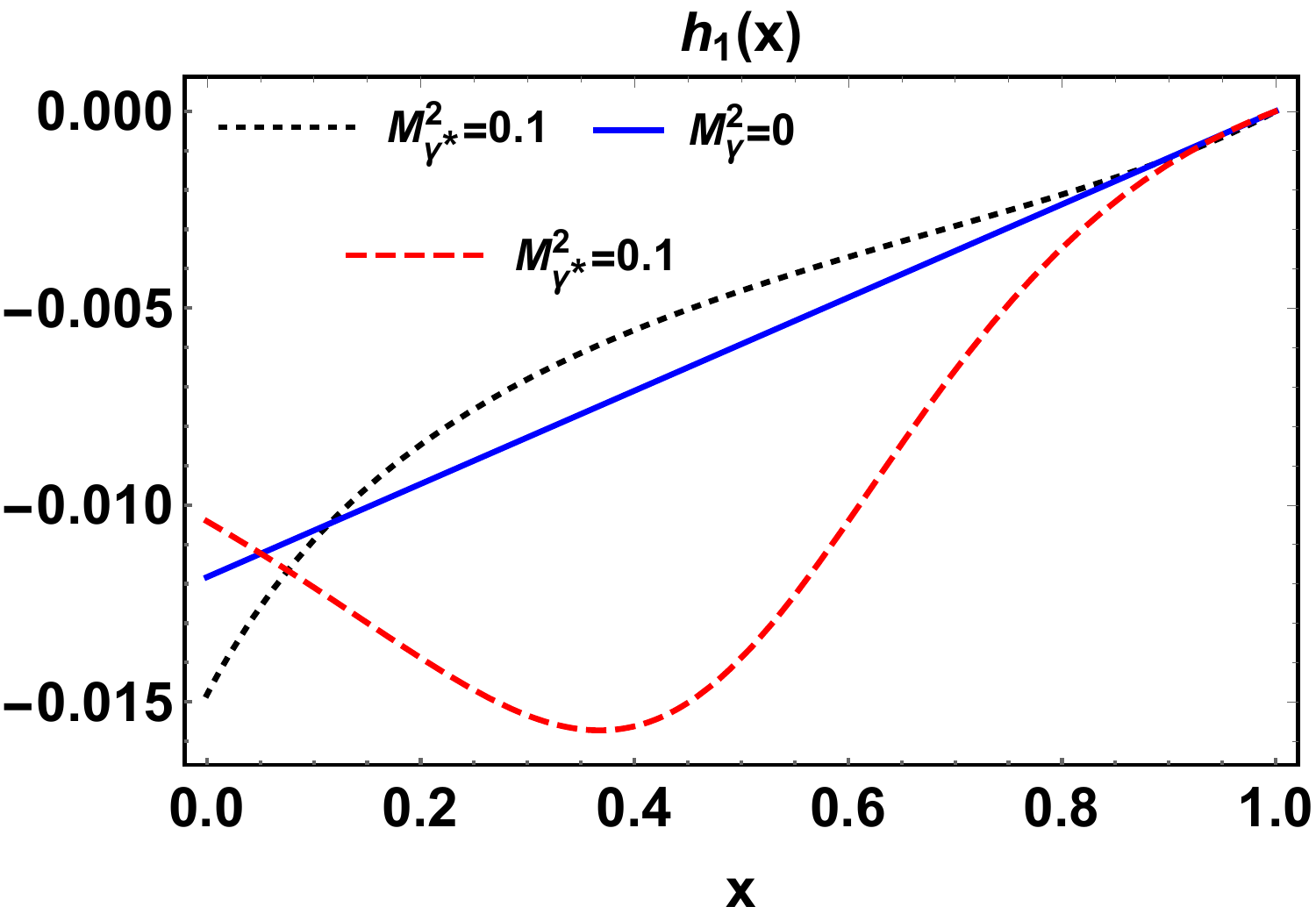}\hspace{1cm}
			(d)\includegraphics[width=.40\textwidth]{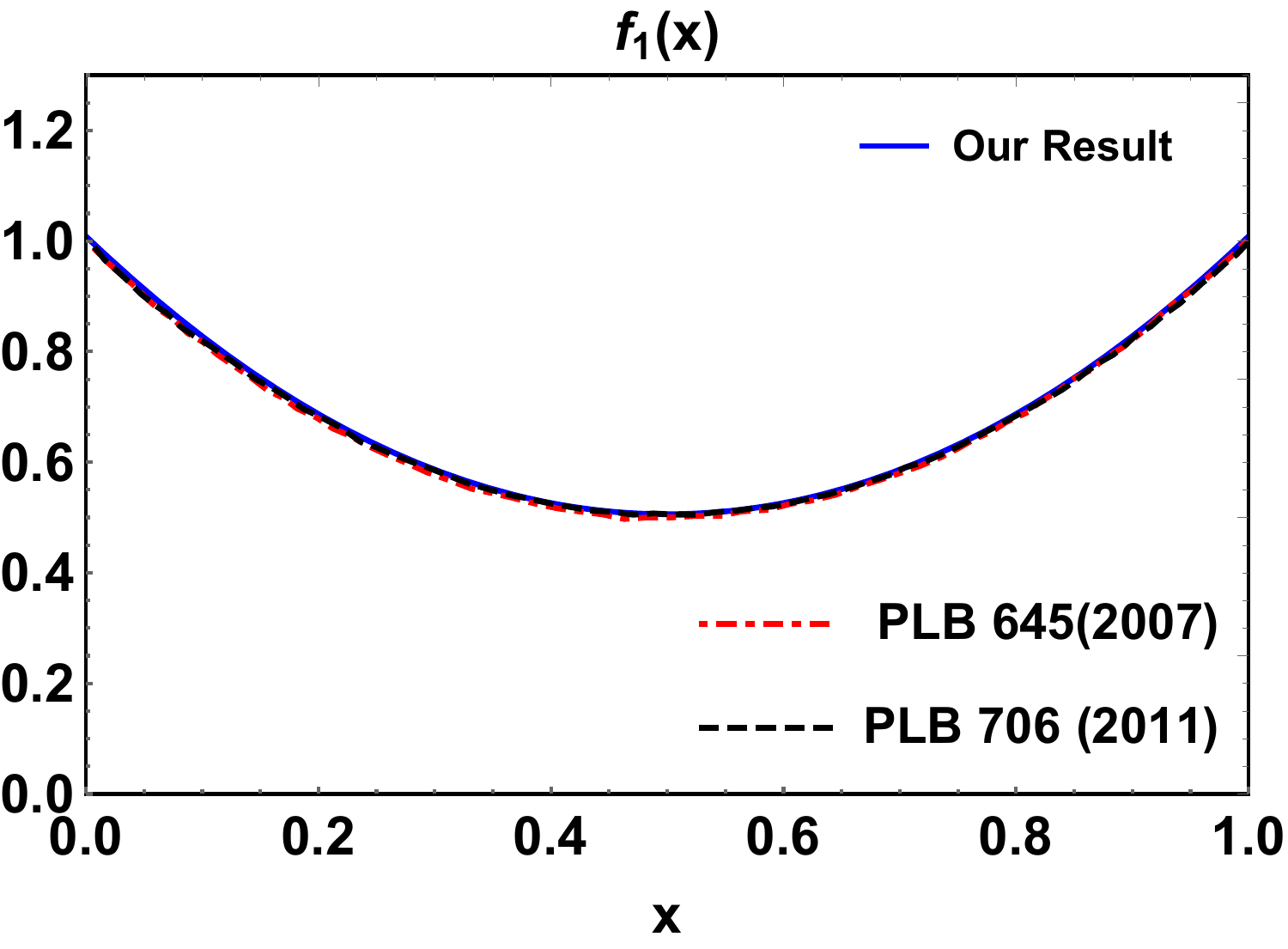}
		\end{center}
	\end{minipage}
 \caption{(Color online) $f_1(x)$, $g_1(x)$ and $h_1(x)$ PDFs have been plotted w.r.t longitudinal momentum fraction $x$ at different photon masses $M^2_\gamma=0$, $0.1$ and $-0.1$ GeV$^2$. The blue, red and black color curves are for real photon, time-like and space-like virtual photon PDFs respectively. We have also compared our unpolarized $f_{1}(x)$ PDF with the result from Refs. \cite{Friot:2006mm,Mukherjee:2013yf}}.
 \label{pdfs}
 \end{figure}

\section{Parton distribution functions}\label{pdF}

PDFs are one-dimensional probability distribution functions encoding  information about polarization and longitudinal momentum fraction $x$ of quark inside a photon \cite{pdf,pdf1,pdf2,pdf3}. Even though the PDFs do not carry any information about the transverse momentum distribution, they have a direct connection with the TMDs and GPDs and can be easily accessed through the DIS experiments \cite{dis}. An extra tensor PDF $f_{1LL}(x)$ exists for the case of spin-$1$ hadron (photon) as compared to the PDFs in the spin-$1/2$ nucleon case \cite{tmd13}.  One can derive these PDFs by integrating $f_1(x,\textbf{k}^2_{\perp})$, $g_{1L}(x,\textbf{k}^2_{\perp})$, $h_1(x,\textbf{k}^2_{\perp})$ and $f_{1LL}(x,\textbf{k}^2_{\perp})$ TMDs over transverse momentum $k_{\perp}$. In our present case, $f_{1LL}(x,\textbf{k}^2_{\perp})$ TMD is zero, therefore, we left with three PDFs. These PDFs can be expressed as
\begin{eqnarray}
f_1(x) & = & \int d^2\mathbf{k}_{\perp} f_1(x,\textbf{k}^2_{\perp}),\\
g_{1}(x) & = & \int d^2\mathbf{k}_{\perp} g_{1L}(x,\textbf{k}^2_{\perp}),\\
h_1(x) & = & \int d^2\mathbf{k}_{\perp} h_{1}(x,\textbf{k}^2_{\perp}). 
\end{eqnarray}
On the other hand, one can solve the PDF correlator of Eq. \ref{phi1}, where the PDFs can be expressed by choosing the suitable form of Dirac matrices as
\begin{eqnarray}
\langle \gamma^+ \rangle_{\boldsymbol{ \mathcal{S}}}^{(\Lambda)} (x) 
&\equiv & f_1(x) + \mathcal{S}_{LL} \, 
f_{1LL}(x) \,, \\
\label{form1-pdf} 
\langle\gamma^{+} \gamma_{5}\rangle^{(\Lambda)}_{\boldsymbol{ \mathcal{S}}} (x) 
&\equiv  & \mathcal{S}_L\,g_{1}(x), \\
\label{form2-pdf} 
\langle\gamma^+\gamma^i\gamma_{5}\rangle^{(\Lambda)}_{\boldsymbol{ \mathcal{S}}} (x) 
& \equiv & \mathcal{S}_\perp^i h_1(x)\,.
\label{form3-pdf}
\end{eqnarray}
For numerical results and to verify the PDF sum rule, we have taken a normalization constant of $2.87496$ for each of PDFs for the case of real photon. We have presented $f_{1}(x)$, $g_{1}(x)$ and $h_{1}(x)$ PDFs with respect to longitudinal momentum fraction $x$ in Fig.  \ref{pdfs} for different photon masses. The $f_{1}(x)$ PDF shows symmetry under $x$ having peaks at $x=0$ and $1$. The magnitude of $f_{1}(x)$ can be arranged in the order of mass as $M^2_{\gamma*}$(space-like virtual photon) $>$ $M^2_{\gamma}$(real photon) $>$ $M^2_{\gamma*}$(time-like virtual photon). While looking into $g_1(x)$ PDF in Fig. \ref{pdfs} (b), we can see that it has both positive and negative distributions at $x>0.5$ and $x<0.5$ respectively. In Fig. \ref{pdfs}(c), we have plotted the $h_{1}(x)$ PDF, which shows a negative distribution at low values of $x$ and vanishes at $x=1$. We have also compared our  result for $f_{1}(x)$ with the PDF extracted from GPDs in Ref. \cite{Friot:2006mm,Mukherjee:2013yf}. We observe that our results overlap with their predictions as  shown in Fig. \ref{pdfs}(d).
The normalized real photon PDFs of present models satisfy the following sum rules 
\begin{eqnarray}
\int_0^1 {\rm d}x~f_1(x)&=&1,\\
\int_0^1 {\rm d}x~x\,f_1(x) + \int_0^1 {\rm d}x~(1-x)\,f_1(x)&=&1,\\
\int_0^1 {\rm d}x~g_{1L}(x)&=&0.
\end{eqnarray}

\section{Spin densities in momentum space for photon}\label{transverse}

 \begin{figure}[t]
	\centering
	\begin{minipage}[c]{1\textwidth}\begin{center}
			(a)\includegraphics[width=.43\textwidth]{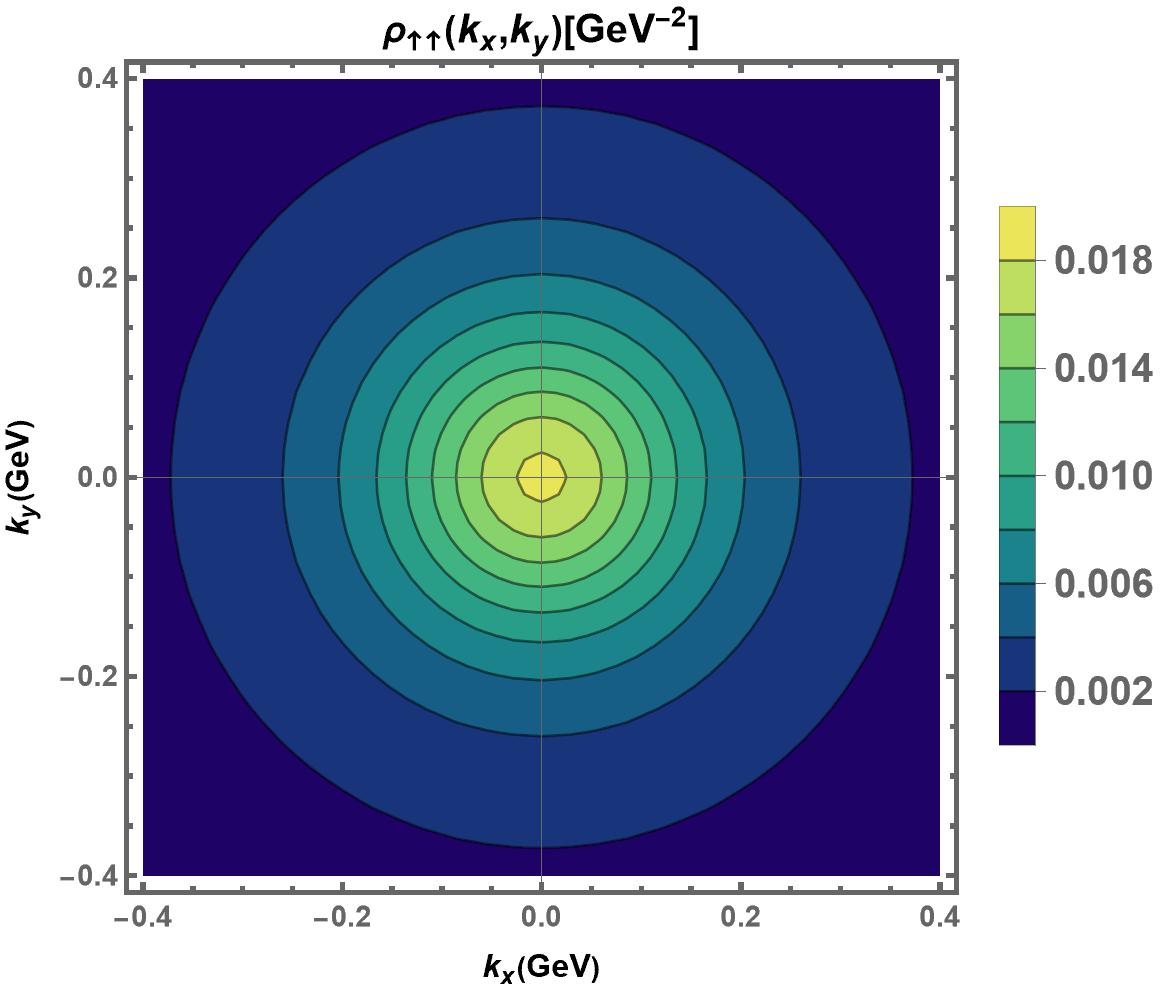}
			(b)\includegraphics[width=.445\textwidth]{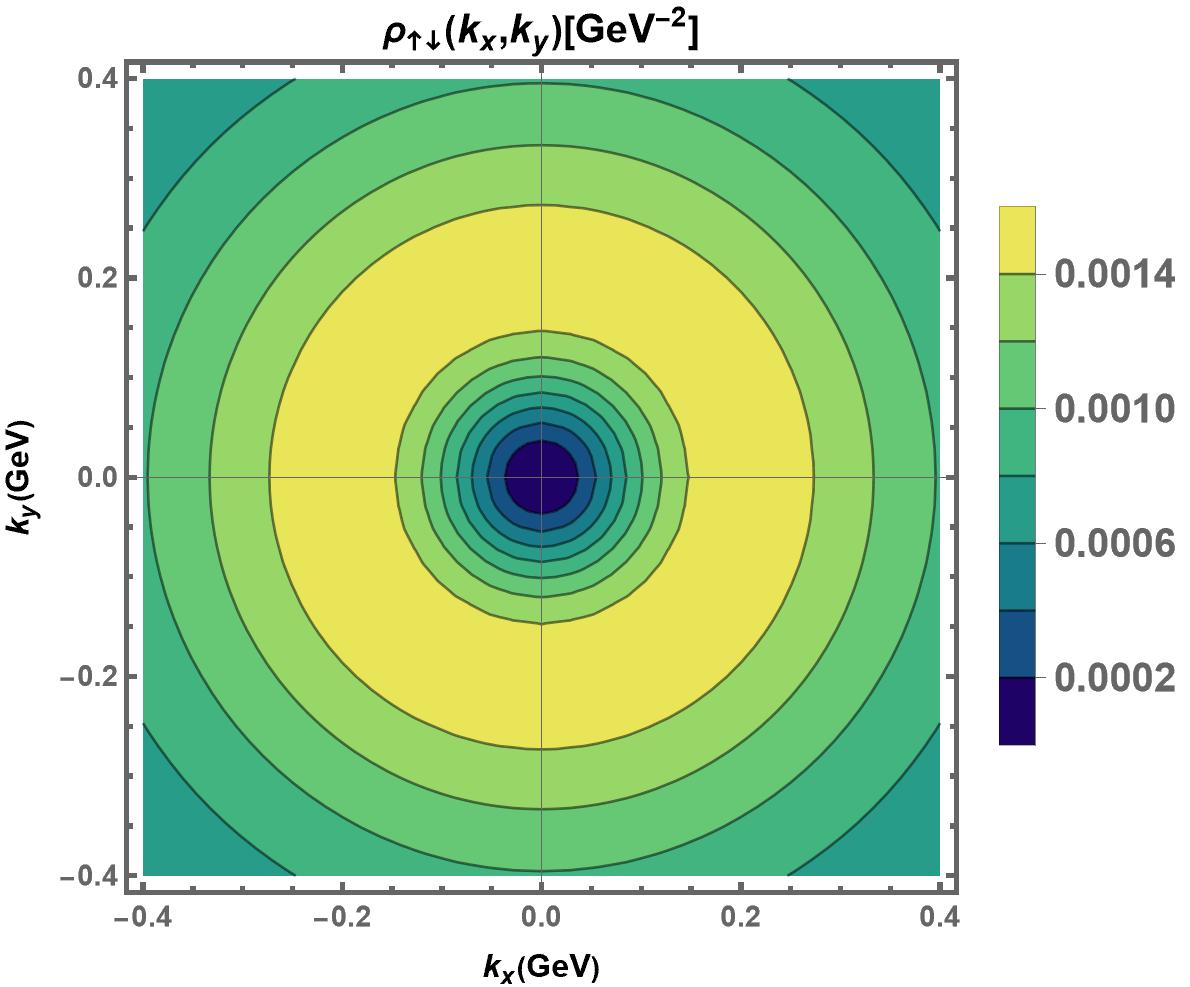}\end{center}
	\end{minipage}
	\begin{minipage}[c]{1\textwidth}\begin{center}
			(c)\includegraphics[width=.43\textwidth]{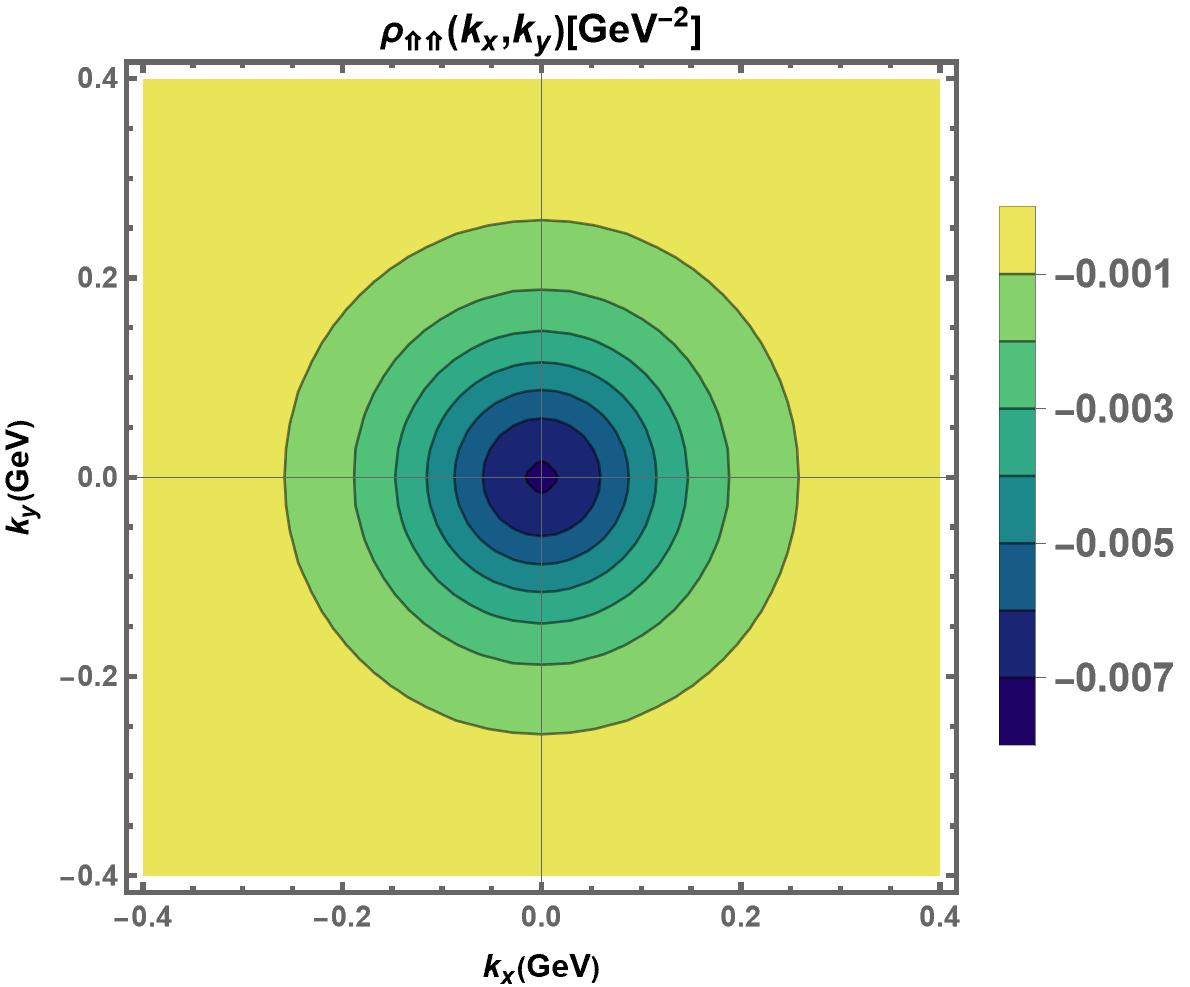}
			(d)\includegraphics[width=.445\textwidth]{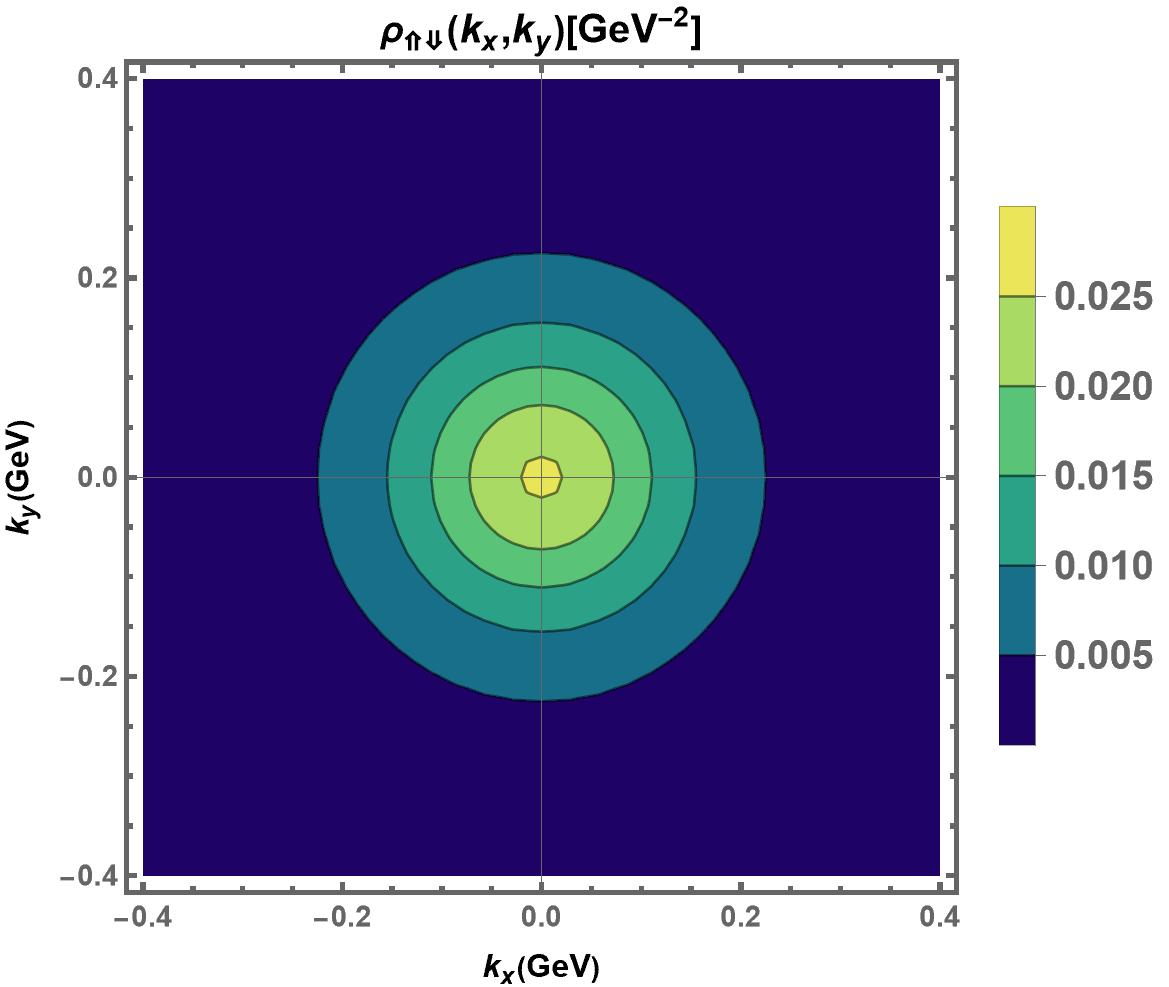}\end{center}
	\end{minipage}
 \caption{(Color line) Longitudinal spin densities $\rho_{\uparrow \uparrow}$ and $\rho_{\uparrow \downarrow}$ in the upper panel in momentum space for real photon. The transverse spin densities $\rho_{\Uparrow \Uparrow}$ and $\rho_{\Uparrow \Downarrow}$ for real photon in the momentum space.} 
 \label{photonspin}
 \end{figure}
The spin-spin correlation between the quark and photon can be described through the spin densities and they can be obtained from the TMDs. Using the TMDs, the quark momentum distributions inside the target with various polarization combinations may be determined. To calculate the spin densities, we have integrated the TMDs over longitudinal momentum fraction ($x$). By considering the different polarization configurations of the quark and the photon spin in the longitudinal  and transverse direction, the spin densities can be expressed as \cite{Kaur:2020emh}
\begin{eqnarray}
    \rho_{\uparrow \uparrow}(k_x,k_y) &=& f_1(k_x,k_y)-\frac{1}{3} f_{1LL} (k_x,k_y) + g_{1L} (k_x,k_y),
     \\
     \rho_{\uparrow \downarrow}(k_x,k_y) &=& f_1(k_x,k_y)-\frac{1}{3} f_{1LL} (k_x,k_y) - g_{1L} (k_x,k_y), 
     \end{eqnarray}
\begin{eqnarray}
     \rho_{\Uparrow \Uparrow}(k_x,k_y) &=& f_1(k_x,k_y)-\frac{\textbf{k}^2_\perp}{M^2_{\gamma}} f_{1TT} (k_x,k_y) + h_{1} (k_x,k_y), \\
      \rho_{\Uparrow \Downarrow}(k_x,k_y) &=& f_1(k_x,k_y)-\frac{\textbf{k}^2_\perp}{M^2_{\gamma}} f_{1TT} (k_x,k_y) - h_{1} (k_x,k_y), 
     \\
     \rho_{\uparrow \Uparrow (\uparrow \Downarrow)}(k_x,k_y) &=& f_1(k_x,k_y) \pm \frac{k_x}{M_\gamma} g_{1T} (k_x,k_y), 
     \\
     \rho_{\Uparrow \uparrow (\Uparrow \downarrow)}(k_x,k_y) &=& f_1(k_x,k_y) \pm \frac{k_x}{M_\gamma} h^{\perp}_{1L} (k_x,k_y). \label{spinden}
\end{eqnarray}
\begin{figure}[t]
	\centering
 \begin{minipage}[c]{1\textwidth}\begin{center}
			(a)\includegraphics[width=.43\textwidth]{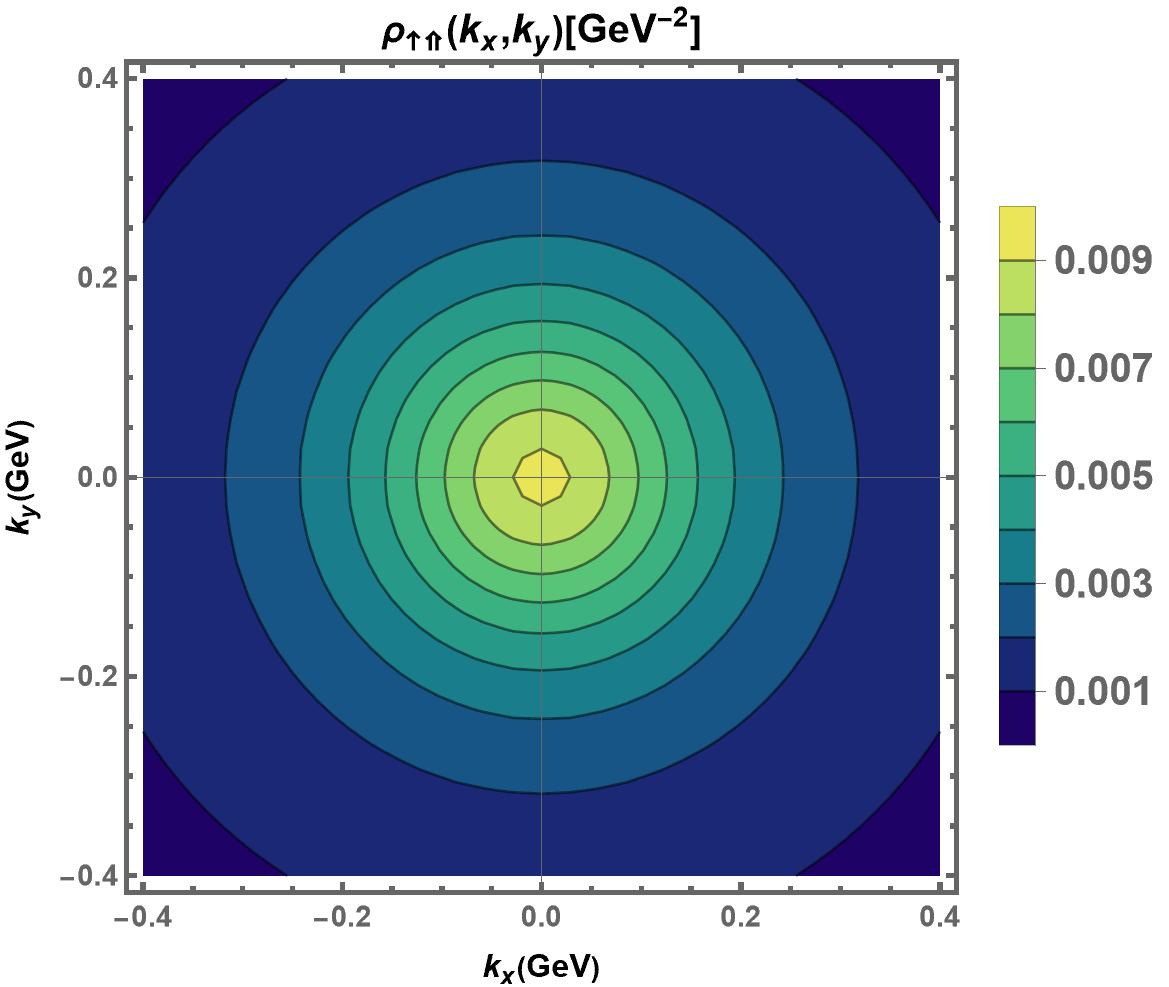}
			(b)\includegraphics[width=.445\textwidth]{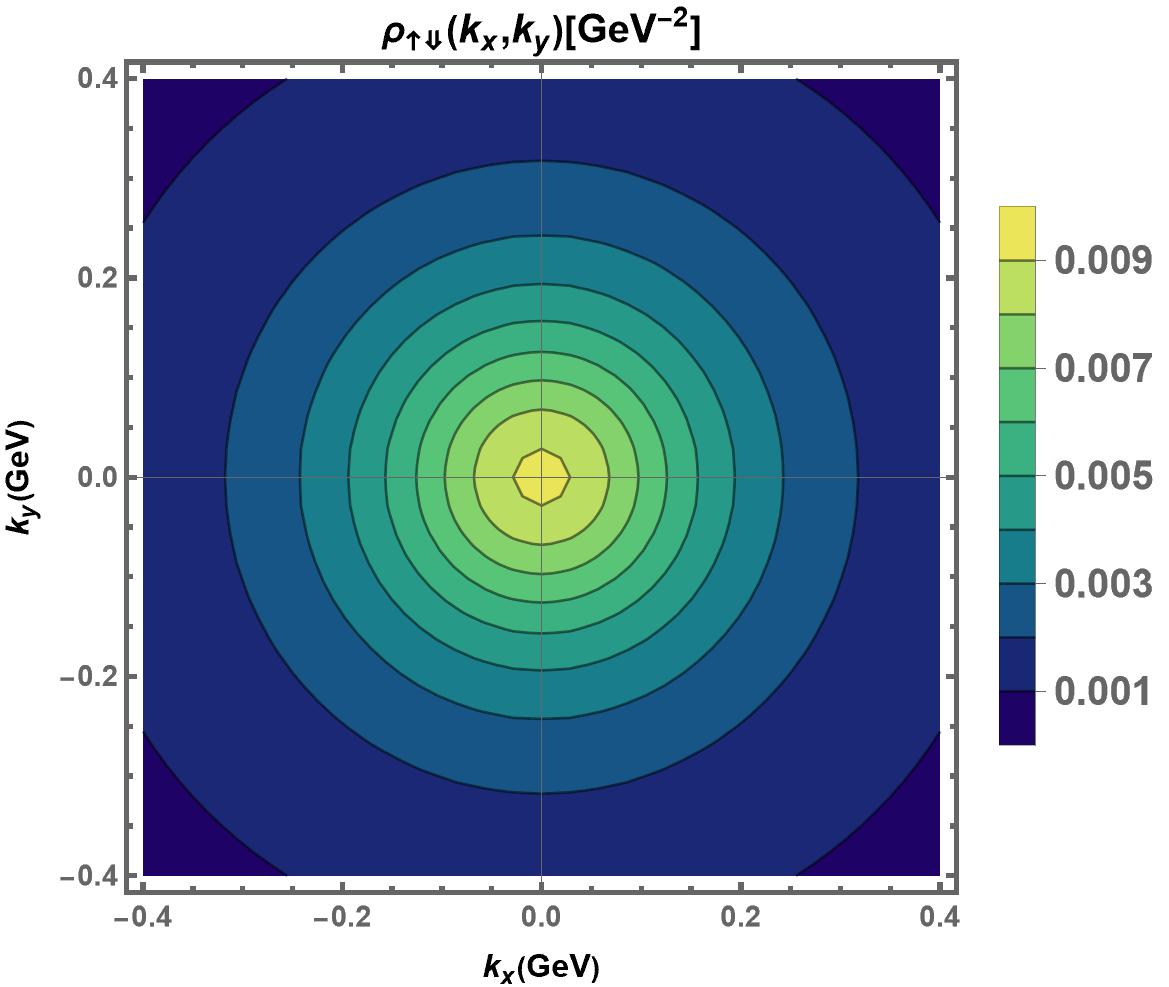}\end{center}
	\end{minipage}
	\begin{minipage}[c]{1\textwidth}\begin{center}
			(c)\includegraphics[width=.43\textwidth]{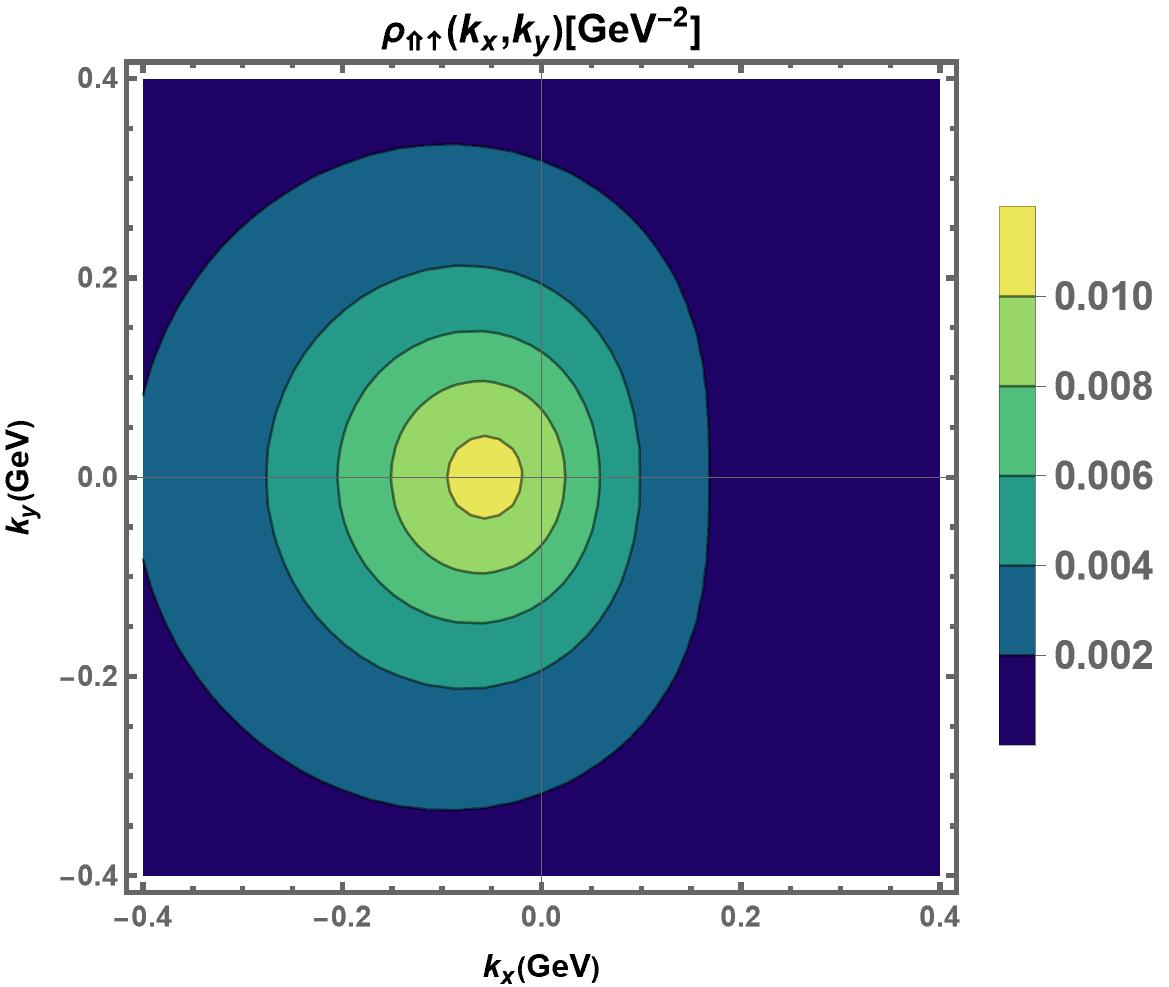}
			(d)\includegraphics[width=.445\textwidth]{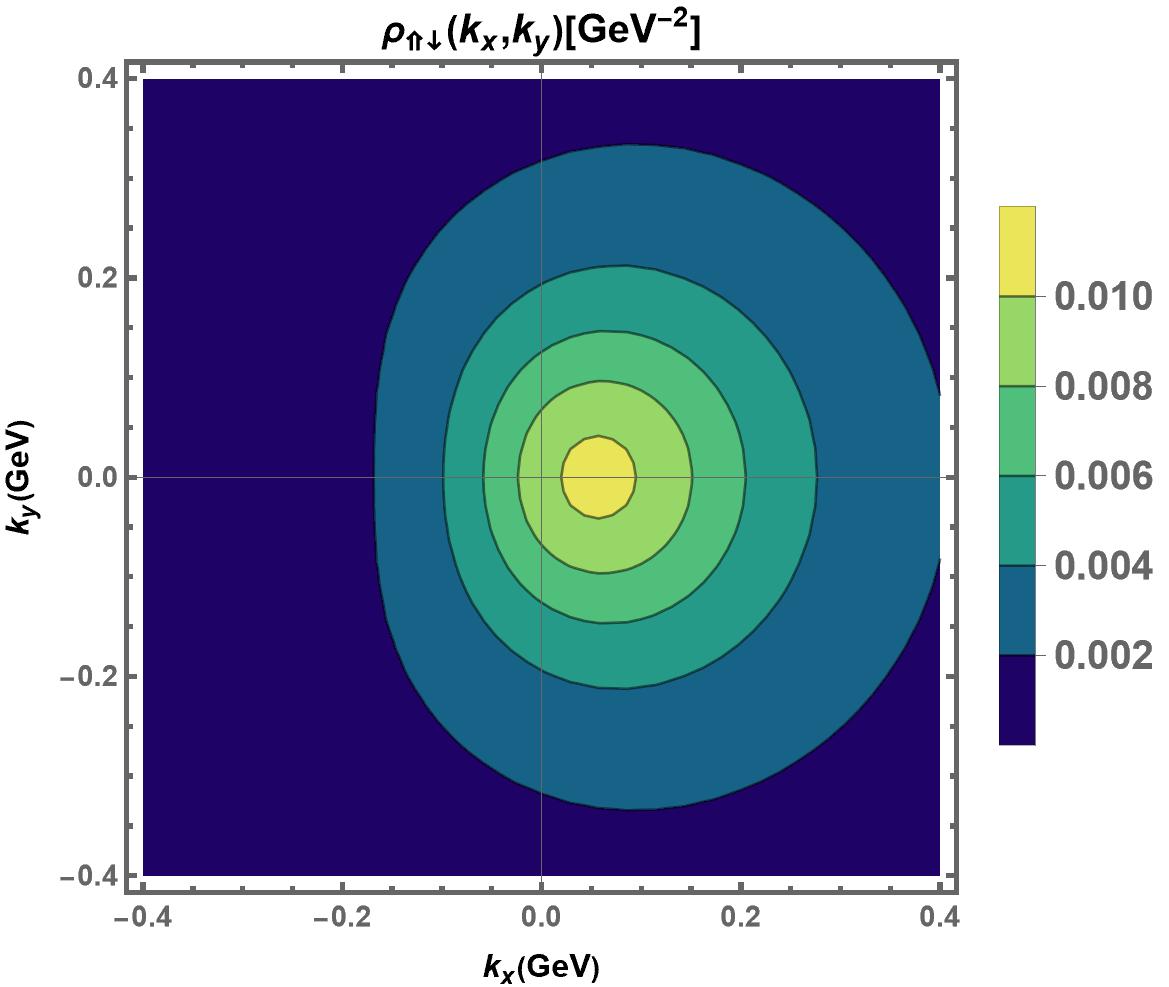}\end{center}
	\end{minipage}
	
	\caption{(Color online)  The spin densities of longitudinal quark spin and transverse photon spin $\rho_{\uparrow \Uparrow (\uparrow \Downarrow)}$ in the upper panel for real photon, while $\rho_{\Uparrow \uparrow (\Uparrow \downarrow)}$ in the lower panel.}
	\label{photonspin1}
\end{figure}
Here, $\rho_{\uparrow \uparrow}(k_x,k_y)$ is the spin density when both the quark and photon have longitudinal spin. The first and second arrow in $\rho$ indicates the spin of quark and photon respectively. $\uparrow(\downarrow)$ represents the longitudinal and transverse spin of the quark  whereas $\Uparrow(\Downarrow)$ represents the longitudinal and transverse spin of the photon. In the present work, we have considered only the real photon case for the calculation of spin densities. In Fig. \ref{photonspin}(a) and (b), we have presented the results for $\rho_{\uparrow \uparrow}$ and $\rho_{\uparrow \downarrow}$ showing the spin-spin correlations when both quark and photon are longitudinally polarized in same and opposite direction respectively. Similarly, in Fig. \ref{photonspin}(c) and (d), we have presented the results for  $\rho_{\Uparrow \Uparrow} $ and $\rho_{\Uparrow \Downarrow}$ showing the spin-spin correlations when both quark and photon are transversely polarized in same and opposite direction respectively. We observe that all the spin densities are symmetric around the center of momentum space. 

In Fig. \ref{photonspin1}(a) and (b) we have presented the results for $\rho_{\uparrow \Uparrow}$$\rho_{\uparrow \Downarrow}$ spin densities which again shows spin-spin correlation when the photon is transversely polarized. We observe that the distribution is again symmetric around the center of momentum space. However, this distribution receives a contribution from the $g_{1T}$ whose actual contribution is very small in magnitude and therefore it didn't distort the distribution. However, in Fig. \ref{photonspin1}(c) and (d), results are presented when the quark is transversely polarized. It is clear from the plots and from Eq. \ref{spinden} that $h_{1L}^{\perp}$ contributes significantly which results in distortion of the distribution in momentum space. 


\section{conclusion}\label{conclude}
In this work, we have presented the results for the leading twist T-even TMDs of the photon using the helicity amplitudes which are basically the overlap form of LFWFs. We consider photon as a composite system of quark anti-quark pair for these calculations. We have taken both the cases of photon being real (massless)  and virtual (space-like and time-like). For the case of real photon, the mass of the photon is taken as zero whereas for the time-like (space-like) virtual photon, the mass square is taken as positive (negative).

In case of real photon, only unpolarized $f_1(x,\textbf{k}^2_{\perp})$, longitudinally polarized $g_{1L}(x,\textbf{k}^2_{\perp})$ and transversely polarized $h_1(x,\textbf{k}^2_{\perp})$ photon TMDs exists. For both real and virtual photon, all the T-even TMDs show maximum distribution around small $\textbf{k}^2_{\perp}$ and decreases with increasing $\textbf{k}^2_{\perp}$. The TMDs show peak distributions around $x=0.5$ for real and space-like virtual photon. The unpolarized $f_1(x,\textbf{k}^2_{\perp})$ TMD shows positive distributions and are symmetric under $x \leftrightarrow (1-x)$ for both real and virtual photon. This is agreement with the BLFQ results  also. The $g_{1L}(x,\textbf{k}^2_{\perp})$ TMD shows some negative distribution at some small range of $x$ and $\textbf{k}^2_{\perp}$. This kind of behavior was also seen in the BLFQ model. We have also observed that the time-like virtual photons have higher distributions than the real and space-like virtual photon. Since no other work except BLFQ has been reported for the case of photon TMDs, it is not possible to compare our result with other models.

\par  The unpolarized $f_1(x)$ PDF of our calculation is found to be exactly overlapping with other model predictions. The collinear photon PDFs however, have been extracted for the first time in our work. The PDFs for space-like virtual photon have higher distributions when compared with the real and time-like virtual photon. The unpolarized $f_1(x)$ PDF shows peak distributions at minimum (maximum) value of $x=0(1)$ and have minimum distributions at $x=0.5$ for all the photons. The longitudinal $g_1(x)$ PDF has both positive and negative distributions at $x>0.5$ and $x<0.5$ respectively. It vanishes at $x=0.5$ for all the photon. The transverse $h_1(x)$ PDF shows negative distributions for all.

\par The spin densities in momentum space have also been discussed for real photon in both longitudinal and transverse case. The spin densities ware found to be symmetric and concentric about the origin. The transverse spin densities of transversely quark spin and longitudinally photon spin structure have some distortion along $k_{x}$ direction due to the interference of $f_1$ and $h^\perp_{1L}$ TMDs. However, no distortion has been seen for $\rho_{\uparrow \Uparrow (\uparrow \Downarrow)}$ spin densities.

\par This work is very essential to study the fundamental structure of photon. In addition, exclusive vector meson generation in virtual photon-proton or photon-nucleus scattering can be studied using the virtual photon LFWFs. Since the photon LFWFs contain all the information on the photon structure, they may also be used to assess other observable, such spin-orbit correlations, Wigner distributions, and electromagnetic and gravitational form factors, that quantify the photon structure in QED.

\section{Acknowledgement}
NK and HD would like to thank Science and Engineering Research Board, Government of India under Teachers Associateship for Research Excellence Award (TARE), Project: Exploring the Internal Structure of Nucleon with Generalized Parton Distributions and Transverse Momentum Parton Distributions (Ref No. TAR/2021/000157) for financial support.


\end{document}